\documentclass[twocolumn,trackchanges,floatfix]{aastex63}

\newcommand{\hei}{{He \sc i}}
\newcommand{\heii}{{He \sc ii}}
\newcommand{\kms}{{$\rm km\ s^{-1}$}}
\graphicspath{{./}{figures/}}
\usepackage{float}
\usepackage{amssymb}
\usepackage{dirtytalk}
\usepackage{longtable}

\received{December 13, 2024}
\accepted{June 13, 2025}
\submitjournal{ApJ}

\shorttitle{SMC Field OB and OBe Binaries}
\shortauthors{Vargas-Salazar et al.}

\begin{document}

\title{New Field OB and OBe Binaries of the SMC Wing:  \\ 
Observational Properties and Population Modeling}

\correspondingauthor{Irene Vargas-Salazar}
\email{ivargasa@umich.edu}

\author[0000-0001-7046-6517]{Irene Vargas-Salazar}
\affiliation{University of Michigan, Department of Astronomy, 1085 South University Ave., Ann Arbor, MI 48109, USA}

\author[0000-0002-5808-1320]{M. S. Oey}
\affiliation{University of Michigan, Department of Astronomy, 1085 South University Ave., Ann Arbor, MI 48109, USA}

\author[0000-0002-1722-6343]{Jan J. Eldridge}
\affiliation{Department of Physics, The University of Auckland, Private Bag 92019, Auckland, New Zealand}

\author[0000-0002-7992-469X]{Drew Weisserman}
\affiliation{University of Michigan, Department of Astronomy, 1085 South University Ave., Ann Arbor, MI 48109, USA}
\affiliation{Department of Physics \& Astronomy, McMaster University, Hamilton, Ontario, L8S 4L8, Canada}

\author{Helen C. Januszewski}
\affiliation{University of Michigan, Department of Astronomy, 1085 South University Ave., Ann Arbor, MI 48109, USA}
\affiliation{Present address: Private}

\author[0000-0002-7733-4522]{Juliette C. Becker}
\affiliation{University of Michigan, Department of Astronomy, 1085 South University Ave., Ann Arbor, MI 48109, USA}
\affiliation{Present address:  Department of Astronomy, University of Wisconsin--Madison, Madison, WI 53706}

\author[0000-0001-5897-9221]{Stefano Zazzera}
\affiliation{University of Michigan, Department of Astronomy, 1085 South University Ave., Ann Arbor, MI 48109, USA}
\affiliation{Present address: Department of Physics \& Astronomy, Queen Mary University of London, Mile End Road, London E1 4NS, UK}

\author[0000-0003-0521-473X]{Norberto Castro}
\affiliation{Institut f\"ur Astrophysik und Geophysik, Friedrich-Hund-Platz 1, 37077 G\"ottingen, Germany}
\affiliation{Leibniz-Institut für Astrophysik An der Sternwarte, 16 D-14482, Potsdam, Germany}

\author[0000-0003-1647-3286]{Yongjung Kim}
\affiliation{School of Liberal Studies, Sejong University, 209 Neungdong-ro, Gwangjin-Gu, Seoul 05006, Republic of Korea}
\affiliation{Korea Astronomy and Space Science Institute, Daejeon 34055, Republic of Korea}

\author[0000-0001-5253-1338]{Kaitlin M. Kratter}
\affiliation{Department of Astronomy and Steward Observatory, University of Arizona, Tucson, AZ 85721, USA}

\author[0000-0002-3856-232X]{Mario Mateo}
\affiliation{University of Michigan, Department of Astronomy, 1085 South University Ave., Ann Arbor, MI 48109, USA}

\author[0000-0002-4272-263X]{John I. Bailey III}
\affiliation{Physics Department, Broida Hall, Santa Barbara, CA, 93106, USA}

\begin{abstract}

We present a radial velocity (RV) survey of the field OB and OBe stars of the SMC Wing. We use multi-epoch observations of 55 targets 
obtained with the Magellan IMACS and M2FS multi-object spectrographs
to identify single- and double-lined spectroscopic binaries.  We also use TESS light curves to identify new eclipsing binary  candidates. 
We find that 10 each of our 34 OB (29\%) and 21 OBe (48\%) stars are confirmed binaries, and at least $\sim$ 6 more are candidates.
Using our RV measurements, we set constraints on the companion masses, and in some cases, on
periods, eccentricities and inclinations. 
The RV data 
suggest
that OB binaries favor more circular orbits (mean eccentricity $\langle e\rangle = 0.08\pm 0.02$) while OBe binaries are eccentric ($\langle e\rangle = 0.45\pm 0.04$).
We identify 
2 candidate black hole binaries, [M2002] 77616, and 81941.
We use BPASS to predict the frequencies of ejected OB and OBe stars and binaries, assuming OBe stars are binary mass gainers ejected by the companion supernova.  We also predict the
frequencies of black-hole, neutron-star, and stripped-star companions, and we model the distributions of primary and secondary masses, periods, eccentricities, and velocity distributions.
The models are broadly consistent with the binary origin scenario for OBe stars, and predict an even larger number of post-supernova OB binaries.  Comparison with the kinematics supports a significant contribution from dynamical ejections for both OB and OBe stars, although less so for binaries.
\end{abstract}


\section{Introduction} \label{sec:intro}


Binarity among massive stars has profound consequences for stellar evolution and massive-star feedback effects.  Binary mass transfer is responsible for the exchange of mass and angular momentum that can generate a wide range of post-interaction products, including
rapid rotators, emission-line stars, merger products, X-ray binaries, stripped stars, and gravitational wave progenitor systems. 
Binarity also affects the stellar evolutionary end stages, and thus the frequency of different types of supernovae, gamma-ray bursters, and other explosive transients.
Due to the complexity of massive star evolution, it is 
vital
to observe and characterize the massive binary population to understand the relationship between these products and their progenitor binary systems.

Observed binary fractions for OB stars 
have lower limits of 
$\sim 50 - 60\%$ \citep{Sana2009,KiminkiKobulnicky2012,Kobulnicky2014,Banyard2022} in cluster environments. Observations have also determined that massive binaries are dominated by short periods ($P\lesssim20$ days), small eccentricities $e\lesssim0.4$, and modest mass ratios ($\langle q\rangle \approx 0.5$) \citep{MoDiStefano2017}. However, massive stars undergo a variety of interactions throughout their lifetime. This extends the 
range of these binary parameters 
and complicates their distributions.

Field OB stars offer important insight to the binary population affected by these interactions. Field stars comprise up to a third of the population of OB stars \citep{Oey2004,Gies1987} and are
are closely linked to binary interactions, being
primarily comprised of 
stars ejected from clusters
\citep{Oey2018,VargasSalazar2020} via two ejection mechanisms:
the binary supernova scenario (BSS) and the dynamical ejection scenario (DES) \citep{Hoogerwerf2000}. 
For BSS, a core-collapse supernova (SN) generates a recoil ``kick" to the companion, 
which, together with its orbital velocity, ejects it into the field \citep{Blaauw1961}.  
Classical OBe stars are now believed to generally correspond to post-interaction, BSS products \citep[e.g.,][]{Rocha2024,Dallas2022, DorigoJones2020, BoubertEvans2018,Shao2014}.
In this work, we consider only field stars, which are ejected from their parent clusters as runaways, which have space velocities $> 30$ \kms, and ``walkaways", which are unbound from their clusters at lower velocities.

An undisrupted binary ejected 
into the field
through the BSS process could include a 
neutron star or black hole companion, 
which may result in a a high-mass X-ray binary (HMXB). 
On the other hand, the DES mechanism 
originates from binary-binary 
or other multi-star
interactions \citep{Poveda1967,LeonardDuncan1988}. This is the only process that 
can
eject a noncompact binary.
These can be observed as
eclipsing binaries (EBs) or double-lined spectroscopic binaries (SB2s).
Additionally, binaries could also be ejected through a mechanism that combines
both DES and BSS, known as a two-step ejection \citep{Pflamm2010}. 
Since such binaries go through a dynamical ejection and then a SN, they may be also appear as post-interaction OBe stars. These types of ejections could be a significant subset of the BSS population \citep{DorigoJones2020, Grant2024}.

\cite{Grant2024} estimated 
the ratio of DES/BSS ejections into the field
for the massive stars of the SMC. This study found that runaways favored DES ejections over BSS ejections by a factor of $\sim 1.7$. However, there are several assumptions that go into this result. One of these is that they assume that all OB stars are DES objects and all OBe stars are BSS objects. This breakdown may be too simplistic since there could be OB objects that are accelerated through a SN kick, and OBe objects that are dynamically ejected.
Additionally, some OBe stars may have
acquired circumstellar disks through a means other than binary mass transfer, and thus would not
necessarily be BSS objects.
A more detailed examination of both DES and BSS ejections for both OB and OBe populations is necessary for a more 
accurate determination
of the DES vs BSS ratio.  
This in turn is needed to clarify the
parameters of the parent stellar populations in clusters.

Thus, the field OB population is fundamentally linked to the massive binary population via their origins as ejected stars.  
Obtaining the frequencies and binary parameters for field massive binaries 
is a difficult observational task, and only a
few studies have attempted to carry this out
\citep[e.g.,][]{Mason2009,Lamb2016}. 
In this work, we extend these efforts by 
focusing on more comprehensively extracting binary parameters for the SMC OB field star population, focusing primarily on the SMC Wing region.  
We carry out radial velocity monitoring of OB and OBe stars in this region to identify new binaries and set constraints on companion masses and orbital parameters.

In this work, we present comprehensive constraints on 
the binary fraction and binary parameters of the field massive stars of the SMC Wing. In Section \ref{sec:sample}, we detail our field star sample as well as our various methods used to identify new radial velocity (RV) binaries, EBs and SB2s. We also combine these results with previous identifications of RV, EB, SB2 and HMXBs to provide a more complete fraction of field massive binaries. In Section \ref{sec:binprop}, we 
set constraints on binary properties and in particular, companion masses. 
In Section \ref{sec:BPASS}, we present BPASS binary population synthesis models of the OB and OBe BSS field population that we use to compare with our observations. In Section \ref{sec:OBebinaries}, we discuss the characteristics of our observed BSS binaries and compare their eccentricities, velocity distributions and companions against our models. In Section \ref{sec:OBbinaries}, we discuss the characteristics of our observed DES binaries
and examine the observed branching ratio between DES and BSS binaries relative to that of the entire SMC OB population.

\begin{figure*}[ht!]
\begin{center}
\includegraphics[scale=0.5,angle=0]{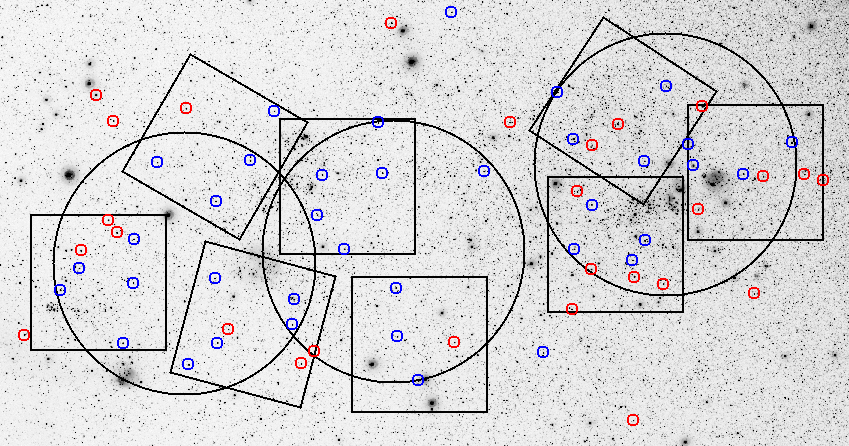}
\caption{
Green (513 nm) image of the SMC Wing region 
from the Magellanic Clouds Emission-Line Survey \citep[e.g.,][]{Paredes2015},
showing our IMACS fields (squares) and M2FS fields (large circles).  Targets indicated by small blue and red circles correspond to OB and OBe stars, respectively.  For reference, the IMACS fields are $15\arcmin.46\times 15\arcmin.46$; north is up and east to the left.
\label{fig:FindingChart}
}
\end{center}
\end{figure*}

\begin{figure*}[ht!]
\begin{center}
\gridline{
\includegraphics[scale=0.4,angle=0]{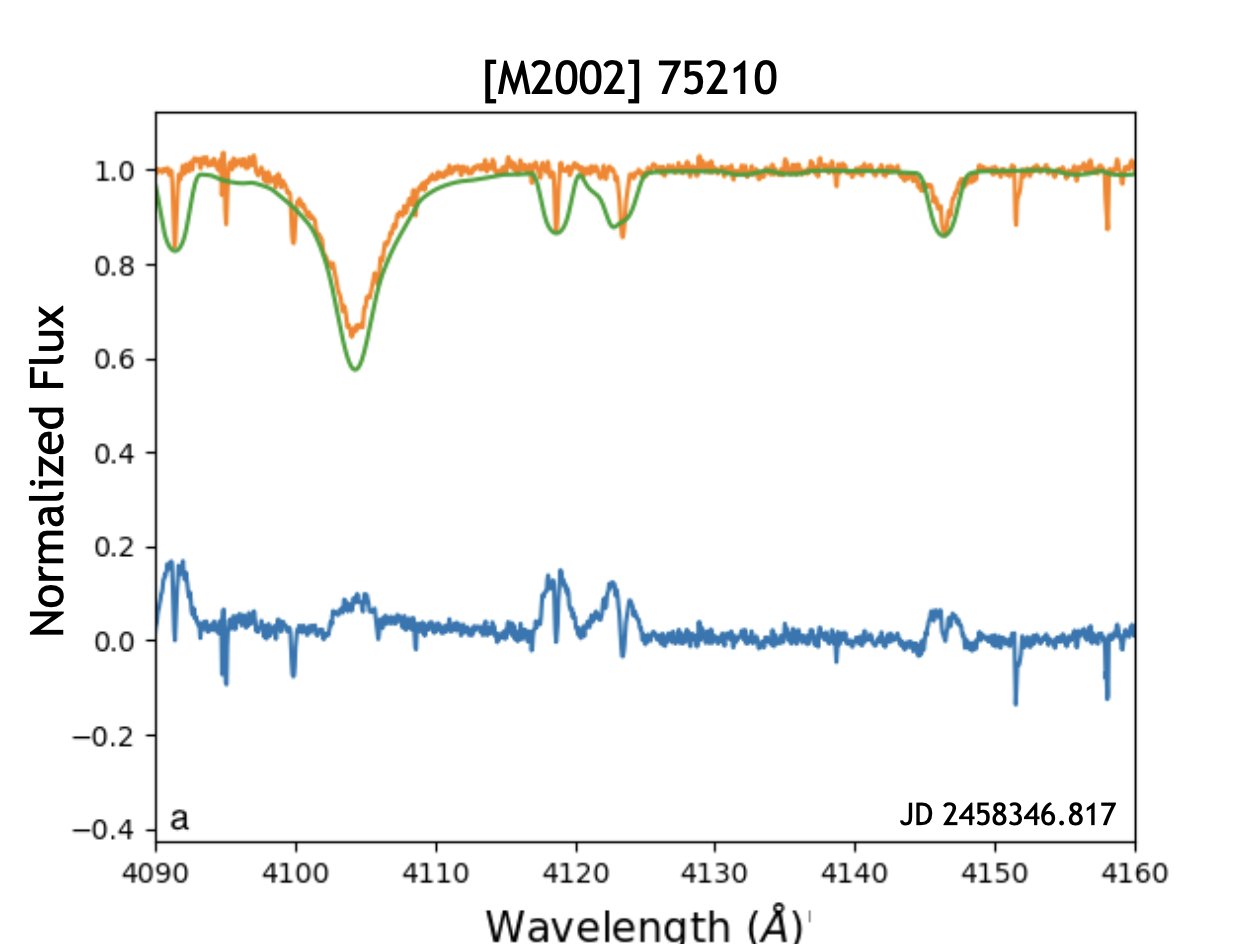}
\includegraphics[scale=0.401,angle=0]{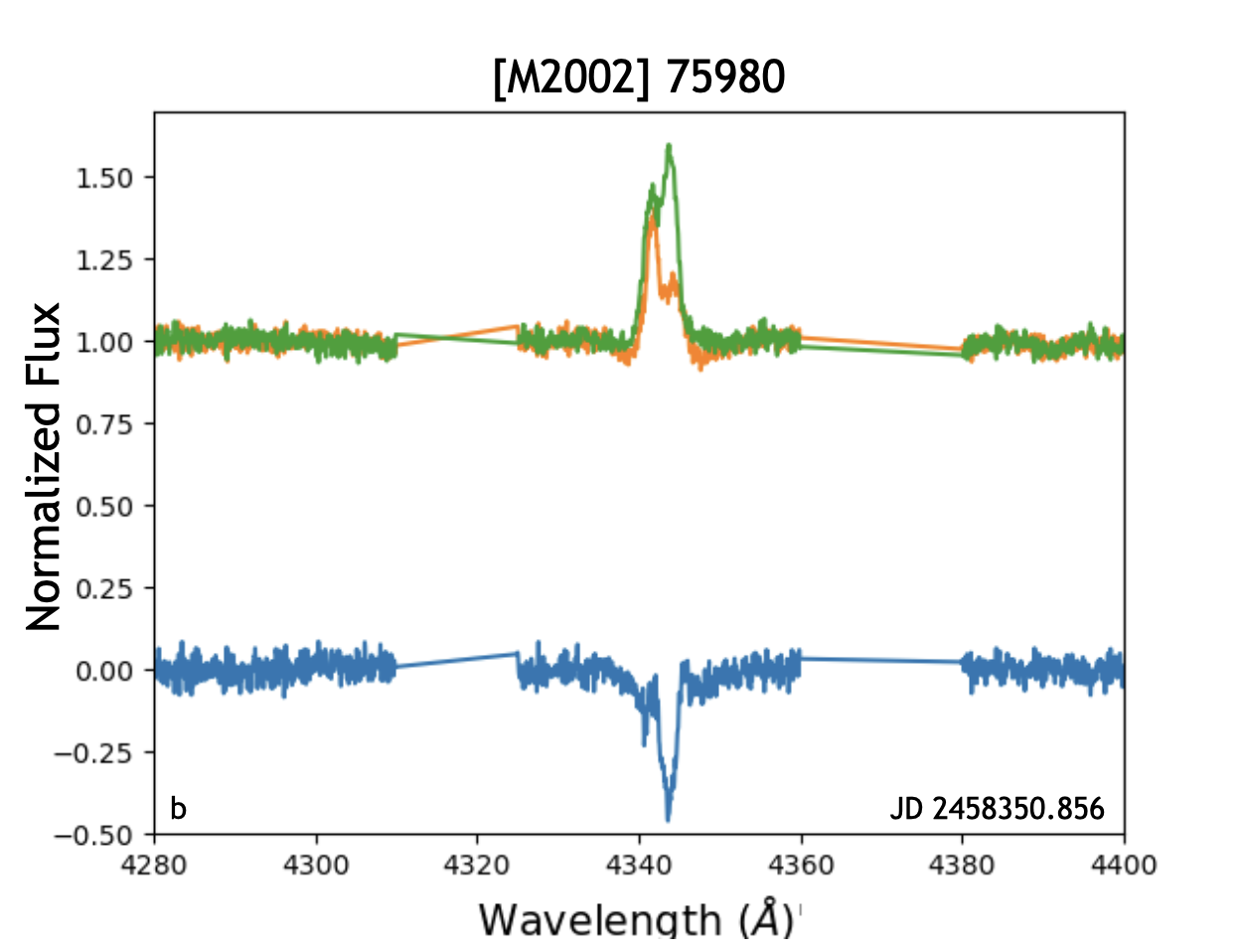}
}
\gridline{
\includegraphics[scale=0.4,angle=0]{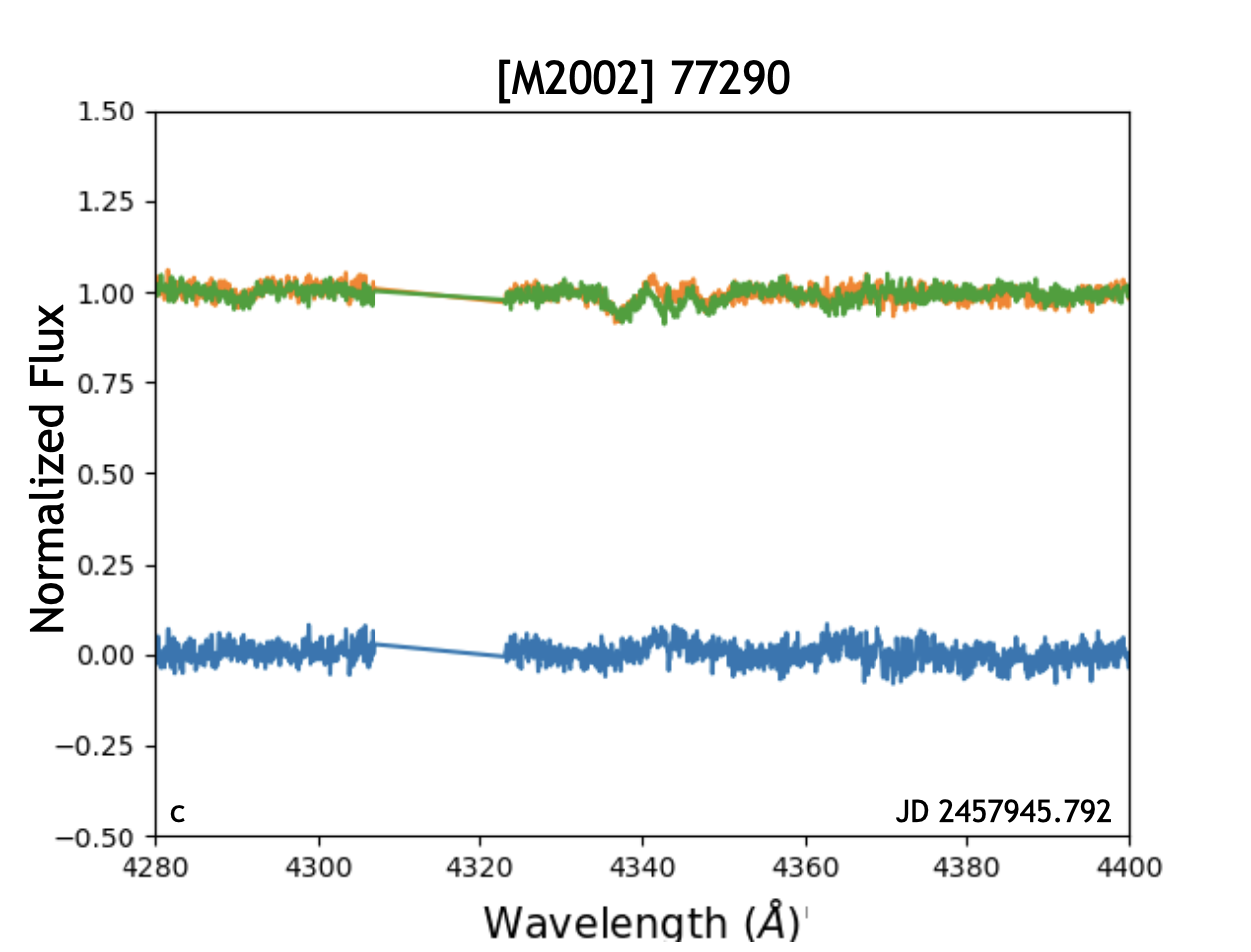}
\includegraphics[scale=0.405,angle=0]{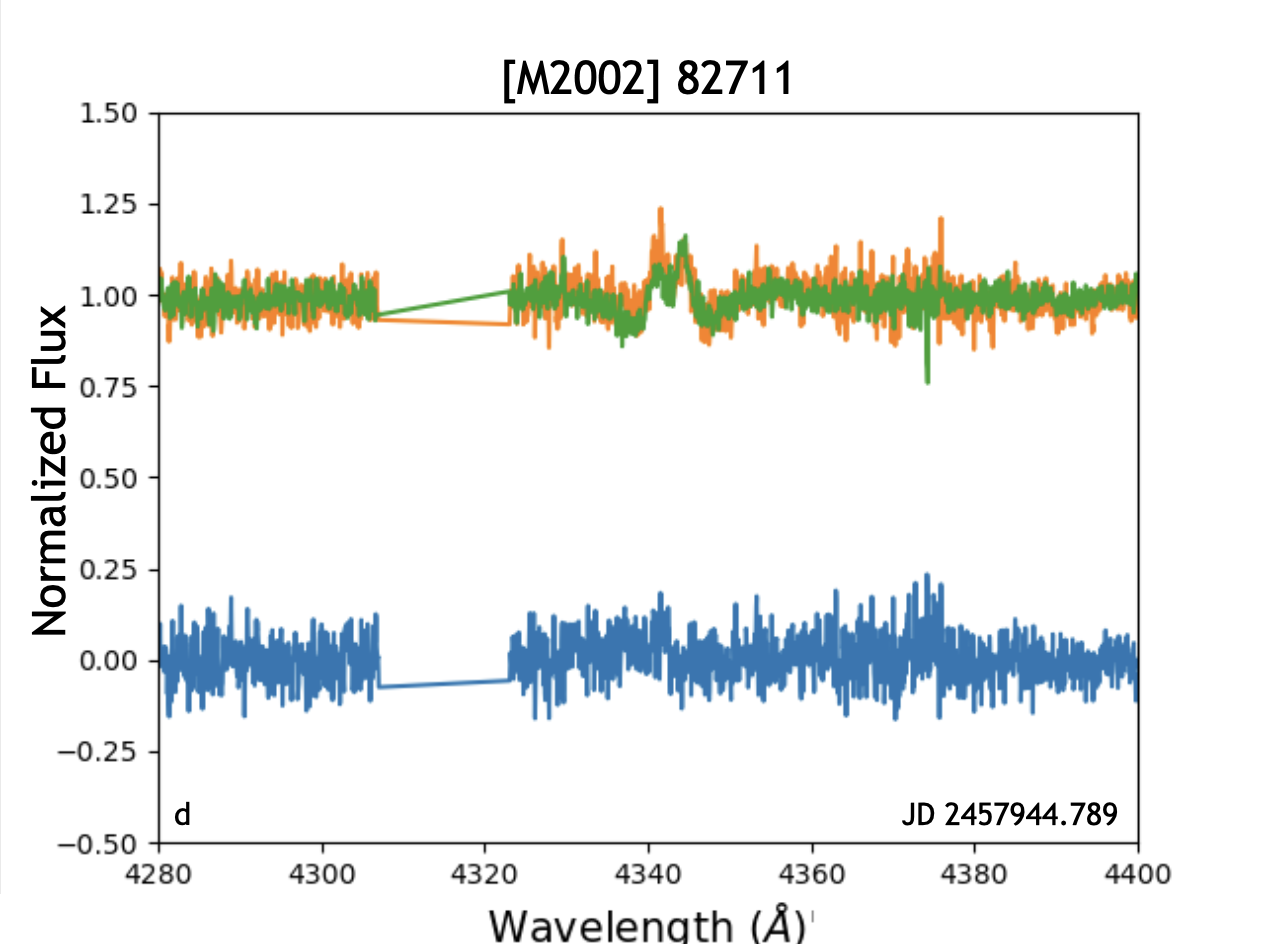}
}
\caption{
Cross-correlation fits to obtain RVs for a normal OB star 
in the top left panel and OBe stars in the remaining panels. 
The data and template are shown in orange and green, respectively, and
the lower, blue line shows the residuals.
The template for the OB star is a synthetic PoWR spectrum, and those for the OBe stars are
the observations of those objects
with the highest signal-to-noise.
\label{fig:specfit}}
\end{center}
\end{figure*}

\section{Field Binary Fraction} \label{sec:sample}

Previous observations yield a binary fraction for field stars in the Galaxy of 51\% \citep{Mason2009}, which is similar to that of clusters \citep[$\sim 50 - 60 \%$;][]{Sana2009,KiminkiKobulnicky2012,Kobulnicky2014,Banyard2022}.  As part of the Runaways and Isolated O-Type Star Spectroscopic Survey of the SMC (RIOTS4), \citet{Lamb2016} carried out radial velocity monitoring on a subset of field stars in the SMC Bar, and obtained a
binary fraction of $60\%$, also quite consistent with the data for the Galactic field.  
Here, we extend this work to the SMC Wing region, again
obtaining targets from RIOTS4.  That survey provides a uniform, statistically complete sample of 379 SMC field OB stars that are at least 28 pc away from any other OB stars.
It is, in turn, a subset of the
\citet[][hereafter OKP]{Oey2004} survey of SMC OB stars, which generally corresponds to objects with
masses $\gtrsim 10 M_{\odot}$
and spectral types earlier than about B0 V and B0.5 I. 
The RIOTS4 sample represents $\sim 28 \%$ of the total SMC OB population.

Over the 5-year period of the RIOTS4 survey, \cite{Lamb2016} obtained spectroscopic monitoring of 29 objects in the SMC Bar to derive an initial estimate of the binary fraction of the SMC field massive stars 
from radial velocity (RV) variability. These observations were carried out in three multi-slit fields of the Inamori-Magellan Aerial Camera \& Spectrograph \citep[IMACS;][]{IMACS}. Each of these fields was observed for 9 -- 10 epochs at intervals of days, weeks, months and years apart. \cite{Lamb2016} 
focused their RV binary analysis on the 17 non-OBe targets,
obtaining a lower limit to the field OB
binary fraction of 60\%.  
They did not consider the OBe stars, and their sample consisted solely of RV binaries.

In this paper, we expand on this work by identifying the binaries among the 55 RIOTS4 targets 
of the SMC Wing. This is a separate, lower-density region of the SMC that is distinct from the SMC Bar.
Our sample includes not just OB targets, but also OBe stars. We again obtain multi-epoch spectroscopic observations 
to identify RV binaries using methods from \cite{Lamb2016}.
We furthermore
also identify new EB and SB2 candidates. 
Together with 
previously confirmed EBs, SB2s and HMXBs, we obtain a more complete estimate of the binary frequency in the SMC Wing.
We then use all of these data to set constraints on the masses and eccentricities of the binary companions for identified binaries and candidates.  We discuss the implications of our results on the nature of OB stars and OBe stars in the context of a binary population synthesis model and links to supernova versus dynamical ejection mechanisms.

\begin{deluxetable*}{ccccccccc}
\tablecaption{RIOTS4 OB Wing Binaries\label{table:OBbinaryresults}}
\tablewidth{700pt}
\tabletypesize{\scriptsize}
\tablehead{
\colhead{Massey ID} & \colhead{SpT \tablenotemark{a}} & 
\colhead{Classification \tablenotemark{b}} & \colhead{RV Measurements} & 
\colhead{RV Method \tablenotemark{c}} & \colhead{$M_1$ \tablenotemark{d}} & 
\colhead{$M_{\rm 2,min}$} & \colhead{$M_{\rm 2,max}$ \tablenotemark{e}} & 
\colhead{Period}  \\ 
\colhead{[M2002] SMC} & \colhead{} & \colhead{} & \colhead{$N$} & 
\colhead{} & \colhead{$M_{\odot}$} & \colhead{$M_{\odot}$} &
\colhead{$M_{\odot}$} & \colhead{(days)}
} 
\startdata
73913 & O9.5 I & , , ,R & 10 & 3 & 34.2 & 1.21 & 5.80 & $\lesssim 40$\\
75210 & O8.5 V& , , ,Rc& 5 & 2 & 29.3 & 0.76 & 3.80 & $\lesssim 50$ \\
76253 & B0.2 V + B& , ,S,& 8 & \nodata & 13.4 & 2.70 & \nodata & \nodata \\
76371  & O9.5 III & , , ,R &9 & 3 & 26.0 & 1.23 & 4.20 \tablenotemark{f}& $\lesssim 16$ \\
77368 & O6 V &  , ,Sc,R & 9 & 3 & 39.3 & 1.20 & 3.90 & $< 16$ \\
77734 & B1-3 II & , , ,Rc &6 & 1 & 14.5 $\ast$& 0.51 & 6.10 & \nodata \\
77816 & B0.2 III & , , Sc,Rc &5 & 2 & 18.7 & 0.27 & 3.50 & $\lesssim 100$ \\
80573 & B1 V & , , ,Rc &6 & 3 & 15.0 & 1.20 & 5.60 \tablenotemark{g}& $\lesssim 40$ \\
81258 & B0-1.5 V &,E,Sc,R& 6 & 3 & 11.9 & 3.63 & 5.55 & 2.70 \tablenotemark{h} \\
81646 & O8 V & , , ,R &13 & 3 & 28.0 & 3.41 & 3.70 & $\lesssim 1$ \\
81696 & B1 V $\star$& , , ,R& 10 & 3 & 15.0 $\ast$ & 3.06 & 4.00 & $\lesssim 1$ \\
81941 & O9.5 III & , , , R &10 & 3 & 25.0 & 5.29 & 24.9\tablenotemark{i} & \nodata \\
82444 & B0 V & , , ,Rc& 9 & 1 & 16.9 & 0.62 & 3.20 \tablenotemark{g}& $\lesssim 50$ \\
83073 & B0.7 V & , , Sc, &6 & \nodata & 13.1 & 3.10 & \nodata & \nodata \\
83232 & B1.5 III & ,Ec, ,R &6 & 3 & 12.8 & 1.32 & 2.81 & 1.68 \\
83510 & O8 V & , , ,R &9 & 3 & 20.9 & 1.85 &2.90 \tablenotemark{g}& $\lesssim 2$\\
\enddata
\tablenotetext{a}{From \cite{Grant2024}, except for those marked with a $\star$, which are from \cite{Dallas2022}.}
\tablenotetext{b}{``E", ``Ec", ``H", ``R", ``Rc", ``S", and ``Sc" indicate EB, EB candidate, HMXB, RV binary, RV binary candidate, SB2, and SB2 candidate, respectively (see text).}
\tablenotetext{c}{``1" indicates binary identification using the maximum $\Delta$RV method, ``2" indicates identification with the $F$-test, ``3" indicates identification from both methods}
\tablenotetext{d}{From \cite{DorigoJones2020}, except for those marked with $\ast$, which are from \cite{Dallas2022}. Masses from \cite{Dallas2022} may be the average of two values given in that paper; details
can be found in the corresponding target's section in the Appendix \ref{App:starnotes}.}
\tablenotetext{e}{
Upper limit of the secondary mass assuming it is not a compact object, obtained from our detection threshold.  For SB2 candidates, this value corresponds to the estimated mass of the possibly detected companion.
Details for each target are found in Appendix \ref{App:starnotes}.
}
\tablenotetext{f}{This star is a BSS candidate due to it being an eccentric binary. 
Thus $M_{\rm 2,max}$ is unconstrained 
if the companion is
a BH.}
\tablenotetext{g}{This star is a BSS candidate due to its being a fast rotator. 
Thus $M_{\rm 2,max}$ is unconstrained if the companion is a BH.}
\tablenotetext{h}{From the OGLE survey \citep{OGLE}.}
\tablenotemark{i}{This object may also have a BH companion (Appendix \ref{App:81941}), in which case $M_{\rm2,max}$ is unconstrained.}
\end{deluxetable*}

\subsection{RV Binaries} \label{sec:RVbin}
\subsubsection{Observations and RV measurements} \label{subsec:RVmeas}

To identify RV binaries, we obtain multi-epoch observations of our sample using two instruments on the 6.5-m Magellan telescopes at Las Campanas 
Observatory. Following \citet{Lamb2016}, we use
IMACS on the Baade telescope in f/4, multi-slit mode. 
We use 8 custom, multi-slit masks with slit widths of $0\arcsec.7$ or $1\arcsec.0$ and the 1200 mm$^{-1}$ grating to produce data with spectral resolution $R \sim 3000$ over a wavelength range of
4000 -- 4700 \AA. Our default exposure time is 
1 hour, taken in three exposures of 20 minutes each to achieve a S/N $> 30$ for our fainter targets. 
We also obtain observations in 3 fields with the Michigan/Magellan Fiber System (M2FS) at the Magellan Clay telescope \citep{M2FS}.  
With M2FS we used the HiRes spectral configuration that provides a resolution of $R \sim 20,000$ over a wavelength range of 
4050 -- 4450 \AA.   This made use of a custom interference filter to isolate echelle orders 80 -- 87.
IMACS observations were taken from 2016 June to 2017 July, and M2FS observations were obtained from 2016 June to 2018 August;
Figure~\ref{fig:FindingChart} shows the positions and targets covered by the IMACS and M2FS fields.

We reduce the IMACS spectra using the 
IMACS data reduction package 
COSMOS \citep{COSMOS1,COSMOS2},
which we use for
bias subtraction, flat-fielding, wavelength calibration, and extraction of 2D spectra. The IRAF\footnote{IRAF was developed by the National Optical Astronomy Observatory \citep{IRAF} and has been maintained by the IRAF community
since 2017.}
package \texttt{apextract} and 
other tasks
are used for the extraction and rectification of 1D spectra. We reduce the M2FS spectra primarily with IRAF. 
We use
the packages \texttt{imred} and \texttt{ccdred} 
for bias correction, and the 
\texttt{hydra} package for flat fielding, wavelength calibration, and spectral extraction.
The M2FS spectra are rectified using Python routines to select and fit low order
polynomials to the continuum (see \cite{Walker2015}).

\begin{deluxetable*}{ccccccccc}
\tablecaption{RIOTS4 OBe Wing Binaries\label{table:OBebinaryresults}}
\tablewidth{700pt}
\tabletypesize{\scriptsize}
\tablehead{
\colhead{Massey ID} & \colhead{SpT \tablenotemark{a}} & 
\colhead{Classification \tablenotemark{b}} & \colhead{RV Measurements} & 
\colhead{RV Method \tablenotemark{c}} & \colhead{$M_1$ \tablenotemark{d}} & 
\colhead{$M_{\rm 2,min}$} & \colhead{$M_{\rm 2,max}$ \tablenotemark{e}} & 
\colhead{Period}  \\ 
\colhead{[M2002] SMC} & \colhead{} & \colhead{} & \colhead{$N$} & 
\colhead{} & \colhead{$M_{\odot}$} & \colhead{$M_{\odot}$} &
\colhead{$M_{\odot}$} & \colhead{(days)}
} 
\startdata
71652 & B0.5$e_2$ & , , ,Rc & 4 & 1 & 16.1 & 0.71 & 5.40 & \nodata  \\
72535 & O8-9: IIIp$e_1$ & , ,Sc, & 8 & \nodata & 46.5 $\ast$& \nodata & 7.30 & \nodata  \\
73355 & B0$e_2$ & ,E, ,R &9 & 3 & 24.9 & 2.12 & 4.30 & 9.37 \tablenotemark{f} \\
75061 & B1$e_2$\ $\star$ & , , ,R &8 & 3 & 20.0 $\ast$ & 0.88 & \nodata & \nodata \\
75980 & B0$e_3$ & , , ,R & 7 & 3 & 26.9 & 1.10 & 9.20 & \nodata \\
76654 & B$e_3$ & , , ,Rc &5  & 2 & 17.6 & 0.53 & 7.70 & \nodata \\
76773 & Be & , , ,R &6 & 3 & 23.5 $\ast$ & 8.95 & \nodata & \nodata \\
77290 & B0.5$e_2$+ & , , ,R & 8 & 3 & 18.8 & 0.74 & 5.10 & \nodata \\
77458 & B0.2$e_1$ & H,,,R& 9 & 3 & 15.4 & 1.47 & 2.14 \tablenotemark{h} & 3.89 \\
77616 & O3-5p$e_3$pec & , , ,Rc & 4 & 1 & 50 \tablenotemark{g} & 1.00 & \nodata & \nodata \\
77851 & B0.2-1$e_3$+ &H, , , & 4  & \nodata & 23.4 & 1.50 & 3.00\tablenotemark{h} & \nodata \\
81465 & B$e_3$ & , , ,Rc &2 & 2 & 15.7 & 0.05 & 3.00 \tablenotemark{i} & \nodata \\
81634 & B1.5 V$e_3$ & , Ec, , & 2 & \nodata & 13.1 & \nodata & 9.30 & 0.99 \\
82328 & B0$e_2$ & , , ,Rc &9 & 1 & 19.5 & 0.57 & 4.20 & \nodata \\
82711 & B1 Ve & H, , ,R& 8 & 3 & 17.3 & 1.50 & 3.00 \tablenotemark{h} & $\lesssim 6$ \\
83171 & B0$e_2$ &  ,Ec, ,R &8 & 3 & 24.4 & 0.92 & 3.18 & 2.02 \\
83224 & B1$e_3$ & , ,Sc,R & 9 & 3 & 15.8 & 0.53 & 4.00 & $\lesssim 100$ \\
\enddata
\tablenotetext{a}{From \cite{Grant2024}, except for those marked with $\star$, which are from \cite{Dallas2022}.}
\tablenotetext{b}{``E", ``Ec", ``H", ``R", ``Rc", ``S", and ``Sc" indicate EB, EB candidate, HMXB, RV binary, RV binary candidate, SB2, and SB2 candidate, respectively (see text).}
\tablenotetext{c}{``1" indicates binary identification using the maximum $\Delta$RV method, ``2" indicates identification with the $F$-test, ``3" indicates identification by both methods}
\tablenotetext{d}{From \cite{DorigoJones2020}, except for those marked with $\ast$, which are from \cite{Dallas2022}. Masses from \cite{Dallas2022} 
may be the average of two values given in that paper; details can be found in the corresponding target's section in Appendix \ref{App:starnotes}.}
\tablenotetext{e}{
Upper limit of the secondary mass assuming it is not a compact object, obtained from our detection threshold.  For SB2 candidates, this value corresponds to the estimated mass of the possibly detected companion.
Details for each target are found in Appendix \ref{App:starnotes}.
}
\tablenotetext{f}{From the OGLE survey \citep{OGLE}.}
\tablenotetext{g}{From \cite{77616paper}.}
\tablenotetext{h}{These objects are believed to have NS companions,
therefore $M_{\rm 2,max}$ limits were calculated or set to NS mass limits
(see Appendices \ref{App:77458}, \ref{App:77851}, and \ref{App:82711}).}
\tablenotetext{i}{The $M_{\rm 2,max}$ value of this target is the value of when $P = 2000$ days.}
\end{deluxetable*}

For IMACS data, the RVs are obtained by fitting gaussians to all available spectral lines by least-squares fitting. 
We eliminate line measurements that have reduced $\chi^2 > 1$ and those whose gaussian amplitudes are $< 1.5 \sigma$ from the continuum. 
Individual line measurements 
are also discarded if they differ from the median RV by more than 35 \kms\ for a given spectrum. 

IMACS was known around this time to have a possible instability in wavelength calibration.
We therefore include a further analysis and correction procedure to minimize this effect.
Within each multi-epoch field, we identify non-varying RV standard stars whose systemic, median RV over all epochs is taken to be constant. These standard stars are defined to be those that have a standard deviation for all measurements $< 20$ \kms, based on the average IMACS errors on individual RV measurements ($\sim 13$ \kms). For each IMACS epoch, the difference between a given RV standard star's measured RV and its median RV over all epochs is applied as a correction to all the stars in that  
multi-object field.  On average, 
these offsets are around $\sim 19.5$ \kms. 
Final individual RV measurements for all targets are given in Appendix \ref{App:RVtable}.

For M2FS measurements, we use the cross-correlation code of
\citet{Becker2015}
to extract RV measurements
through a Markov Chain Monte Carlo (MCMC) analysis. This code performs the MCMC on cross-correlations of our observed data with synthetic spectral templates and 
gives the best-fit values for the RV.
We use spectral templates from the PoWR grid of stellar atmospheres \citep{PoWR} for spectral types ranging from early B stars to early O stars $(T_{\rm eff} = 15\rm\ kK - 50\ kK)$. We select the models at SMC metallicity that best match our data,
which are
those with high mass-loss rates and $\log g = 2.0\rm /(cm\ s^{-2}) - 4.4 /(cm\ s^{-2})$.
Additionally, we apply barycentric corrections to 
obtain heliocentric velocities.
An example of such a cross-correlation 
is shown
in Figure \ref{fig:specfit}a.

For fast rotators, the template spectra
are convolved with the measured $v\sin{i}$ \citep{DorigoJones2020} at the corresponding resolution to take into account both instrumental and rotational broadening.
For such targets with no measured $v\sin{i}$,
the value is estimated by eye using other stars with measured values as a guide. 
Finally, bad columns and other artifacts in the observed spectra are masked out.

The OBe-star emission lines pose a significant problem for the cross-correlations.
For \citet{Lesh1968} class $e_1$ targets, the emission is weak enough that the spectra can be fitted using the same procedure as for non-OBe targets. For class $e_2$ targets,  
the emission produces significant infill in
the H absorption lines, 
but the He absorption lines remain clean enough to be used in the fitting process. For these objects, we mask the observed spectra so that only the He lines are 
correlated against the model template. In class $e_3$ targets, the emission lines are even stronger, 
so that even the He lines cannot be used. 
For each of these targets, we select the spectral epoch with the best signal-to-noise and use this as the template spectrum instead of a PoWR model (Figure \ref{fig:specfit}b). This allows us to identify relative RV variations, but not the systemic velocities, 
which we obtain by averaging Gaussian fits to H$\delta$ and H$\gamma$ emission lines.

We caution that the substantial variability in OBe emission-line profiles makes our RV measurements more uncertain than for OB stars.  
We first test how well the measurements agree when obtained from model versus observed templates for the $e_1/e_2$ targets [M2002] SMC-77458 and 82328, which each have 5 M2FS observations.  They both show some emission in the Balmer line cores, but not above the continuum, and thus their default RV measurements are obtained by using the model OB star templates.  For stars with stronger OBe emission where only an observed template is used, our RV cross-correlations make use of only the H Balmer lines, since other features are often unseen in such stars.  In our test comparison, we therefore mask all features except the Balmer lines in the observations of the two targets.  We note that minimizing the available data in this way for the cross-correlations cause them to fail for 2 of the 5 observations of [M2002] SMC-82328.

For the remaining observations, we find average measured differences for $\rm RV_{obs} - RV_{model}$ of --6.4 \kms\ and --1.7 \kms\ for [M2002] SMC-77458 and 82328, respectively.  These can be compared to the respective average measurement errors of 10.1 \kms\ and 10.4 \kms\ obtained by combining the measurement errors for these average differences in quadrature.  These results demonstrate that the observed template works as well as the model template for objects with weak OBe emission.  This is the case even though only H$\gamma$ and H$\delta$ are used for the observed-template fits, whereas the model-based fits also include He absorption lines.  
We carry out further testing of the measured RVs for OBe stars in Section~\ref{subsec:RV}.

\subsubsection{Radial-Velocity Binaries} \label{subsec:RV}

\begin{figure*}[ht!]
\begin{center}
\gridline{
\includegraphics[scale=0.55,angle=0]{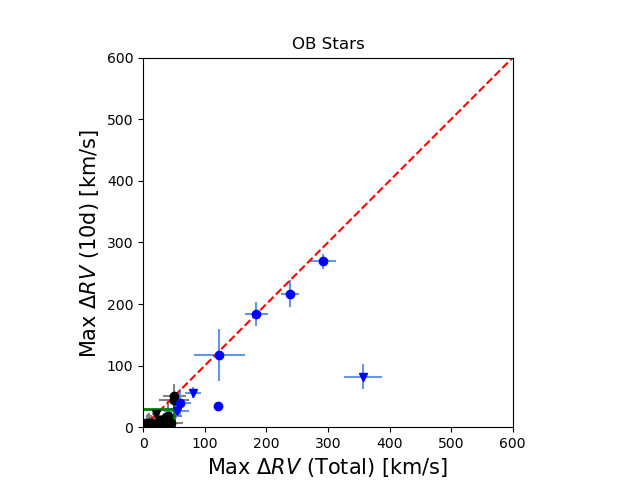}
\includegraphics[scale=0.55,angle=0]{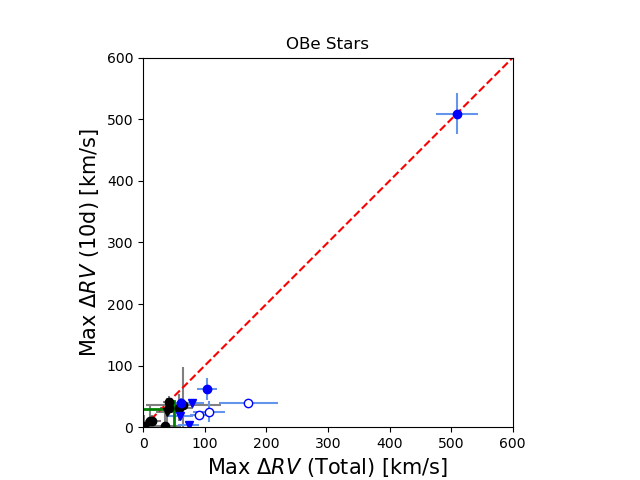}
}
\gridline{
\includegraphics[scale=0.55,angle=0]{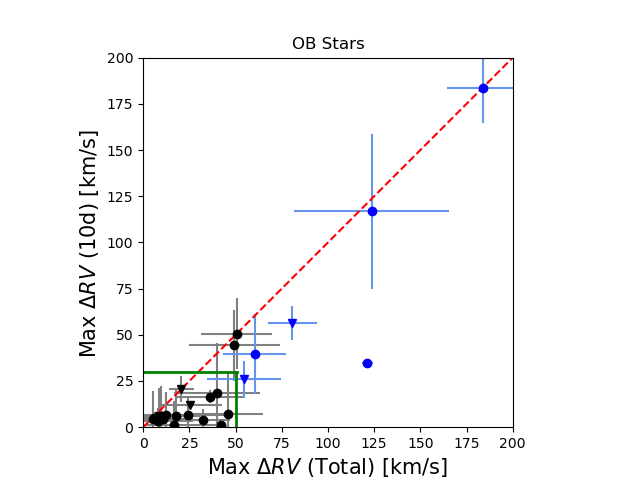}
\includegraphics[scale=0.55,angle=0]{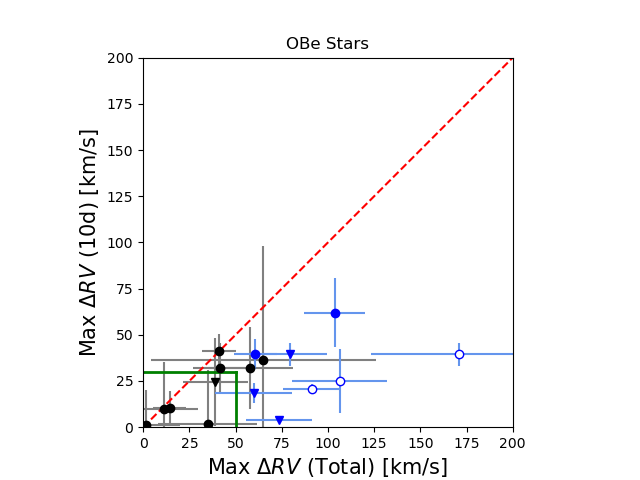}
}
\gridline{
\includegraphics[scale=0.55,angle=0]{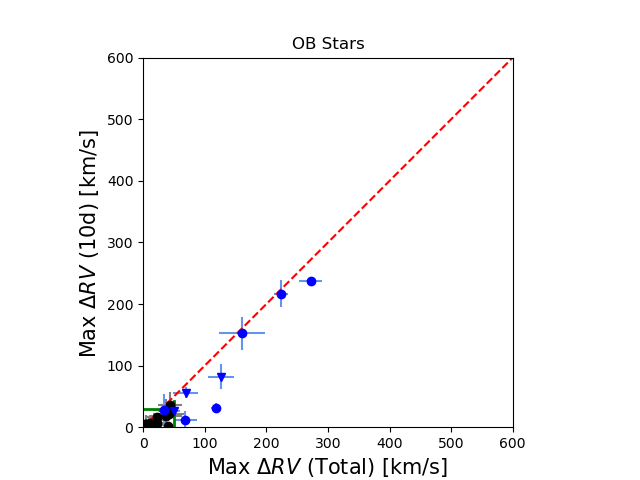}
\includegraphics[scale=0.55,angle=0]{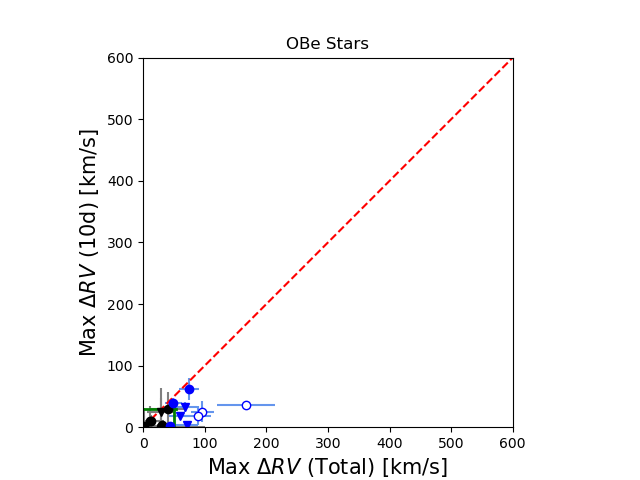}
}
\caption{
The largest absolute-value RV variation in \kms\ within 10 days vs within the entire survey, for OB stars (left panels) and OBe stars (right panels) of the SMC Wing.  The top plots show our full sample, the middle plots are a zoom of the top panels to 200 \kms, and the bottom plots are the same as the top but now with the epoch with the largest RV outlier removed for each target. 
Blue points indicate objects identified as binaries by RV both methods and
triangles indicate EBs and/or SB2s. The open circles are our most eccentric OBe binaries (see text).
The red dashed line is the locus where $\Delta$RV(10d) $=\Delta$RV(tot), which is more generally expected for binaries with circular orbits.
Targets with velocities within the green boundaries are considered to be non-detections for binary status.
 \label{fig:binaryID1}}
\end{center}
\end{figure*}

To identify single-lined spectroscopic binaries (SB1), which we refer to in this work as RV binaries, we use two of the same methods used by \cite{Lamb2016}. 
The first method, shown in Figure \ref{fig:binaryID1}, compares absolute values of 
the largest RV variation within 10 days, $\Delta$RV(10d), and the largest variation overall, $\Delta$RV(tot).  We use a 10-day interval since binaries whose periods are shorter than this tend to have circularized their orbits \citep{Renzo2024}. 
Objects are identified as binaries if $\Delta\rm RV(10d) > 30$ \kms, and/or if $\Delta\rm RV(tot) > 50$ \kms.  While, in general, our measurement uncertainties appear to be  substantially lower (Figure~\ref{fig:binaryID1}), we set these generous thresholds due to the IMACS field corrections for the possible wavelength calibration issue described above; these are more likely to affect longer-period intervals.  We also caution that stellar pulsations can also mimic radial velocity variations up to $\sim30$ \kms\ \citep{SimonDiaz2024}.

The
 middle
panels of Figure \ref{fig:binaryID1} show a zoom of the upper panels, and we see that the data appear well-behaved down to 10 -- 15 \kms.
We suggest that the clump of objects 
at the lowest $\delta$RV values
corresponds to true single objects.  Therefore, it may be likely that the number of binaries, especially for the OB stars, is significantly underestimated by this method.  We see at least 3 non-compact binaries (triangles) identified below the detection threshold region.

The error bars in Figure \ref{fig:binaryID1} are calculated as $\sigma = \sqrt{\sigma_1^2 + \sigma_2^2}$, where 
$\sigma_1$ and $\sigma_2$ are the individual errors from 
the two RV measurements used to calculate the plotted $\Delta$RV values.
These RV errors are calculated through the MCMC cross-correlation fit and take into consideration the standard deviation of the continuum of each observation which make up our
observational errors of our individual RV measurements.

{\it Figure~\ref{fig:binaryID1} 
suggests remarkable differences 
between the orbital behavior of 
the OB and OBe targets in our sample.} The majority of our OB binaries (left panels) 
show RVs consistent with being in
tight, circular orbits. 
Note that stochastic deviation from the one-to-one relation in the figure is still expected for circular orbits.  
The fitted slope to the data in the upper left panel of Figure~\ref{fig:binaryID1} is $e=0.08 \pm 0.02$, which we can take as their average eccentricities.

In contrast to the OB binaries, the 
OBe binaries in Figure~\ref{fig:binaryID1} (right panels) show a much stronger tendency toward eccentric orbits.
A least-squares fit to the data in the upper right panel yields an average
$e = 0.45 \pm 0.04$.  
As discussed below, this supports the interpretation that they are predominantly binaries that have survived their first SN explosions.
The middle
panels of Figure~\ref{fig:binaryID1} also show that OBe stars generally have greater RV variations suggesting both a higher binary frequency and tighter and/or higher-mass companions.
The bottom panels of the figure are the same as the top, but now calculating the $\Delta$RV values after dropping the epoch with the largest outlier RV measurement for each object.  We see that the trends for the two populations persist, showing that they are not based solely on singular data points for the individual stars. 
It may be unsurprising that the two OBe systems with $\Delta\rm RV \sim 500$ \kms\ 
in the top OBe panel 
show circular orbits since, as noted above, tight systems with short periods will quickly circularize. However, since these two systems are no longer found at these extreme values 
in the bottom panel, their $\Delta\rm RV \sim 500$ \kms\ are more likely spurious.

\begin{figure*}[ht!]
\begin{center}
\gridline{
\includegraphics[scale=0.5,angle=0]{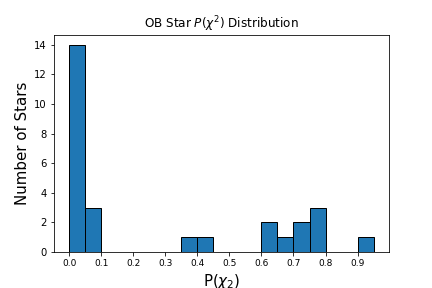}
\includegraphics[scale=0.5,angle=0]{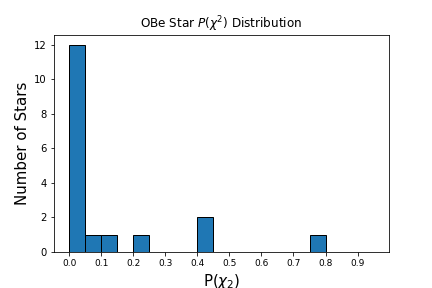}
}
\caption{
Distribution of probabilities 
that the observed RV variations for 
individual targets
are due to statistical noise. Targets in the first 
bin
($P(\chi_{2})<0.01$) are identified as binaries. \label{fig:binaryID2}}
\end{center}
\end{figure*}

The second method
to identify RV binaries
is known as the $F$-test which considers the observed variations and examines the probability that they are due to statistical noise. Following \cite{DuquennoyMayor1991}, we first calculate $\chi^2$, accounting for number of observations per target $n$, by:
\begin{equation} \label{eqn:Ftest_chi}
\chi^2 =  (n - 1) (\sigma_{\rm obs}/\sigma_{\rm avg})^2 ,
\end{equation}
where $\sigma_{\rm obs}$ is the standard deviation of the measured RVs and $\sigma_{\rm avg}$ is the mean of the statistical measurement errors associated with each radial velocity measurement.
Using the cumulative $\chi^2$ distribution given by:
\begin{equation} \label{eqn:Ftest_dist}
F_k (\chi^2) = G(k/2 , \chi^2/2) ,
\end{equation}
where $G$ is the regularized Gamma function for a given degree of freedom $k = n-1$, we can then calculate $P(\chi^2) = 1 - F_k (\chi^2)$.
Targets where this probability $P(\chi_{2}) < 0.01$ are identified as binaries
(Figure \ref{fig:binaryID2}), since their RV variations are not due to statistical noise.

\begin{figure}[ht!]
\begin{center}
\gridline{
\includegraphics[scale=0.55,angle=0]{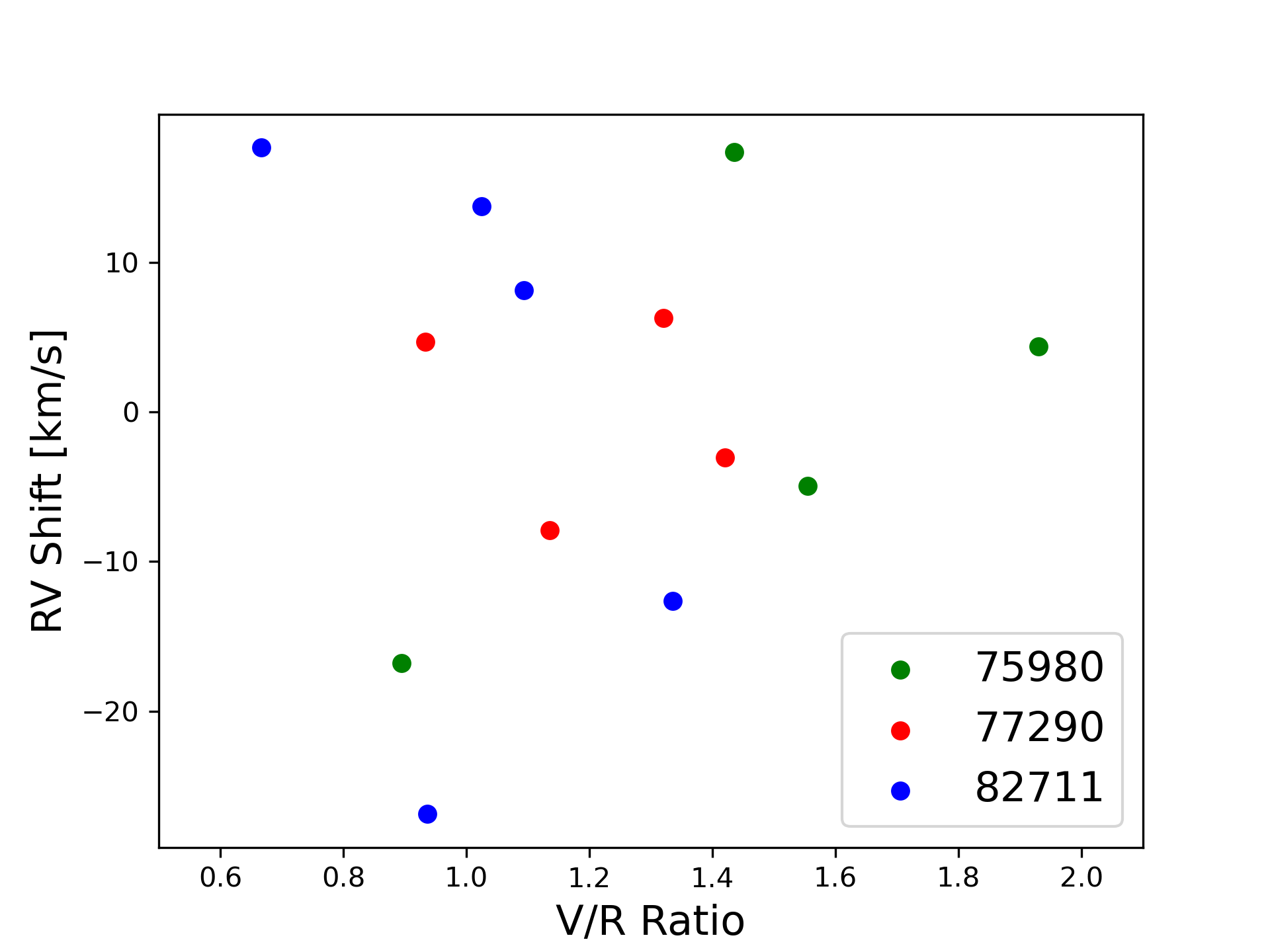}
}
\caption{
RV shift relative to the template spectrum versus average V/R ratio of H$\delta$ and H$\gamma$ emission for the M2FS observations of our three most eccentric OBe RV binaries when fitted as described in the text.  The data are normalized to the average RV shift for each object.
\label{fig:shift_data}}
\end{center}
\end{figure}

As noted above, we strongly caution that the RV measurements for the OBe stars may be affected by their emission-line variability in unknown ways.  
The trend seen in the OBe panels of Figure~\ref{fig:binaryID1} is  dominated by 3 stars that the figure suggests have the highest eccentricities ([M2002] SMC-75980, 77290, and 82711; marked with open circles).
Therefore, we further test the extent to which the Balmer variability affects the M2FS RV measurements for these stars, which are based on observed templates.  We mask all the spectral features except the two Balmer lines H$\gamma$ and H$\delta$ and we refit the RV.
Figure~\ref{fig:specfit} shows examples of the fit at H$\gamma$ for each of these three stars, in which the emission-line profiles vary significantly from the template. 
We then calculate the violet-to-red (V/R) ratios for the double-peaked emission seen in these stars, i.e., the ratio of the two peak fluxes.  This provides a measure of the extent to which the line profile is skewed toward low or high velocity.  As shown in Figure~\ref{fig:shift_data}, the measured RV relative to the template observation shows no obvious trend with V/R ratio, despite that fact that the latter varies significantly for all three stars.

Still, it is difficult to evaluate the effect of variable emission-line profiles for objects with stronger OBe emission.  We note that while variations in the emission-line core can be significant, there is often stability in other components such as the line wings and absorption components.  Much depends on which components are most affected by variable disk kinematics.  We do find that the Balmer emission cores often dominate the RV fit, and so further work is essential to understand the details of the disk kinematics and extent to which they affect the measured RVs.
Nevertheless, our limited analyses here and in the preceding section may suggest that OBe RV variabilities reported below could be real; however, they should be treated with caution.

Our identified binaries
are given in Tables \ref{table:OBbinaryresults} and \ref{table:OBebinaryresults}
for OB and OBe targets, respectively.
RV binaries are considered confirmed when 
identified as binaries using both of the methods 
above;
or when identified in the literature as an HMXB, EB, or SB2.
We stress that the identification of new OBe binaries that we regard as confirmed only by RV variability applies to only 4 objects out of the 14 OBe RV binaries in our sample; the remaining OBe binaries identified in this way are also confirmed by other methods.
There are 18 confirmed RV binaries and 10 candidates, for
a total of 28 in the SMC Wing. 
This yields a total frequency of $51 \pm 12 \%$ for confirmed and candidate binaries; and $33 \pm 9 \%$ if we consider only confirmed binaries
(Table \ref{table:Binfreq}).

\subsection{Eclipsing Binaries}\label{sec:TESSEB}

\begin{figure}[ht!]
\gridline{
\includegraphics[scale=0.3,angle=0]{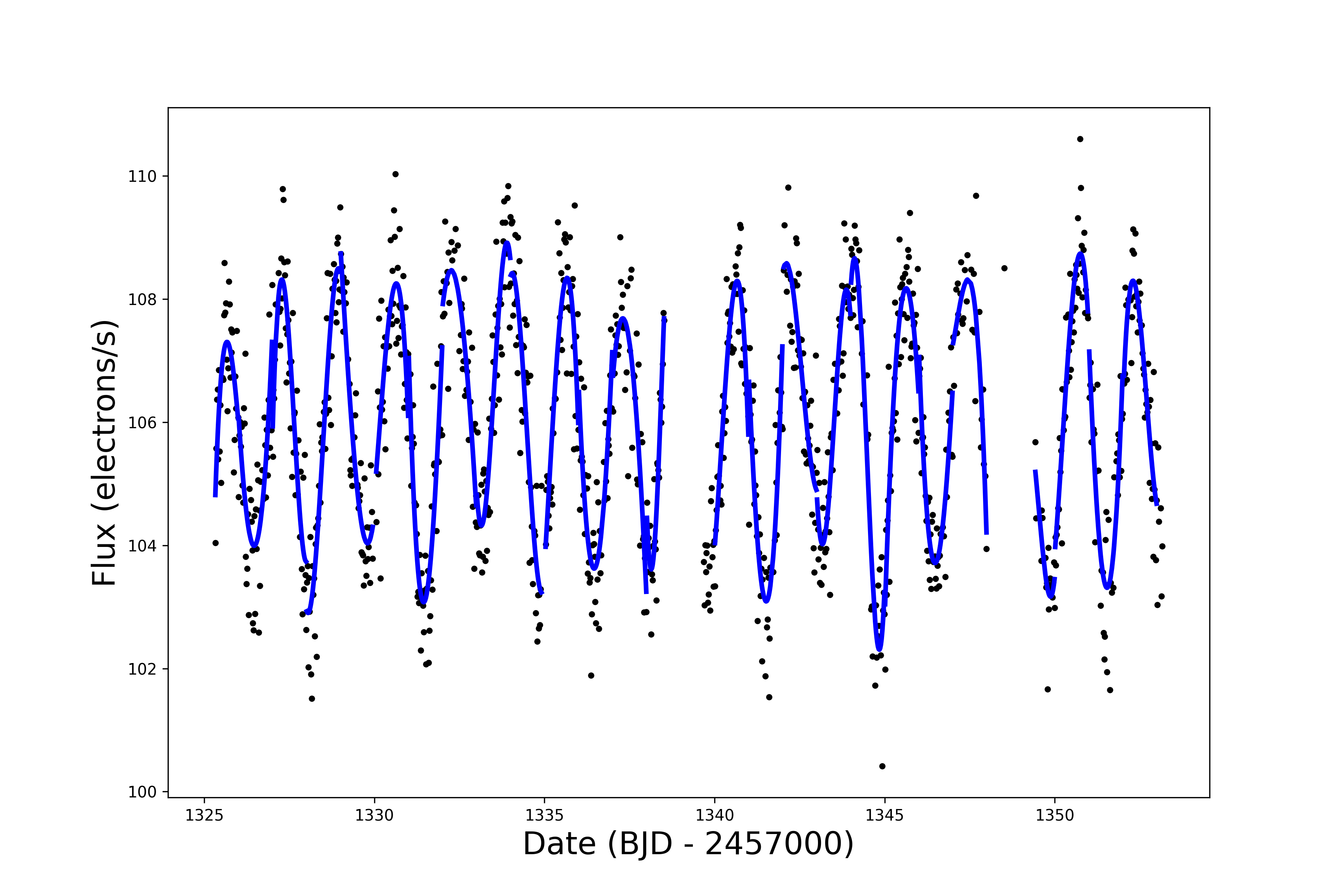}
}
\caption{
TESS light curve of Sector 1 for [M2002] SMC 83232, a source in the OKP sample showing variability that may suggest an EB candidate. The blue curve shows the polynomial fit to our data. 
\label{fig:TESS_R1}}
\end{figure}

Our first source for identifying EBs 
is the OGLE catalog \citep{OGLE}, 
which
provides identifications, light curves, and
measured periods.

We also identify new EBs using TESS \citep{TESS} from the full sample of OKP. 
Of the 1364 stars in the OKP sample, 444 are observed by TESS.
Light curves for these stars were accessed and obtained using the {\tt eleanor} module, which performs background subtraction and removes possible systematic effects on an orbit-by-orbit basis as described 
by \citet{FeinsteinMontet2019}. 
We used data from the TESS full-frame images, all obtained with a 30-minute cadence. Due to the large TESS angular resolution of $21\arcsec$/pixel, 11 
stars
are identified as having overlapping pixel apertures with other stars in our sample, thus contaminating their light curves; these stars are dropped from our sample. All available TESS data for each star were used up to, and including, TESS sector 48, corresponding to all TESS observations for these stars from 2018 July 25 and  2022 February 26.
The photometric precision for our targets, which generally 
have TESS magnitude between 13 and 15,
is on the order of $0.1\% - 1\%$ \citep{RickerWinn2015}.

\begin{figure*}[ht!]
\gridline{
\includegraphics[scale=0.225,angle=0]{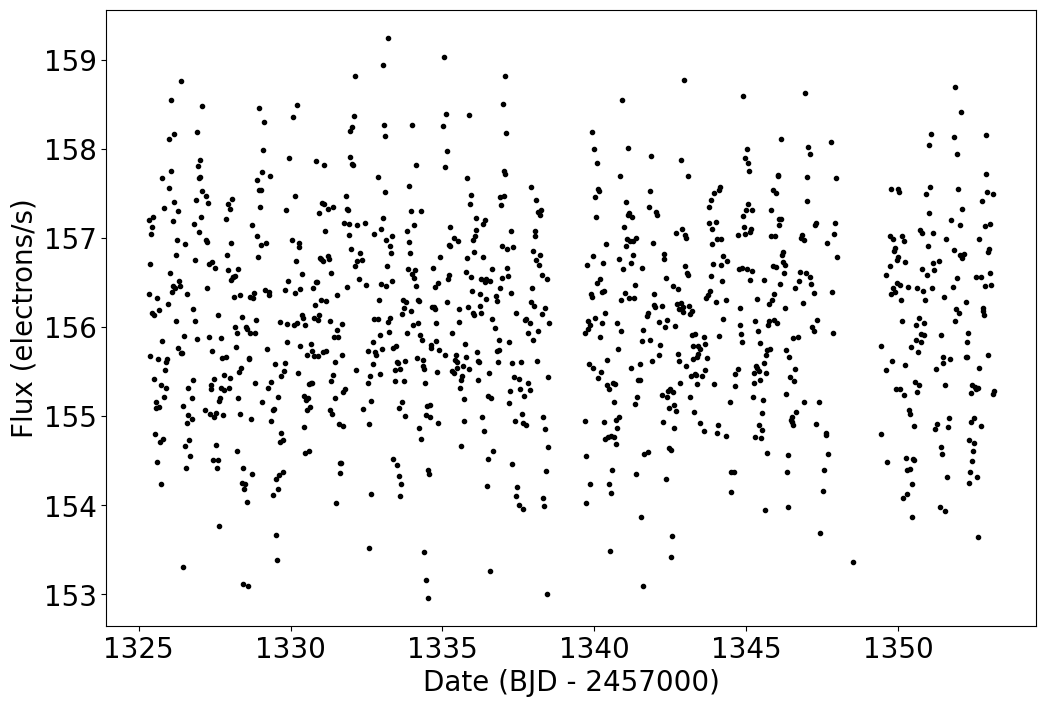}
\includegraphics[scale=0.31,angle=0]{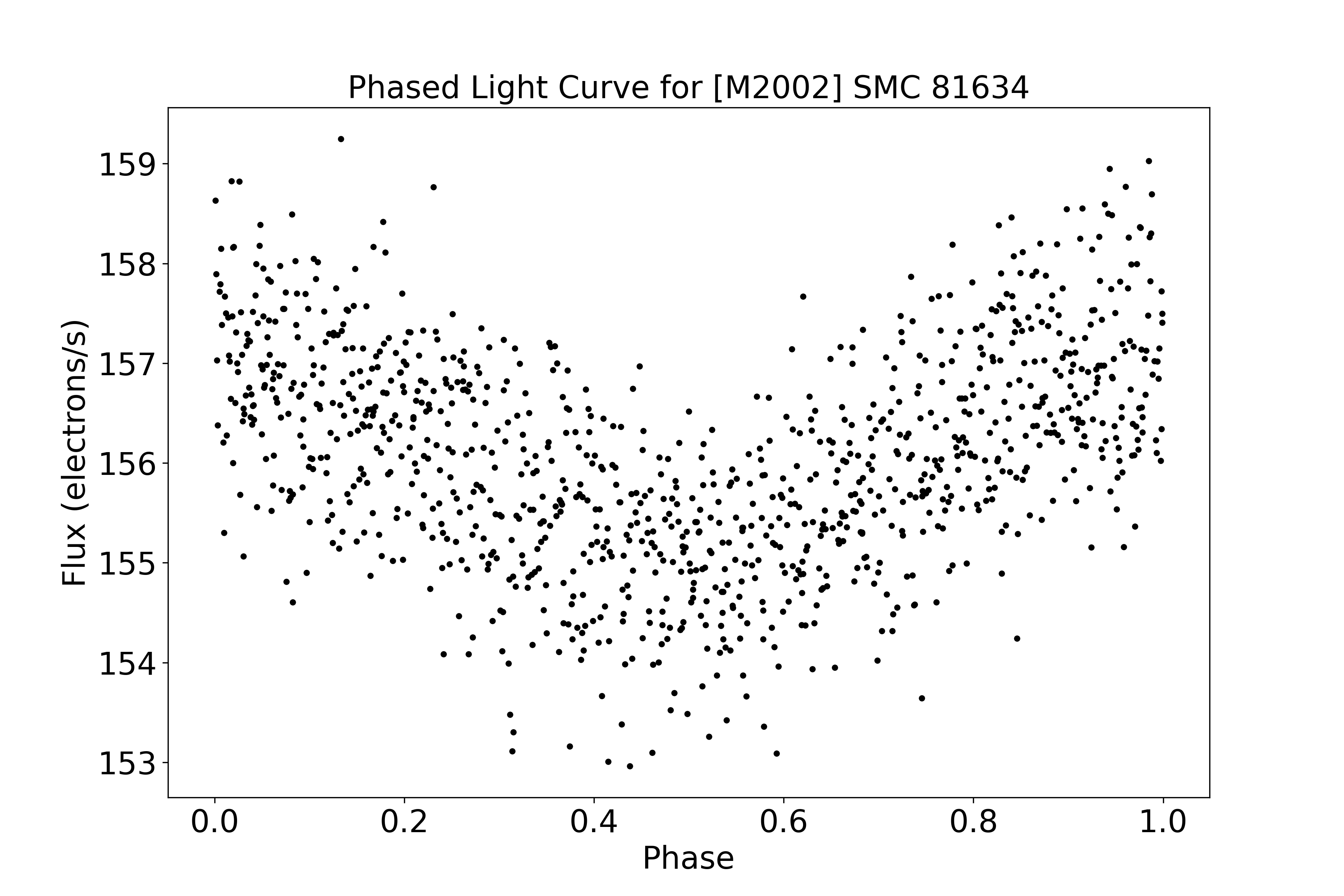}
}
\gridline{
\includegraphics[scale=0.225,angle=0]{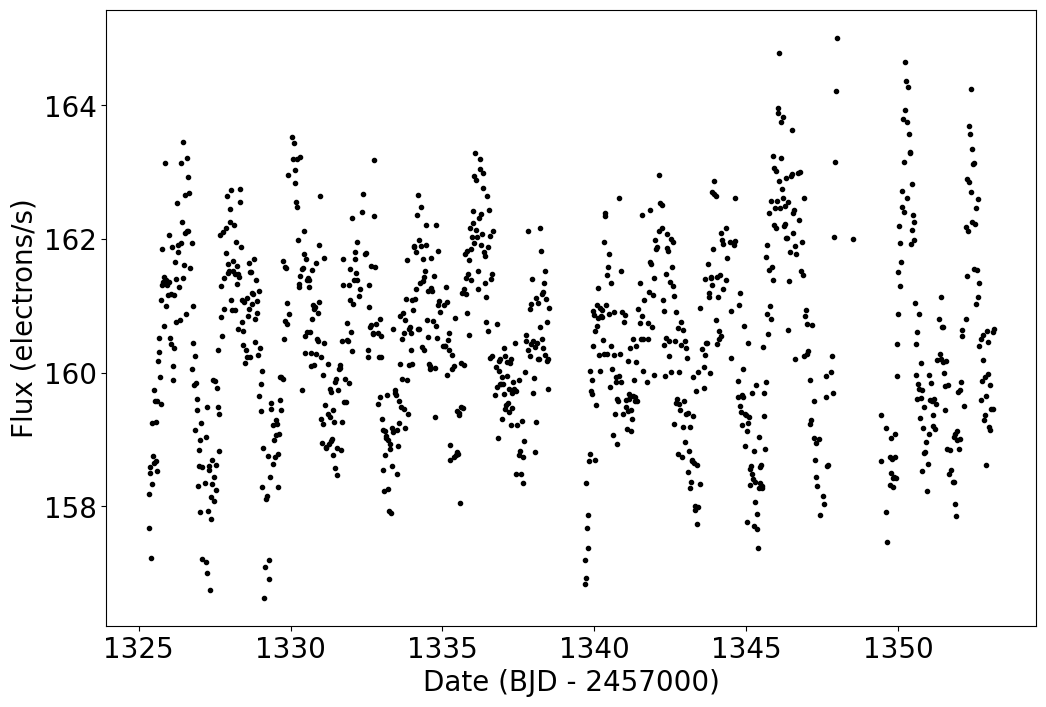}
\includegraphics[scale=0.31,angle=0]{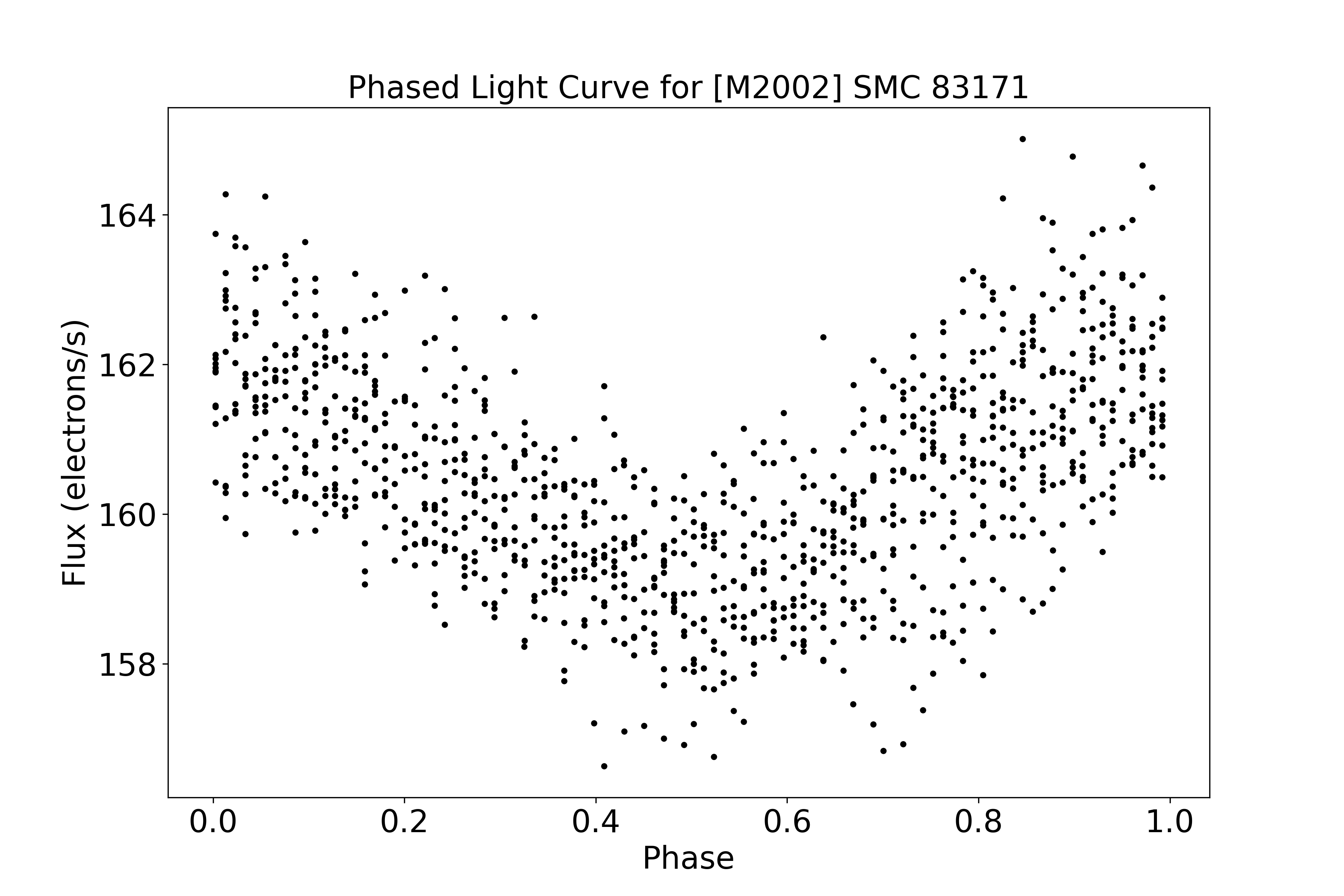}
}
\gridline{
\includegraphics[scale=0.225,angle=0]{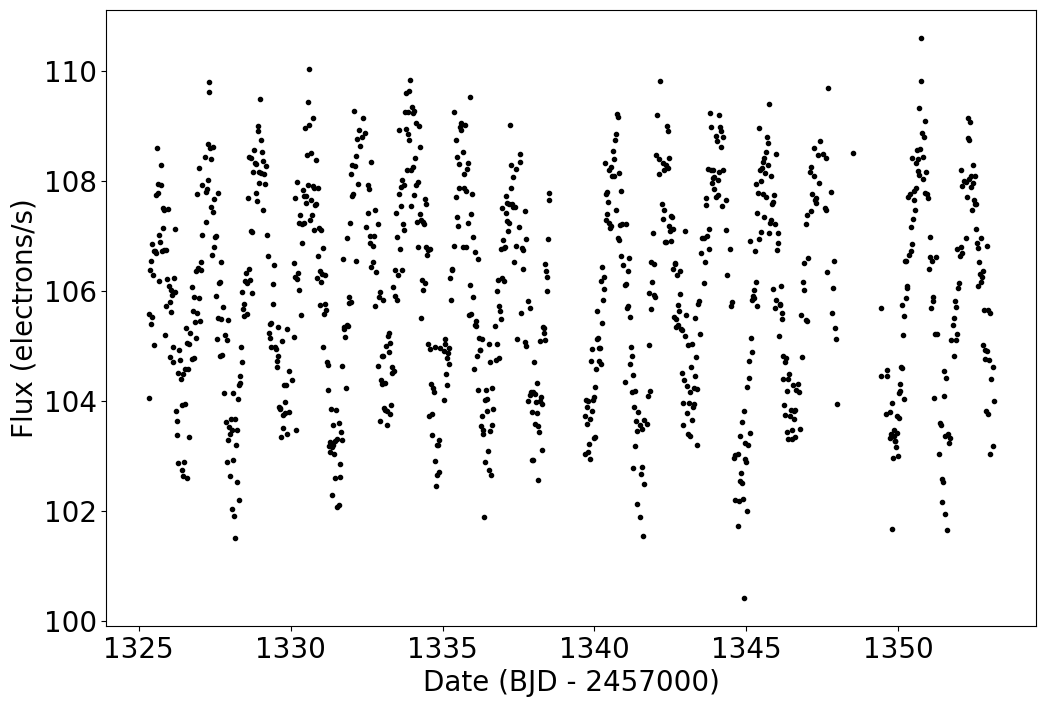}
\includegraphics[scale=0.31,angle=0]{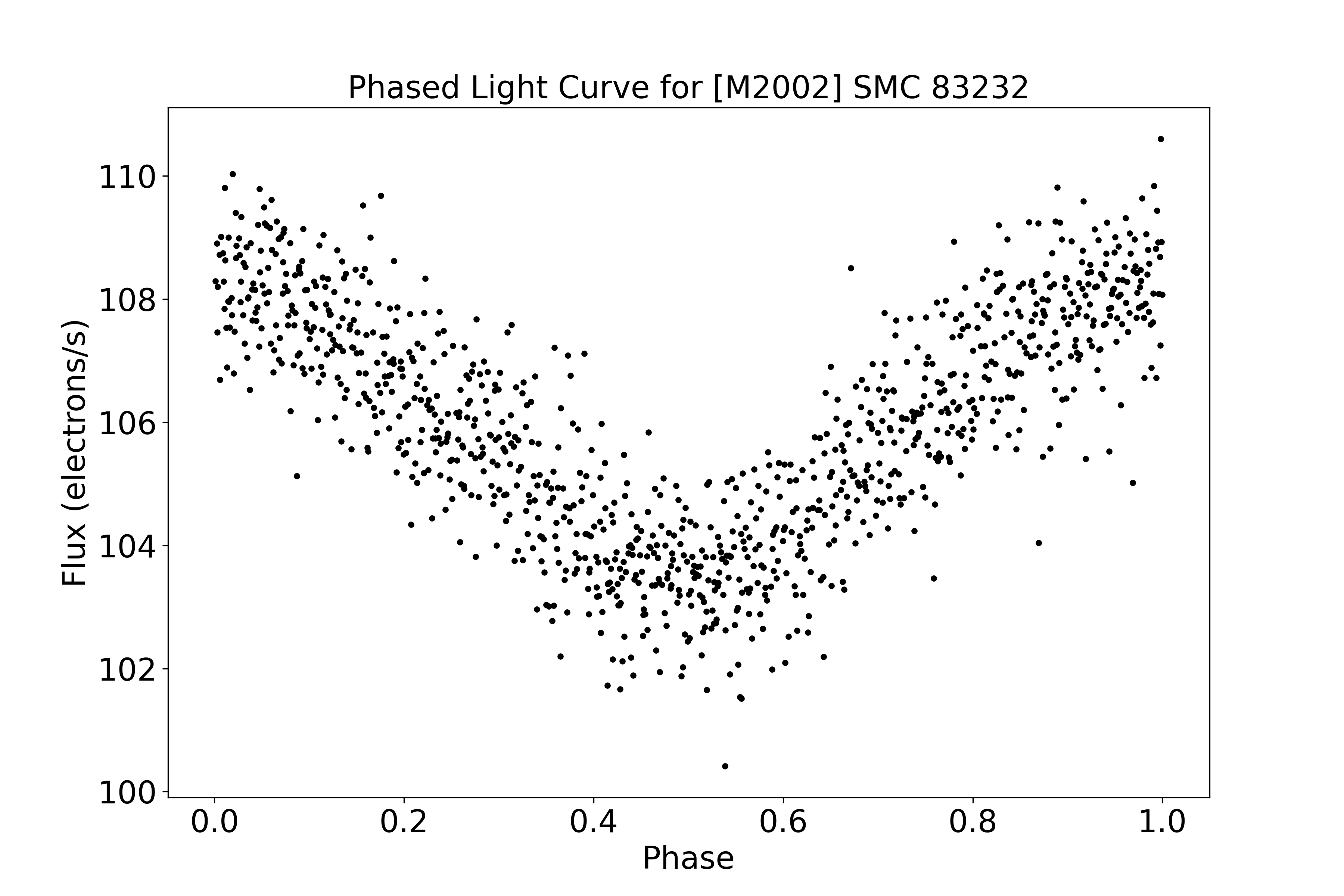}
}
\caption{The TESS light curves (left) and phased light curves (right) for our new EB candidates in the SMC Wing. From top to bottom: targets [M2002] SMC 81634, 83171 and 83232. \label{fig:bestEB}}
\end{figure*}

We search for photometric variability in the light curves 
by using two methods to compare photometric trends to the noise, as follows.
We first fit an analytic function to each light curve by dividing it into 1-day segments and fitting cubic polynomials to them (Figure~\ref{fig:TESS_R1}).
The light curve is then normalized by the analytic function, removing most of the large-scale photometric variation. We define a parameter $R_1$ to be the ratio of the standard deviation of the original light curve $\sigma_0$ to the average standard deviation of the normalized segments
$\langle\sigma_i\rangle$. $R_1 = \sigma_0~/\langle\sigma_i\rangle$ 
is thus a measure of the amplitude of large-scale variations to the noise.

The second method for evaluating photometric variability is based on smoothing the light curve by averaging over every 5 data points. The parameter $R_2$ is defined as the ratio of the standard deviation of the 
smoothed light curve $\sigma_5$ to the standard deviation of the original one $\sigma_0$.
For light curves whose variability is due only to
pure Gaussian noise, 
$R_2=\sigma_5/\sigma_0 = 
1/{\sqrt{5}}$, 
and for those whose variability is due only to real variation with
no noise, $R_2=1$. 
  
We then identified photometric variable stars to be those with $R_1 > 1.15$ or $R_2 > 0.65$.
The periodic variables are identified based on the 
periodograms of their light curves obtained by using the \texttt{lightkurve} package \citep{LightkurveCollaborationCardoso2018}: any light curves with peaks in their periodograms corresponding to a false alarm percentage below $1\%$ were identified as periodic.  
The periodograms are constrained to periods below the TESS orbital period of 14 days. 
Comparing our results to eclipsing binaries in our sample that were previously identified by the OGLE survey, 
we recover 23 of the 26 OGLE EBs.
For the 3 remaining targets, [M2002] 29468 has a period of 18.5 days, which is longer than what TESS can detect; and [M2002] 44984 and 65814 are too faint for detection by TESS. 

EB candidates are then identified from the sample of periodic stars by visual inspection of the light curves, which are shown in Figure \ref{fig:bestEB} for our new TESS candidate EBs in the SMC Wing.  
 We include in this category ellipsoidal variables like contact binaries and common-envelope objects which are not strictly true eclipsing binaries, and which may be suggested by the light curves in Figure~ \ref{fig:bestEB}.
None of the 3 identified objects
are especially strong EB candidates considering the shape and amplitudes of the light curves, 
which may be more consistent with stellar pulsation or starspots,
but a binary origin cannot be ruled out for these objects.
This leads to a total of 3 new
TESS EB candidates and 2 confirmed OGLE EBs for a total of 5 EBs in the Wing overall.  This yields a total frequency of $9 \pm 4 \%$ for confirmed  and candidate binaries, and $4 \pm 3 \%$ considering only confirmed binaries
(Table \ref{table:Binfreq}).

\subsection{Other Binaries} \label{sec:otherbinaries}

SB2s in our sample are identified by
\cite{Lamb2016}. 
In addition, our team identified 6 more SB2 candidates by visually comparing the spectra of each target taken at different epochs.  These new candidates are
[M2002] SMC-72535, 77368, 77816, 81258, 83073, and 83224.
The spectra of these objects suggest variations in Doppler shift corresponding to two different components;
the epochs and absorption-line features for which these shifts are seen are described in Appendix \ref{App:starnotes} for each target. 
We classify these objects as
SB2 candidates;
further observations and analysis \citep[e.g.,][]{Shenar2024} are required to confirm their status.

We identify 1 confirmed SB2 and a total of 6 SB2 candidates. This yields a total frequency of $13 \pm 5 \%$ for confirmed and candidate binaries, and $2 \pm 2 \%$ 
for only
confirmed binaries
(Table \ref{table:Binfreq}).

HMXB identifications are from the catalog of \cite{HaberlSturm2016}. 
Binary parameters such as the
period, orbital eccentricity, and nature of the companion are 
obtained from the literature and provided
in Appendix \ref{App:starnotes} for each target. 
There are 3 confirmed HMXBs in our SMC Wing sample, yielding a frequency of $5 \pm 3 \%$
(Table \ref{table:Binfreq}).

\subsection{Binary Fraction} \label{sec:binfrac}

Overall, we find that there are
20 confirmed binaries and 13 candidates, with a combined total of 33 
out of the 55 targets in the SMC Wing.
Candidates are labeled with a ``c" in Tables \ref{table:OBbinaryresults} and \ref{table:OBebinaryresults}. Some of these targets 
are identified via multiple methods, as
indicated in the third column of these tables. 

\begin{deluxetable*}{lcccccccc}
\tablecaption{RIOTS4 SMC Field Binary Frequencies\tablenotemark{a}
\label{table:Binfreq}}
\tablewidth{700pt}
\tabletypesize{\scriptsize}
\tablehead{
\colhead{Type } & \colhead{Wing Confirmed} & 
\colhead{Frequency\tablenotemark{b}} & \colhead{Wing Confirmed} & 
\colhead{Frequency\tablenotemark{b}} & \colhead{SMC Confirmed} &
\colhead{Frequency\tablenotemark{b,d}}  & \colhead{SMC Confirmed} &
\colhead{Frequency\tablenotemark{b,d}}\\
\colhead{ } & \colhead{+ Candidates} & 
\colhead{} & \colhead{Only} & 
\colhead{} & \colhead{+ Candidates\tablenotemark{d}} &
\colhead{}  & \colhead{Only\tablenotemark{d}} &
\colhead{}
} 
\startdata
RV  & 28 & $51 \pm 12 \%$& 18 & $33 \pm 9 \%$ & 43\tablenotemark{c} & $60 \pm 12 \%$\tablenotemark{c}& 26\tablenotemark{c} & $36 \pm 8 \%$\tablenotemark{c}\\
EB  & 5 & $9 \pm 4 \%$& 2 & $4 \pm 3 \%$  & 27 & $7 \pm 1 \%$ & 24 & $6 \pm 1 \%$\\
SB2  & 7 & $13 \pm 5 \%$& 1 & $2 \pm 2 \%$ & 19 & $5 \pm 1 \%$ & 13 & $3 \pm 1 \%$\\
HMXB  & 3 & $5 \pm 3 \%$& 3 & $5 \pm 3 \%$ & 25 & $7 \pm 1 \%$ & 25 & $7 \pm 1 \%$ \\
All\tablenotemark{e} & 33 & $60 \pm 13 \%$& 20 & $36 \pm 9 \%$ & \nodata & \nodata & \nodata & \nodata \\
\enddata
\tablenotetext{a}{Binary frequencies for the SMC Wing field stars and RIOTS4 SMC field stars.}
\tablenotetext{b}{Errors are derived from Poisson errors and error propagation. However, values should be considered lower limits.}
\tablenotetext{c}
{Determined from the sample in the SMC Bar spectroscopically monitored by \citet[][17 targets]{Lamb2016} and the SMC Wing fields (this work, 55 targets).  Thus the sample total is 72 targets instead of the total number of SMC field stars.
}
\tablenotemark{d}{Determined for entire RIOTS4 sample, with a total of 379 field stars, unless otherwise noted.}
\tablenotetext{e}{Unique binaries only.  Other rows include objects that may
belong to more than one category.}
\end{deluxetable*}

Our observed binary fraction in the Wing is $60\%$ if we include both confirmed and candidate binaries, and $36\%$ if we only consider confirmed binaries. 
We find an even breakdown between OB and OBe binaries in our sample: there are 10 confirmed OB binaries and 6 candidates, and 10 OBe binaries and 7 candidates.
As discussed in Section~\ref{subsec:RV}, our detection threshold for RV binaries may be overly conservative, in which case, the frequency of RV OB binaries could increase by $\gtrsim50$\%.
A breakdown of our binary frequencies is shown in Table \ref{table:Binfreq}, where the last row shows the total number of unique binaries identified. The remaining rows identify the total number of RV, EBs, SB2s and HMXBs, and these may include targets that are found in more than one of these categories. 
Table~\ref{table:Binfreq} also shows values combining with the \citet{Lamb2016} RV sample of 29 SMC field OB stars in the SMC Bar, and for the entire RIOTS4 field OB sample for SB2s, EBs and HMXBs.  

\section{Constraining Binary Properties} \label{sec:binprop}

For each RV binary,
our available
data allow us to set
some constraints on the plausible parameter space for period, orbital inclination, and eccentricity, 
ultimately allowing a rough 
constraint on the possible mass
of the companion star.
There are spectroscopic mass determinations of the target stars, mainly from \cite{DorigoJones2020} and the remainder from \cite{Dallas2022}; 
these values are adopted as the masses of the primary stars ($M_1$).  They are technically upper limits, since a secondary companion may contribute to the luminosity, but they serve as first-order estimates.

To constrain the companion star mass $M_2$,
we use the radial velocity semi-amplitude equation \citep{Fischer2014}:
\begin{equation} \label{eqn:Kepler}
(M_2\sin i)^3 =(M_1 + M_2)^2\frac{PK^3}{2\pi G}(1 - e^2)^{\frac{3}{2}} \quad ,
\end{equation}
where $M_1$ is the mass of the primary,
$i$ is the orbital inclination 
and $e$ is eccentricity. Our semi-amplitude
$K = \frac{1}{2}\Delta\rm RV(tot)$ 
as observed for each target star. 
While $\Delta$RV(tot) is unlikely to sample the actual maximum and minimum of a given RV curve, it 
is sufficient to provide a lower limit to our 
$M_2$ estimates as we know the true amplitude
cannot be less than
the measured value.  
We solve equation~\ref{eqn:Kepler} for a period range of 
0 -- 2000 days for each target.

We can obtain further constraints if additional information is available about the system.  In particular, the period is known for EBs and for some HMXBs.  For the latter, the companions are all known to be neutron stars.
Additionally, as seen in Section~\ref{sec:RVbin}, Figure \ref{fig:binaryID1} 
provides important
information about orbital eccentricities. 
We assume that RV binaries 
whose greatest RV variation within 10 days is also within $30\rm\ km/s$ of
their greatest variation overall (the red line of Figure~\ref{fig:binaryID1})
have $e = 0$ when carrying out our analysis. 
For other systems,
we allow $e$ to be a free parameter.

In the case of non-compact stellar companions, we can impose additional constraints on $M_2$ from our spectral observations. If we do not spectroscopically detect the photosphere of a companion star, the S/N of our observations set an upper limit to $M_2$. 
This is
estimated based on the temperature and luminosity of the primary component \citep{DorigoJones2020, Dallas2022}, and using the evolutionary tracks for SMC metallicity from \cite{Brott2011}. The age of the primary star is derived from the evolutionary tracks, and we assume that the secondary component is less massive but of similar age. By interpolating across the isochrone corresponding to the primary star's age on these evolutionary tracks, we obtain the luminosity of the secondary for various masses. We then match the observed $V$-band flux of the stars \citep{Massey2002} and the predicted flux of the secondary for each mass based on the luminosities from \cite{Brott2011}. To convert these luminosities to visual fluxes, we apply the same bolometric correction, SMC distance modulus, and extinction used by \cite{Lamb2016}.
The limiting value of $M_2$ then corresponds to the one that generates a visual flux corresponding to that at the detection limit for the object considering its S/N.

We caution that if mass transfer has occurred during the evolution of the binary system, the isochrone and expected evolutionary tracks described by \cite{Brott2011} may not be applicable. Mass transfer in close binary systems can drastically alter the evolutionary paths of both components in the Hertzsprung-Russell diagram \citep[e.g.,][]{vanBever1998,Wang2020}. 
However, in all cases, our assumption that the companion star is on the main sequence should set a reasonable upper limit to its mass estimate.

Minimum and maximum $M_2$ estimates are given in 
Tables~\ref{table:OBbinaryresults} and \ref{table:OBebinaryresults}, assuming the secondary is a non-compact star. 
In some cases, we can also set some constraints on the period, eccentricity and inclination of the binary system. 
If it is a black hole, then the mass estimate is unconstrained.
Details for these
calculations for 
individual objects are given in Appendix \ref{App:starnotes}.

\begin{figure}[ht!]
\gridline{
\includegraphics[scale=0.5,angle=0]{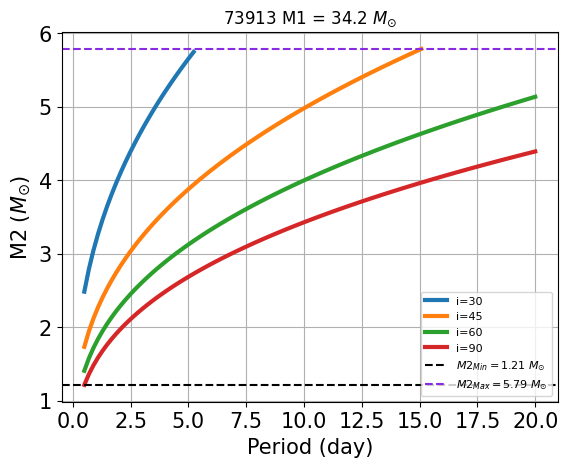}
}
\caption{An example of the parameter space for our $M_2$ estimate of an OB system.
We plot $M_2$ vs $P$ for a range of $i$, shown with different colors.
The black dotted line shows the lower limit for $M_2$ obtained from our $i = 90^{\circ}$ red line.  The purple dotted line shows the upper limit based on our detection threshold. 
This constrains the period to
$P<40$ days (not shown). 
Although $P$ is only plotted up to 20 d, our calculations extend to 2000 d.
\label{fig:OBparamex}}
\end{figure}

\subsection{OB Binaries} \label{sec:OBmasses}

OB binaries are more likely to be pre-SN \citep{Grant2024}, and hence likely to
have non-compact companions.
Our default assumption for these systems is that they have circular orbits unless they are $> 30 \rm\ km/s$ from the 
relation $\Delta\rm RV(10) = \Delta RV(tot)$
(the red line in Figure \ref{fig:binaryID1}). 
As described earlier, for most systems, 
the lower-mass limit is set by our RV constraints via equation~\ref{eqn:Kepler}. 
For SB2 
candidates, we estimate an upper limit for $M_2$ corresponding to the detection threshold as described above, which assumes the object is indeed a non-compact binary.
For non-SB2 systems, this detection threshold
provides an upper limit to $M_2$. 
An example of the parameter space is shown in Figure \ref{fig:OBparamex}.

For some stars, we obtained $M_1$ mass estimates from \citet{Dallas2022}, who provide maximum and minimum values.  We use the minimum $M_1$ value to estimate $M_{\rm 2,min}$ in Tables~\ref{table:OBbinaryresults} and \ref{table:OBebinaryresults}.  
For the OB stars, this applies to only one target ([M2002] 77734).
There are no cases where the maximum values for $M_1$ set constraints for $M_{2,\rm max}$.

There are 10 RV OB binaries that are neither EBs nor SB2. Of these targets, 8 of them are likely
near-circular 
binaries,
so we assign their eccentricities to be $e=0$ for the purpose of modeling; 
one of them, [M2002] 76371, has
$e>0$, and so its eccentricity is a free parameter.  

\subsection{OBe Binaries} \label{sec:OBeMasses}

\begin{figure*}[ht!]
\begin{center}
\gridline{
\includegraphics[scale=0.5,angle=0]{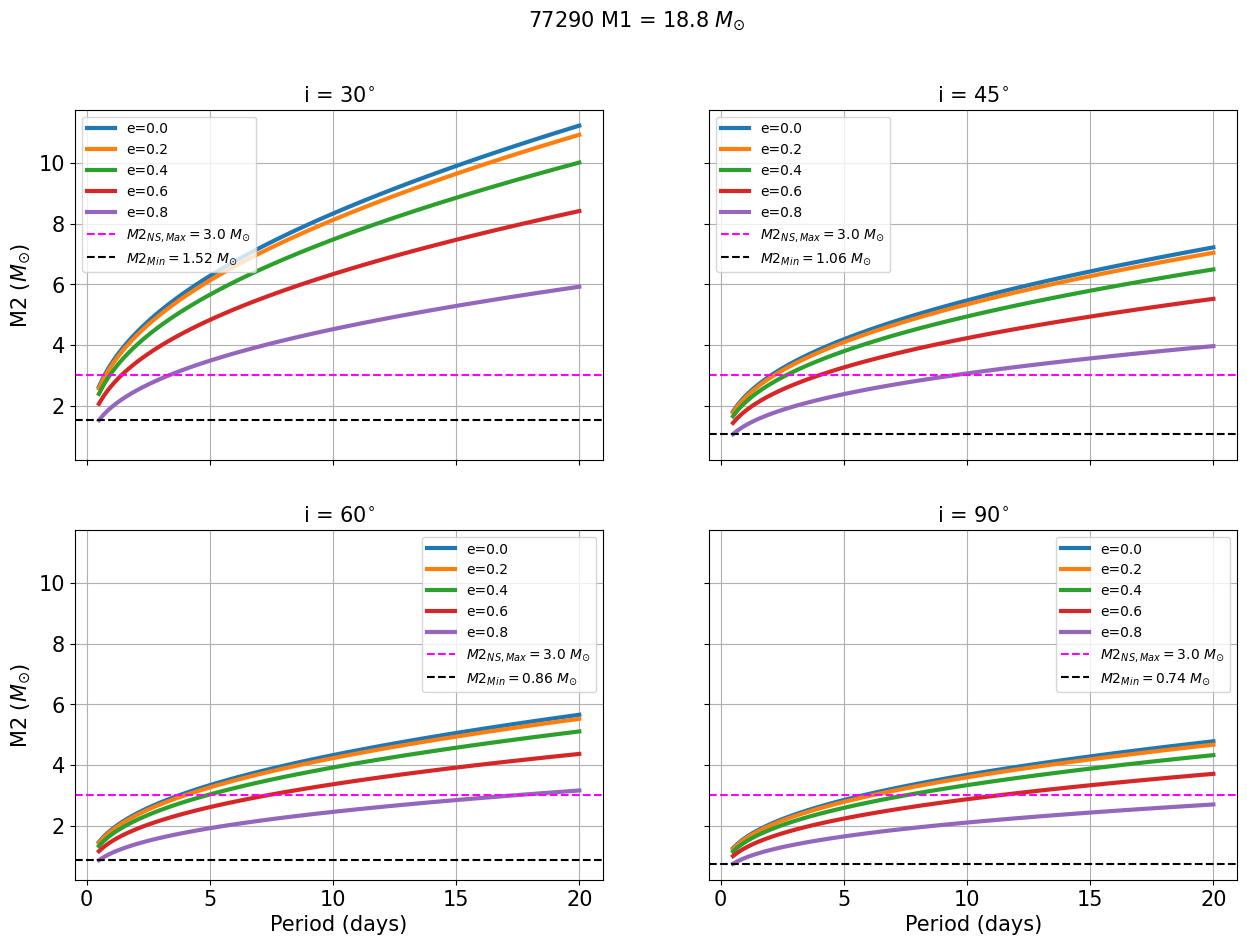}
}
\caption{An example of the parameter space for our $M_2$ estimates for an OBe binary, [M2002] 77290. 
We show $M_2$ vs $P$,
with the colors representing a range of eccentricity and each plot representing a different orbital inclination. The black dotted line shows the lower limit for $M_2$, and
the pink dotted line shows the nominal NS
upper-mass limit of $3\ M_\odot$. 
The plots show periods up to 20 days, but our calculations extend to $P < 2000$ days. \label{fig:OBemassex}}
\end{center}
\end{figure*}

Since OBe binaries are typically 
post-SN systems \citep[e.g.,][]{Dallas2022, DorigoJones2020, Vinciguerra2020, Bodensteiner2020a},
 most are likely in eccentric orbits,
as may be suggested by Figure~\ref{fig:binaryID1}.
Thus
they have among the most unconstrained parameter space.
However, their companions are likely mostly compact objects. 
While black hole masses are unconstrained, we set a limit of $M_2 \leq 3 M_{\odot}$ for 
objects with known 
neutron star companions.
An example of the parameter space for one of our OBe binaries is shown in Figure \ref{fig:OBemassex}.

\begin{figure}[ht!]
\begin{center}
\gridline{
\includegraphics[scale=0.5,angle=0]{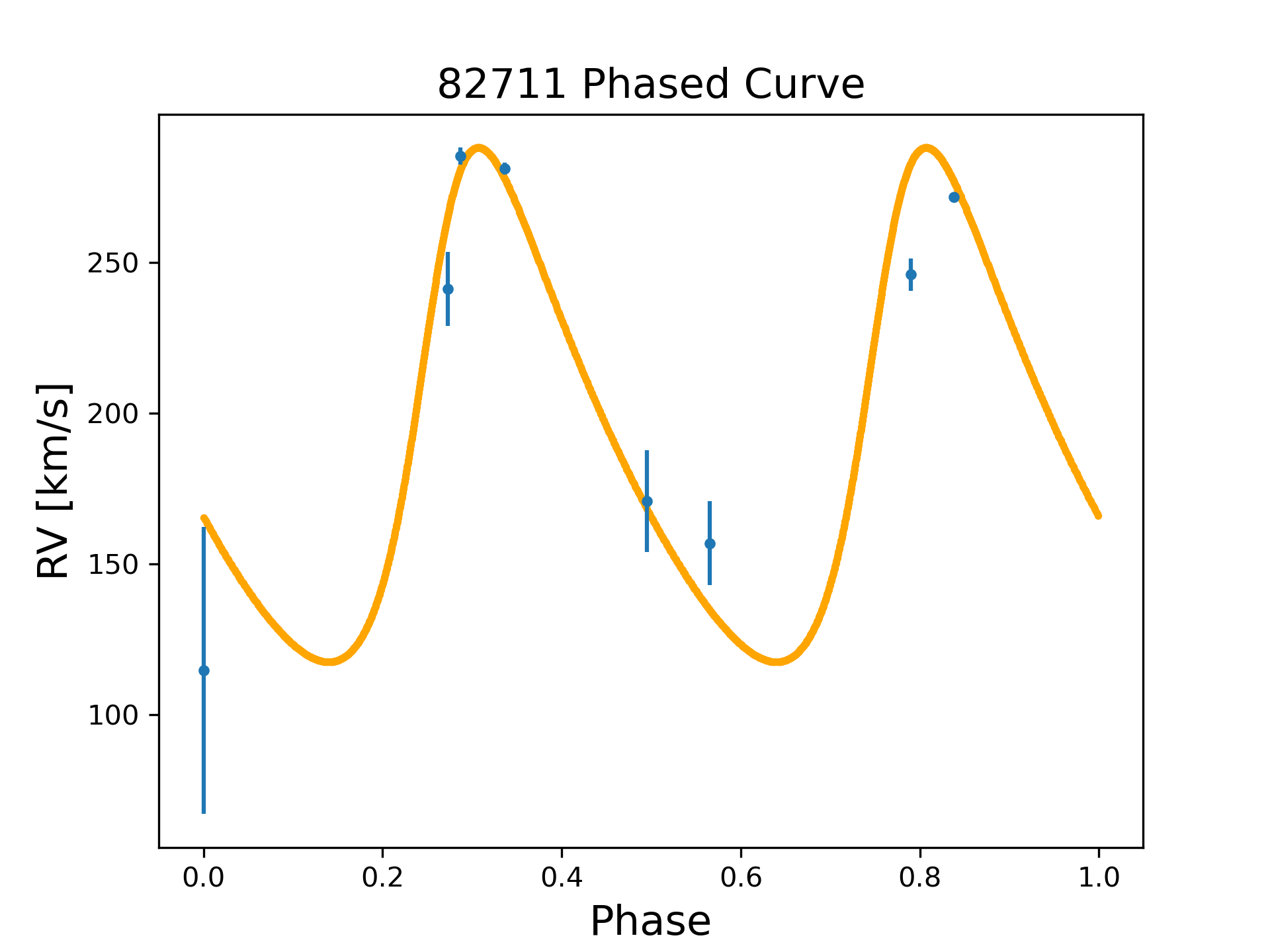}
}
\caption{A possible phased RV curve for [M2002] SMC-82711. This target has its period constrained to $< 6$ days from our RV data. For this particular phased curve we use $P = 2$ days, which yields an eccentricity of 0.3. 
\label{fig:phase82711}}
\end{center}
\end{figure}

As noted earlier, our RV measurements for OBe stars are more uncertain, which extends to the suggestion of systematic eccentricity in Figure~\ref{fig:binaryID1}.  
Figure~\ref{fig:phase82711} shows a possible phased RV curve for [M2002] SMC-82711, which is suggested to be one of the most eccentric objects by Figure~\ref{fig:binaryID1}.  This RV curve is consistent with the period of $< 6$ days. Using that constraint, we generate this potential phased curve using $P = 2$ days, an eccentricity of 0.3, $\omega = 300^{\circ}$, and $0^{\circ} \leq \theta \geq 360^{\circ}$.

There are 9 OBe binaries that are not EBs, SB2s or HMXBs: 4 of them are RV binaries and are within our $\Delta$RV limits for circular binaries
(Section~\ref{sec:binprop}),
and so we assign them to have $e=0$ for our orbital modeling. Additionally, 
5 of them have $e > 0$ and thus have unconstrained eccentricities.

\subsection{Eclipsing Binaries} \label{sec:EBMasses}

For EBs, the period is known either through the OGLE survey or from our own measurements (Section \ref{sec:TESSEB}), and the companion 
is most likely
a non-degenerate star, implying 
that the $M_2 < M_1$ constraint applies. Additionally, if the target is an EB but not an SB2, then the stellar companion must lie below our spectroscopic detection limit.

\begin{figure}[ht!]
\gridline{
\includegraphics[scale=0.5,angle=0]{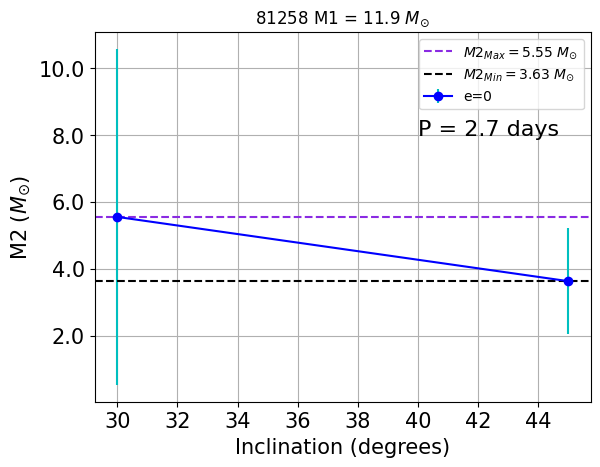}
}
\gridline{
\includegraphics[scale=0.5,angle=0]{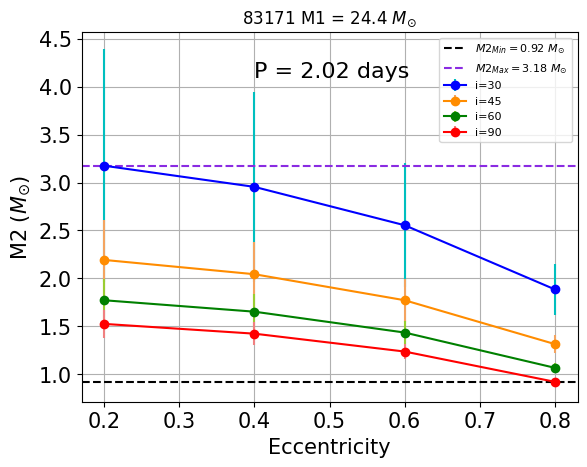}
}
\caption{An example of the parameter space for our $M_2$ estimates for EBs. Target [M2002] 81258 (top panel) likely has a circular orbit and
we plot $M_2$ vs orbital inclination.
Target [M2002] 83171 (bottom panel) is likely to have an eccentric orbit, so
we plot $M_2$ vs $e$ for varying orbital inclination as shown.  The black and purple dotted lines show the lower and upper limits for $M_2$, respectively, as before.
\label{fig:EBCircex}}
\end{figure}

Of the 5 EBs in our sample, there are 4 identified as RV binaries.
There are 2 OB EBs with presumed circular orbits: 1 confirmed OGLE EB ([M2002] 81258) and 1 TESS candidate ([M2002] 83232). 
For these objects, only the inclination is unknown (Figure \ref{fig:EBCircex}, top panel).
There are also 2 OBe EBs likely to have $e > 0$: one confirmed OGLE EB ([M2002] 73355) and one TESS candidate ([M2002] 83171). We vary the inclination and eccentricity for these target models as shown in Figure \ref{fig:EBCircex} (bottom panel). 
We caution that
[M2002] 83171 is only classified as a candidate EB, and its light curve
is complex; the amplitude of variations is only around $2\%$ 
and the maximum amplitude behaves erratically (Figure \ref{fig:bestEB}), 
which may indicate some other kind of variability (Appendix \ref{App:83171}). 

\subsection{SB2 systems} \label{sec:SB2Masses}

\begin{figure}[ht!]
\gridline{
\includegraphics[scale=0.5,angle=0]{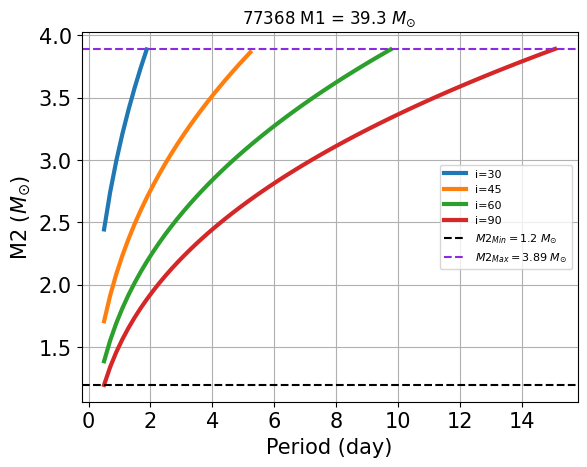}
}
\caption{An example of the parameter space for our $M_2$ estimate for an RV binary that is an SB2 candidate but not a known EB. Target 77368 likely has a circular orbit, but we have no previous information on the period or orbital inclination. We plot $M_2$ vs $P$,
showing values up to 20 days, 
for a range of $i$ as shown.
The black dotted line shows the lower limit for our estimated $M_2$.
The constraint of being near our spectroscopic detection threshold 
determines the estimated upper limit for $M_2$ (purple dotted line).
These constraints imply that $P\lesssim 16$ days. \label{fig:SB2paramex}}
\end{figure}

We currently have no confirmed SB2s in the SMC Wing sample that are
identified as RV binaries.
However, there are 3 SB2 candidates (2 OB and 1 OBe) that have RV data that we can use to 
estimate the mass of a putative companion.  These would be non-degenerate
companions with $M_2 < M_1$, and
$M_2$ also must be such that the companion is near the spectroscopic detection limit,
since we are able to see possible evidence of an SB2.
Figure \ref{fig:SB2paramex} shows an example of this parameter space for the $M_2$ mass estimate. 
If these objects turn out not to be SB2s, but still RV binaries, then the constraints on $M_2$ are as described in Sections~\ref{sec:OBmasses} and \ref{sec:OBeMasses}.

\subsection{HMXBs} \label{sec:HMXBMasses}

\begin{figure}[ht!]
\gridline{
\includegraphics[scale=0.5,angle=0]{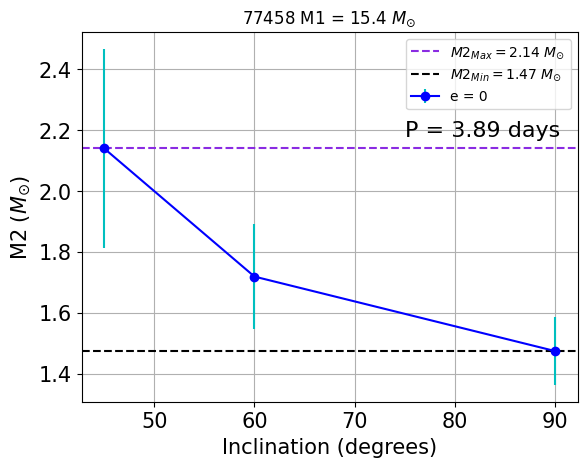}
}
\caption{An example of the parameter space for $M_2$ estimates for HMXBs. Target 77458 (SMC X-1) has a period of 3.89 days \citep{Clark2000} and a low eccentricity \citep{Falanga2015}. \label{fig:HMXBparamex}}
\end{figure}

There are 3 HMXBs ([M2002] 77458, 77851, 82711) in our SMC Wing sample, 
and all are known to have neutron star companions \citep{Reynolds1993,vandermeer2007,Schmidtke2013,Gvaramadze2021}. 
Only 2 of these systems ([M2002] 77458, 82711) have RV monitoring data. We are able to obtain information on their periods, eccentricities and neutron star companion masses from the literature. 
We can use these systems as a check on our method of obtaining $M_2$ estimates, which are carried out as shown in Figure~\ref{fig:HMXBparamex}.  

For [M2002] 77458, its neutron star companion has a mass of 1.04 $M_{\odot}$ which was determined by \cite{Rawls2011} by modeling the eclipse duration of the pulsar to obtain an estimate of the NS mass. Our mass estimates based on our RV measurements do place it at a NS mass ($1.47 - 2.14 M_{\odot}$), which is in reasonable agreement with the value from the literature. 
Additional details about this system are given in Appendix \ref{App:77458}.

Target [M2002] 82711 has a known NS companion with no previous mass measurement. The system is noted to have a long period of 656 days \citep[][]{Schmidtke2013,SchmidtkeCowUd2019,Gvaramadze2021}. However, we note that in our estimates of $M_2$ from our RV measurements, in order to remain within the NS range ($1.5 -3.0 M_{\odot}$), the period would have to be $\lesssim 6$ days.
Additional details about this system are given in Appendix \ref{App:82711}.

\subsection{Sample Biases} \label{sec:bias}

While we consider our field binaries in the SMC Wing to be representative of the entire SMC field OB population, there are various observational biases that must be considered.  A principal issue is that our results are biased toward detecting shorter-period binaries due to both the prevalence of short-period ($<$ 1 week) sampling and the $\sim5$-year survey duration.  The discovery of at least one long-period binary ([M2002] 77616, $P\gtrsim 7.3$ yr) demonstrates that a number of others may be present in our sample, but which are difficult to detect with our current RV survey and RV detection thresholds.

Similarly, our results are biased toward detecting higher-mass companions, which generate stronger signals, in both velocity and light curve variations.  Moreover, the masses of our target OBe stars may be overestimated if they are inflated as fast-rotating mass gainers \citep{Richards2024,Lau2024, Castro2018, Sabin2017, Herrero1992}, thereby enhancing their luminosities.  Such overestimates would also affect the range of binary parameters, including overestimates on the mass of $M_2$.

Thus, we are likely incomplete in the detection of wide binaries and systems with low mass ratios, and there may be a systematic overestimate of total system masses for OBe stars and other potential mass gainers.  Overall, our binary frequencies should therefore be regarded as lower limits.

\section{BPASS Models} \label{sec:BPASS}

Given the richness of the observational data we have obtained for the SMC Wing field binaries, the modelling required for it is non-trivial. To this end, we use the Binary Population and Spectral Synthesis (BPASS) v2.2 results and code suite \citep{2017PASA...34...58E,2018MNRAS.479...75S}. BPASS is a widely used and tested population synthesis code which uses detailed binary evolution models to follow the results of binary interactions.

The model population details include an initial metallicity mass fraction with $Z=0.004$, the IMF of \citet{1993MNRAS.262..545K} with a maximum initial mass of 300~M$_{\odot}$, and the initial binary parameter distributions of \citet{MoDiStefano2017}. A constant star-formation rate is assumed. With these initial parameters, and considering runaways and walkaways from the BSS mechanism only, we can make predictions such as the following for OB and OBe stars:  the expected numbers of field single stars, field compact and non-compact binaries, the secondary masses, and distributions of velocities, binary periods and eccentricities. 

\begin{deluxetable*}{lccccccccccc}
\rotate
\tablecaption{Observed Populations and BPASS BSS Predictions\tablenotemark{a}
\label{table:BPASSpop}}
\tablewidth{700pt}
\tabletypesize{\scriptsize}
\tablehead{
\colhead{Population} & \colhead{Observed} &\colhead{Observed} & \colhead{BPASS} & \colhead{BPASS}   & \colhead{SMC Wing} &\colhead{SMC Wing}  &\colhead{BPASS Binary} & \colhead{BPASS Binary}\\
\colhead{} & \colhead{Number} &\colhead{Freq} & \colhead{Freq} & \colhead{Single}  & \colhead{Total Binaries \tablenotemark{b}} & \colhead{BSS Binaries \tablenotemark{c}} &\colhead{(OB/OBe+star/He)\tablenotemark{d}} & 
\colhead{(OB/OBe+NS/BH)} 
\\
\colhead{} & \colhead{} &\colhead{} & \colhead{} & \colhead{}  & \colhead{(Conf.+Cand./Conf.)} & \colhead{(Conf.+Cand./Conf.) } &\colhead{} & 
\colhead{} 
} 
\startdata
All OB &         773   &  0.649$\pm$0.030    &  0.880 & 0.101&  --  &  --  & 0.714 (0.702/0.012)& 0.064 (0.004/0.059) \\
All OBe &        418    &   0.351$\pm$0.020 &  0.129 & 0.059 &  --  &  -- & 0.019 (0.014/0.005)& 0.042 (0.012/0.025) \\
All OB HMXB &       7     &   0.006$\pm$0.002&  0.003 &  --   &  --  & -- &  --  &  -- \\
All OBe HMXB &      39     &  0.033$\pm$0.005&  0.009 &  --   &  --  &  -- &  --  &  -- \\
Field OB &         255  &   0.214$\pm$0.015  &  0.087 & 0.039 & 16 / 10 & 0.073 $\pm$ 0.038 / 0.055 $\pm$ 0.032& 0    &  0.048 (0.004/0.043) \\
Field OBe &         167    &  0.140$\pm$0.012 & 0.070 & 0.042 & 17 / 10  & 0.200 $\pm$ 0.066 / 0.145 $\pm$ 0.055 & 0    &  0.028 (0.012/0.016) \\
Field OB HMXB &      4     &  0.003$\pm$0.002 & 0.003 &  --   & --   & --  & --  & --\\ 
Field OBe HMXB &     19   &   0.016$\pm$0.004 & 0.007 &  --  & 3 & 0.055 $\pm$ 0.032  & --   & --  \\
\enddata
\tablenotetext{a}{Frequencies out of the total SMC OB/e population.
Data for the observed total OB/e SMC populations are from \citep{Dallas2022} and for observed HMXBs from \citep{HaberlSturm2016}.}
\tablenotetext{c}{Total binaries include both BSS and DES binaries of the SMC Wing.}
\tablenotetext{c}{Frequencies out of 55 total from our SMC Wing population. Errors derived from Poisson errors and error propagation. However, values should be considered lower limits except for HMXBs.}
\tablenotetext{d}{Pre-SN binaries, i.e., OB star + non-compact companion or He-star companion.}
\end{deluxetable*}

Our first step in creating the synthetic populations is to identify the stars that have the following parameters:
\begin{enumerate}
    \item $M_{\rm V} < -3.75$.
    \item Surface hydrogen mass fraction is $>0.3$.
    \item The star must be on the main sequence and not have a hydrogen-exhausted core.
\end{enumerate}
From these we select out OB stars, OBe stars and HMXBs by the following criteria:
\begin{enumerate}
    \item OB stars have effective temperatures $>23$~kK.
    \item OBe stars have effective temperatures $>23$~kK and are taken to be secondary stars that have accreted more than 5\% of their initial mass from the primary star in a binary interaction.
    \item OB HMXBs are selected to be systems that have: an OB star bound to a compact remnant with mass $\ge1.4$~M$_{\odot}$, a donor star $> 5~M_{\odot}$ and $T_{\rm eff}>23$~kK, and donor volume $>80\%$ of the Roche lobe volume. This includes wind-fed HMXBs \citep[e.g.][]{2021PASA...38...56H} as well as those experiencing Roche lobe overflow.
    \item OBe HMXBs are selected to be OBe stars (see \#2 above) that are bound to a compact companion in an orbit with $P< 1000$ d. This is suggested to be the upper period bound for BeXRB stars from observations \citep[e.g.][]{2024MNRAS.527.5023L}.
\end{enumerate}

From these constraints we build two populations of stars. First, the total population of all OB and OBe stars in the SMC; and second, all stars that have a 1-D runaway velocity from the binary supernova scenario of $>15~{\rm km \, s^{-1}}$. 
This roughly identifies stars that have escaped from their birth clusters. We estimate the numbers of stars in these populations and determine the fraction of each type compared to the total population. We then compare these to the corresponding observed fractions from \citet{Dallas2022} and binary frequencies derived in this work.

We show the observed and predicted frequencies relative to the total SMC OB/e population in Table \ref{table:BPASSpop}.  
The first two columns list the subpopulations and observed numbers of stars in each sample.  Columns 3 and 4 give the frequencies of the total observed and predicted populations, respectively, and Column 5 gives the predicted frequency of single stars.  Column 6 gives our observed number of binaries in the field of the SMC Wing, showing totals that include and exclude candidate binaries; Column 7 gives the observed frequencies of the subset of BSS binaries as defined in Section~\ref{sec:OBebinaries}, for the same 55 Wing stars; values are given for both confirmed + candidate binaries and confirmed-only samples.
Column 8 gives the BPASS predicted frequencies for binaries consisting of an OB or OBe star plus a non-compact companion; values are also shown for non-He star and He-star companions.  Column 9 gives the values for binaries hosting a compact companion, with values for NS and BH also shown.

We see that we tend to underpredict the frequency of OBe stars and field stars by a factor of two.  
There are a variety of possible explanations for this.
One reason might be that our accretion threshold for OBe stars and/or velocity threshold for escape into the field are too high.
In addition, DES ejections, which are unaccounted for, are also likely an important contribution to the field population.
Observational biases are also significant; \citet{DorigoJones2020} estimate that our selection of field stars excludes over half of all slow, walkaway ejections (Section~\ref{sec:DESvsBSS}).  
We also may be underestimating the lifetimes of OBe stars due to now accounting for rotational mixing in these stars below 20~M$_{\odot}$; or observed OBe stars may have lower masses than those in the models, due to possible rotational inflation that is not accounted for in the models.  
Furthermore, we are assuming constant star formation in our model; if this assumption is incorrect then this would also change our predicted numbers slightly. 
The values in the table are discussed in more detail in the following sections.

We also underpredict the frequency of HMXBs, 
in particular OBe HMXBs, so this effect may be dominated by the general underprediction of OBe stars.
More sophisticated criteria for identifying
HMXBs could increase the number of predicted systems. Or, some of these systems may have experienced dynamical ejections that may have affected the orbit in such a way as
to enhance the occurrence of HMXBs.
Generally, it is essential to note that our synthetic populations of field stars are assumed to only arise from the BSS and do not include dynamically ejected runaways. We therefore should expect to see some  differences between our predicted and observed populations that are due to DES ejections.  This will be leveraged in Section~\ref{sec:OBbinaries}.

\section{BSS Binaries} \label{sec:OBebinaries}

Classical OBe stars have recently been 
linked to post-interaction origins. The binary model for these objects has the more massive star filling its Roche lobe and becoming a mass donor to its companion. This 
increases the mass gainer's angular momentum enough to generate a decretion disk \citep[e.g.,][]{Pols1991, deMink2013}. The massive donor later explodes as a SN, and can cause these OBe companions to accelerate through BSS ejections \citep[e.g.,][]{Blaauw1961, Brandt1995, Renzo2019},
which may also include retaining a bound compact companion. Studies have shown that OBe stars are consistent with this binary mass transfer origin \citep[e.g.,][]{Bodensteiner2020a, Hastings2021}, linking it to BSS ejections \citep{Grant2024, DorigoJones2020,Dallas2022}. Additionally, HMXBs, 
which are known post-SN
BSS ejections, are often emission-line stars \citep[e.g.,][]{Maravelias2018}.
Their kinematics are also consistent with those of OBe stars in general
\citep{Grant2024, DorigoJones2020}. 

BSS binaries are those that have compact companions. 
Binary population synthesis modeling of BSS systems by, e.g,
\citet{Brandt1995} and \citet{Renzo2019} 
find that about $10-20\%$ of massive binaries remain stably bound, generally with
low runaway velocities. 
This is due to mass transfer from the primary to the secondary,
which widens the orbit; thus, the increase in secondary mass and the larger orbit make it harder to reach faster ejection speeds. 

An important characteristic of a BSS binary is the nature of the compact companion. To 
understand the relative frequencies and parent binary populations for these objects is
complex. Stellar evolution models have shown that,
e.g., black holes do not originate according to a simple threshold lower mass for the progenitor stars.  Instead, there are ``islands of explodability", and black hole formation moreover is stochastic at any given mass
\citep{Ugliano2012,Sukhbold2016}.  Observationally, neutron star companions are more prevalent for OBe stars and HMXBs.
Black hole companions 
are harder to detect, and also
more likely to be misclassified.
For example, the only Be system that has been suspected to have a black hole, MWC 656 \citep{Casares2014}, has since been refuted by higher resolution spectroscopic data and is
now suggested to be
a stripped star instead \citep{Rivinius2022, Janssens2023}. Other examples include LB-1, HR 6819, and NGC 1850 which present cases that were initially OB stars orbiting stellar-mass black holes that have since been convincingly argued to have stripped stars \citep{Bodensteiner2020,Shenar2020,ElBadry2022}.
Thus, detailed observations of field massive binaries with constraints on binary companions and orbital parameters are essential to constrain both parent and descendant populations.

There are clear differences in kinematics between 
objects that we identify as BSS binaries and 
the remainder of the OB binaries in our sample, and
they support the BSS origins of OBe binaries.
As discussed previously, OBe stars 
apparently represent the majority of
BSS ejections.
Therefore, those
that are identified as binaries in our sample are largely systems
with compact remnants
that remain bound after the supernova event;
few are expected to have non-compact companions
\citep[e.g.,][]{Bodensteiner2020a}.

Our binary census increases
the binary frequency of OBe stars in the SMC Wing by $> 50 \%$. We have a total of 17 unique OBe binaries 
and candidates out of 21 targets, 
or $\sim 81 \%$ of our total OBe sample. Of these, 10 are confirmed binaries, which yields a lower limit of $\sim 48\%$. 
Allowing targets to share multiple classifications, they can be divided as: 14 RV binaries (6 circular orbit binaries, 7 eccentric binaries, and 1 identified solely through the $F$-test), all 3 HMXBs, 3 EBs, and 2 SB2s (Table~\ref{table:OBebinaryresults}).

\begin{figure}[ht!]
\includegraphics[width=\columnwidth,angle=0]{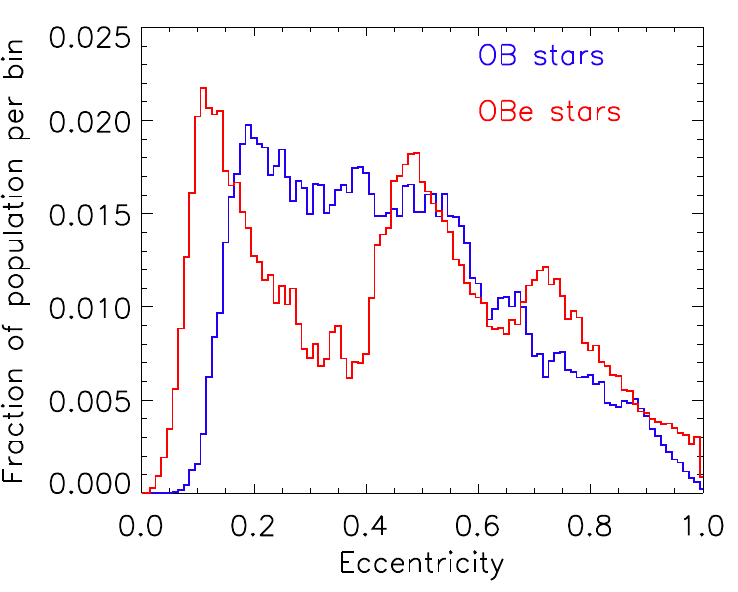}
\caption{Eccentricity distribution for BPASS BSS OB (blue line) and OBe (red line) binary systems post supernova. \label{fig:BPASS_ecc_distribution}}
\end{figure}

\begin{figure*}[ht!]
\gridline{
\includegraphics[width=1.7\columnwidth]{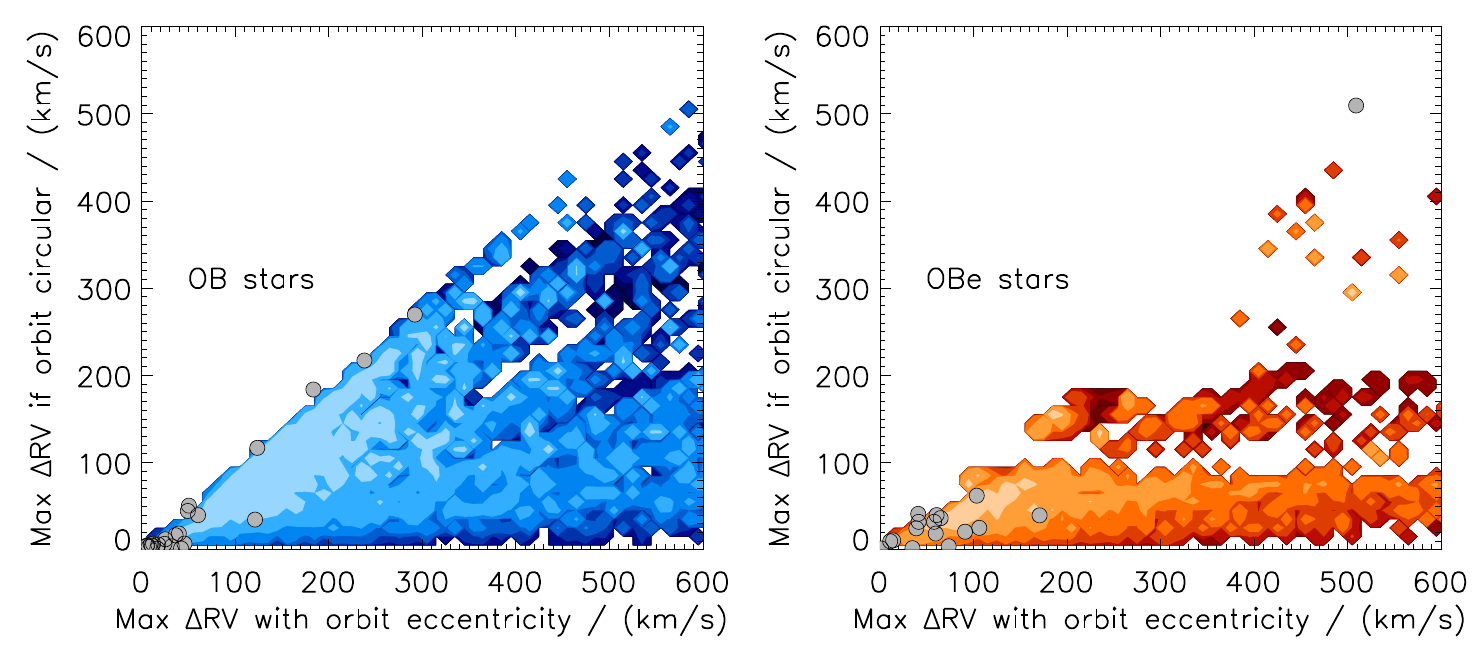}
}
\caption{Estimated maximum $\Delta$RV distributions for BPASS ejected BSS systems, with OB and OBe binaries shown in the left and right panels, respectively. 
The contours are estimated from the model binary population using the maximum $\Delta$RV for the post-SN eccentricities ($x$-axis) and using the equivalent circular orbits ($y$-axis). Each contour indicates an order-of-magnitude difference in probability density, with bin size of 10~${\rm km \, s^{-1}}$ by  10~${\rm km \, s^{-1}}$; lighter colors indicate higher probability.
The circles show the observed systems from Figure~\ref{fig:binaryID1}, where the $x$- and $y$-axes correspond to maximum $\Delta$RV(total) and $\Delta$RV(10d), respectively.  Although the modeled quantities are different,
they give a close approximation by comparing the velocities most likely to be observed from an eccentric orbit, i.e., those similar to an equivalent circular orbit, vs the extremum possible in the eccentric orbit.   
\label{fig:BPASS_RV}}
\end{figure*}

\begin{figure*}[ht!]
\gridline{
\includegraphics[width=1.7\columnwidth,angle=0]{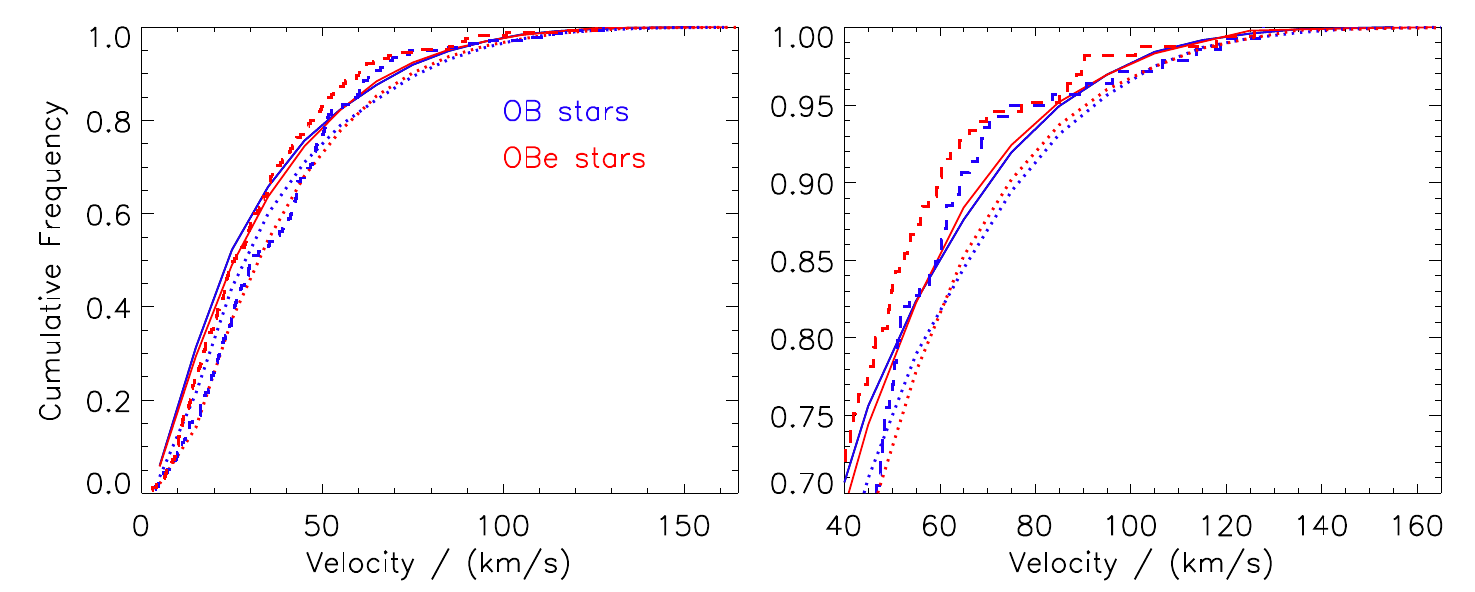}
}
\gridline{
\includegraphics[width=1.7\columnwidth,angle=0]{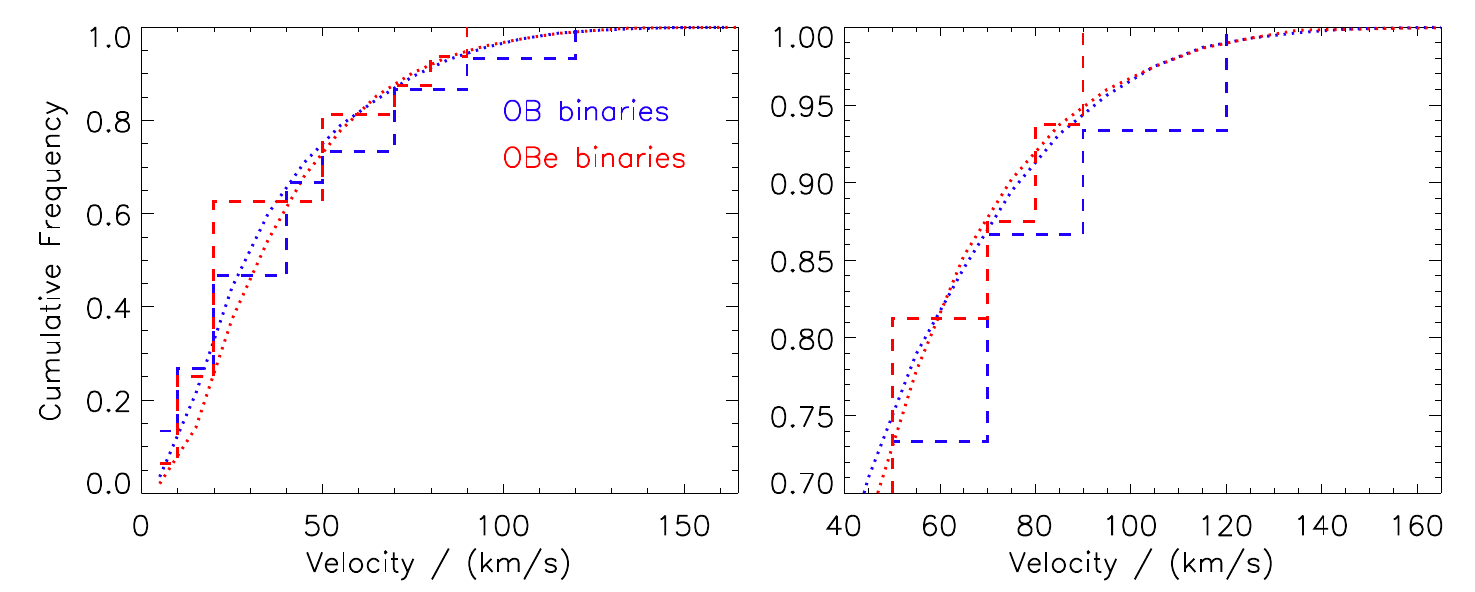}
}
\caption{Cumulative 2D velocity distributions for BPASS
predictions compared to the field OB and OBe velocities from \citet{Grant2024}. The dashed lines are the observed sample, the solid lines are the BPASS distribution for all stars with a BSS velocity greater than 15~${\rm km \,s^{-1}}$, and the dotted lines are the BPASS distribution for binary stars only. The upper panels include the full OB and OBe populations from their sample, while the lower panels show only the binary distributions for our field SMC Wing sample and BPASS models. The left panels show the full distributions and the right panels show a zoom for velocities $> 40\ {\rm km\ \,s^{-1}}$.  
\label{fig:BPASS_VelHist}}
\end{figure*}

\begin{figure*}[ht!]

\includegraphics[width=\columnwidth,angle=0]{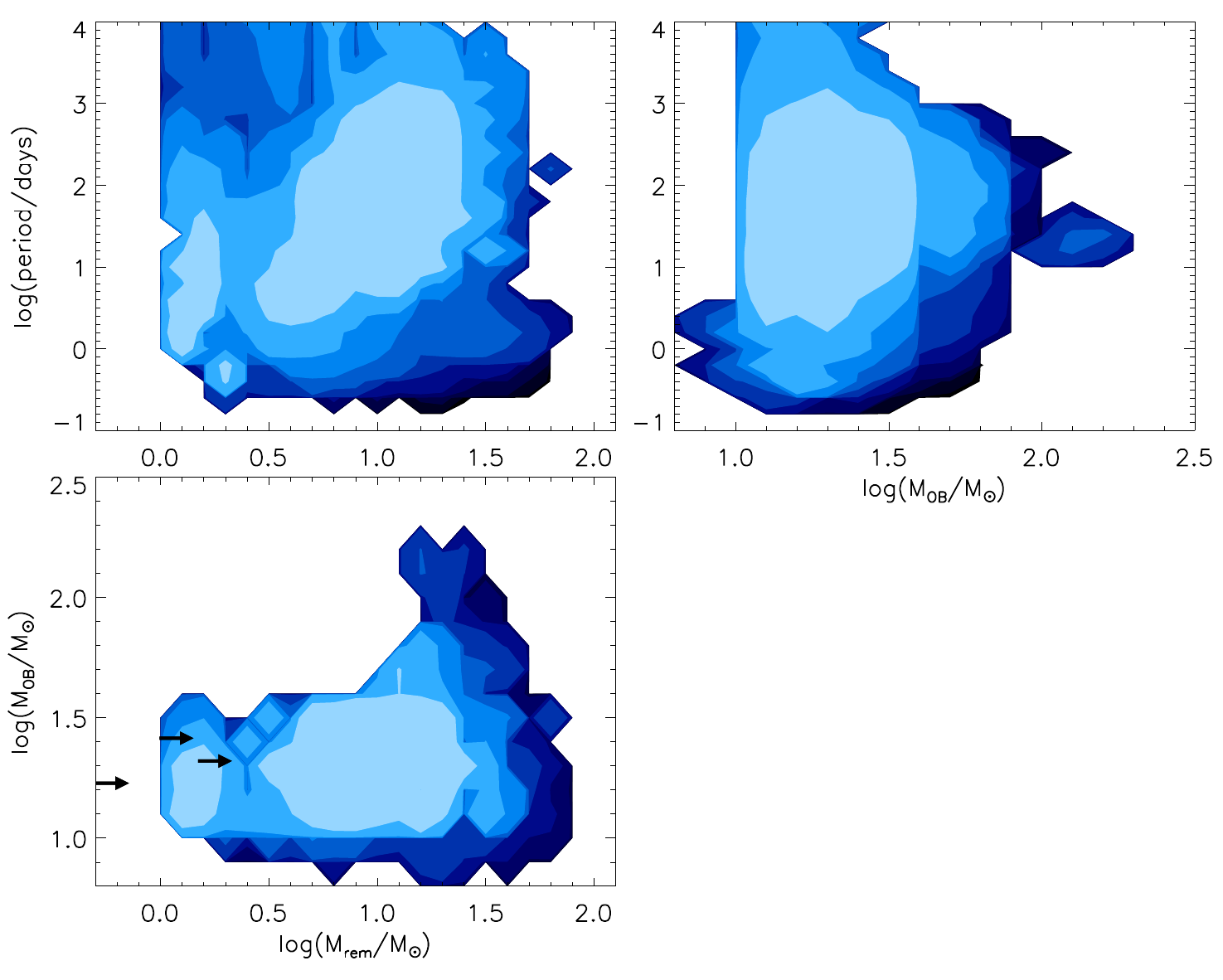}
\includegraphics[width=\columnwidth,angle=0]{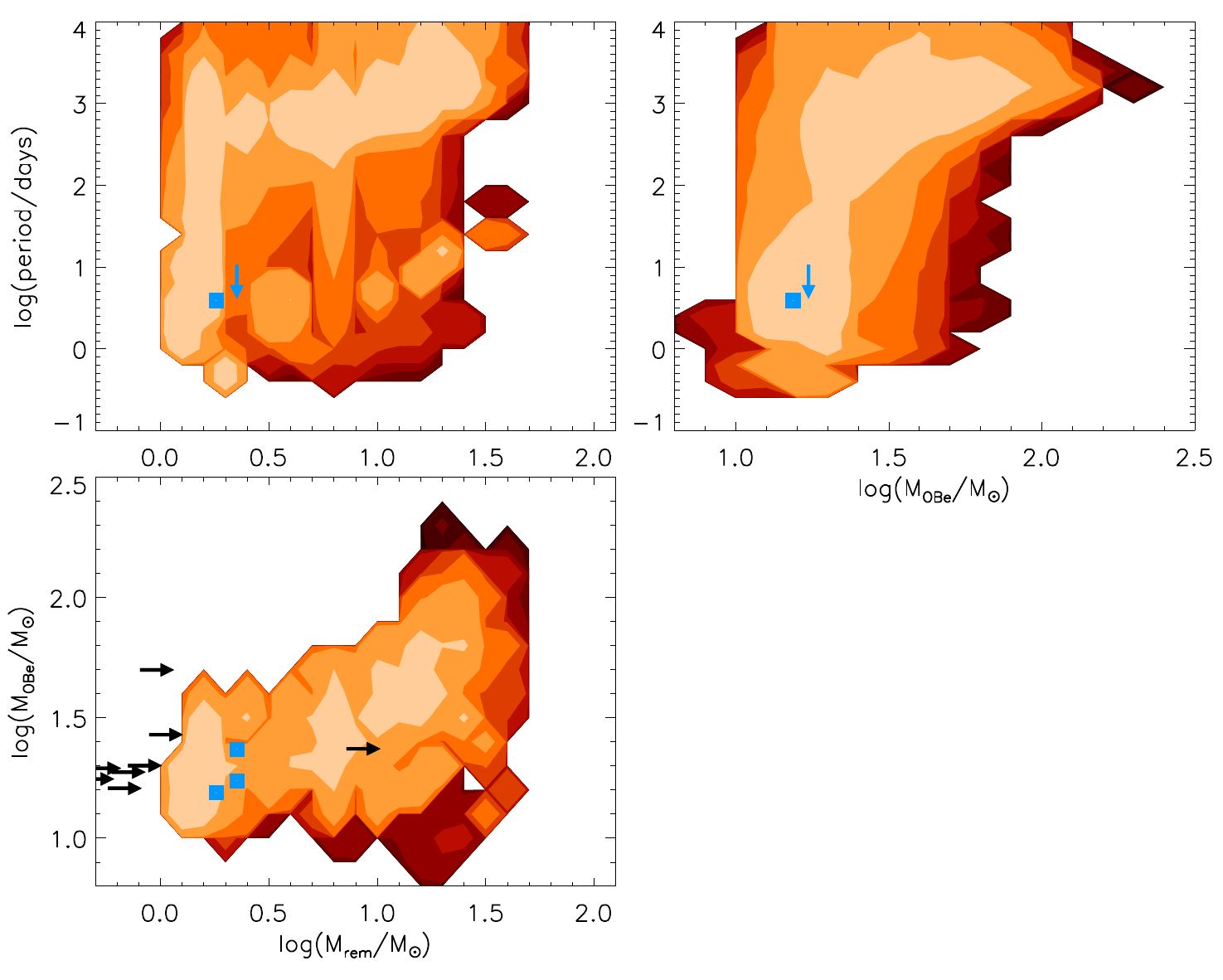}

\caption{Mass and period distributions of BPASS BSS ejected compact remnant binaries.  OB and OBe stars are shown in the left and right panels, respectively, with
contours shown as in Figure~\ref{fig:BPASS_RV}.
Black arrows indicate the lower-mass limits on $M_2$ for observed binaries in our sample, blue squares show our HMXBs, and blue arrows indicate the upper limit on $P$ for the HMXB [M2002] 82711.
AzV 493 ([M2002] 77616) is the object with the highest-mass OBe primary, and [M2002] 
76773 is the object
with the largest $M_{\rm2,min}$ value; both
are BH candidates,
although 
76773 may be spurious (see text).
\label{fig:BPASS_mass_distribution1}}
\end{figure*}

\begin{figure*}[ht!]
\includegraphics[width=2\columnwidth,angle=0]{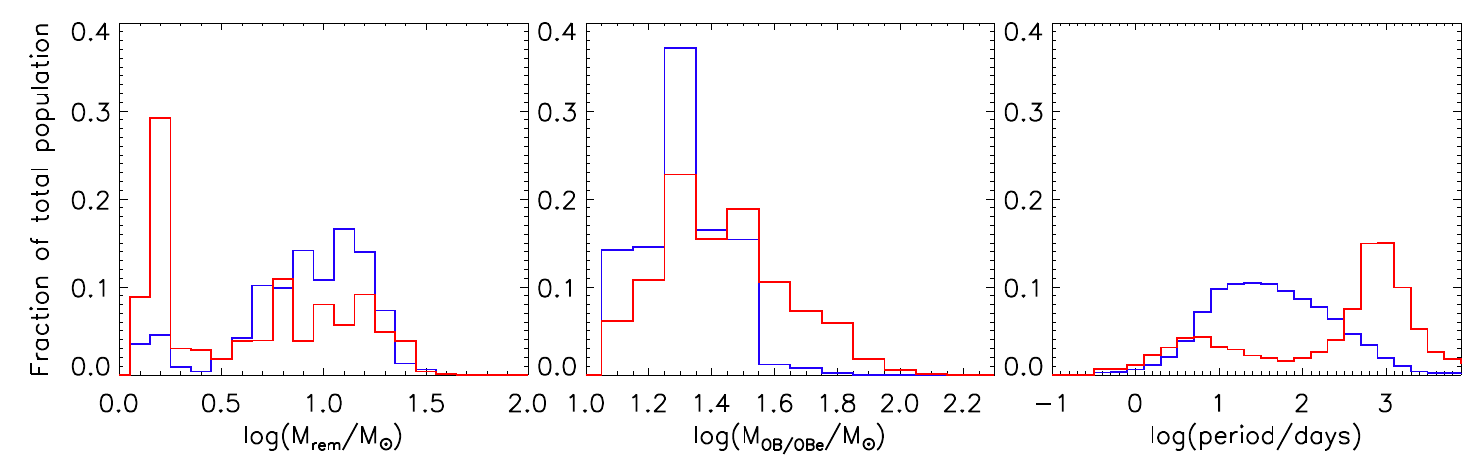}

\caption{The same mass and period distributions for BSS binary systems as in Figure \ref{fig:BPASS_mass_distribution1} but now shown as 1D distributions across each of the different parameters, OB/OBe star mass, compact remnant mass, and period. OB stars are in blue and OBe stars in red. \label{fig:BPASS_mass_distribution2}}
\end{figure*}

However, 
some OB binaries should also share BSS origins.
These are objects where a circumstellar disk has not survived, or not formed, after the binary interaction and SN.
In fact, Table \ref{table:BPASSpop} shows that our BPASS models 
predict 
more BSS OB binaries than
OBe systems in the field sample, 
thus there should be a subsample of OB binaries with compact companions that are ejected through BSS. We now examine this further.

\subsection{Fast rotators}\label{sec:BSSrot}

One potential indication of OB-star BSS objects can be fast rotation; 
like OBe stars, the surviving stars in
OB binaries may be spun up due to mass transfer.
Following \citet{Grant2024}, we use a threshold of 150 \kms\ to identify such objects.
In our SMC Wing sample, there are 4 RV OB binaries that are fast rotators: [M2002] 77368, 80573, 82444, and 83510 (Appendix~\ref{App:77368}, \ref{App:80573}, \ref{App:82444}, \ref{App:83510}). 
We therefore consider these OB binaries to be BSS products.
We caution that 
[M2002] 77368 is an SB2 candidate, indicating that this is a potential DES ejection of an object that is still pre-SN; 
thus we consider this object to be a candidate BSS object.
However, as a general caveat,
\cite{Grant2024} find that the velocity distribution of 
fast-rotating OB stars is more consistent with that of other OB stars, rather than OBe stars.

If we consider all 4 of these objects (3 fast rotators and 1 eccentric binary) to be BSS binaries, these comprise $25\%$ of the OB population in our sample. Our BPASS results show that $\sim 55\%$ of the total OB field population in the model corresponds to BSS binaries, indicating that our 4 binaries are well within expectations of the model, and in fact there should be many more.

\subsection{BSS Eccentricities \label{sec:BSSecc}}

The OBe binaries tend to have eccentric orbits,
as demonstrated above in Section~\ref{subsec:RV}, where we found that
of the 14 RV binary OBe stars, 7 of them appear to have $e > 0$;
on average their eccentricity $e = 0.45 \pm 0.04$, compared to $e=0.08 \pm 0.02$ found in non-BSS OB binaries (Section~\ref{subsec:RV}).
This is consistent with expectations that BSS binaries have experienced SN kicks.  As shown in Section~\ref{sec:DESOBBin} below, pre-SN binaries likely have circularized orbits, and therefore the eccentricities of OBe orbits are most likely caused by the SN event.

For the BPASS model, we record the eccentricity of each surviving binary after the primary star's supernova. 
Figure \ref{fig:BPASS_ecc_distribution} shows the predicted distribution of eccentricities for BSS objects, both OBe and OB stars. The OBe eccentricity distribution (right panel) has three peaks around $ e \sim 0.1$, 0.5, and 0.7. 
The low eccentricity peak around 0.1 tends to have wider binaries $(\log P > 2)$ with massive black hole (BH) remnants $>10 M_{\odot}$ and mostly massive OBe stars with mass $> 20 M_{\odot}$. 
These originate in our models from efficient mass transfer in quasi-chemically homogenous evolution (QHE) systems \citep{Eldridge2011} which are presumed to occur at $Z \le 0.004$ and initial masses $> 20 M_{\odot}$.  These stars are rotationally fully mixed during the main sequence lifetime if they accrete more than 5\% of their initial masses, 
which is our model criterion for identifying OBe stars. 
The peak at $e \sim 0.5$ includes more moderate-mass BHs, some in very wide periods on the order of decades to centuries, and some shorter periods.  At $e \sim 0.7$, the population is dominated by NS systems with a range of periods.  However, there is significant overlap between the binary parameters for these three groups.

We see in Figure~\ref{fig:BPASS_ecc_distribution}
that OB BSS binaries are also expected to have $e > 0$. Our observational results in Figure \ref{fig:binaryID1} show
one target with apparently significantly higher eccentricity, which therefore may correspond to an OB BSS object
([M2002] 76371; Appendix~\ref{App:76371}); it 
also is not an EB or SB2.
Although this object is not a fast rotator, we consider this star to be a BSS object based on its eccentricity.
We note that for longer periods and correspondingly lower RV variations, it becomes harder to evaluate eccentricities (Figure~\ref{fig:binaryID1}).

We use our BPASS models to compare with the observed
information about eccentricities seen in Figure~\ref{fig:binaryID1}. In Figure \ref{fig:BPASS_RV} we present contour plots of the expected radial velocity patterns expected in the population.
We calculate the maximum $\Delta$RV expected for the binary in its eccentric orbit, obtained from the velocities at periastron and apastron. 
To approximate $\Delta$RV(10d), we adopt the $\Delta$RV that we would find for an equivalent circular orbit. This estimate is robust enough to make a useful comparison with the observations,
which are obtained by semi-random sampling.

Figure \ref{fig:BPASS_RV} shows that our radial velocity data show the same trends as the BPASS model predictions for the OBe population originating as post-interaction mass-gainers that are also post-SN BSS objects. We can expect that the observed data will underestimate the maximum $\Delta$RV values used by the models, and this effect is seen for both OB and OBe stars in Figure~\ref{fig:BPASS_RV}.  In both panels, we also see a subset of observed objects that are clustered at very low $\Delta$RV values.  These may be inconsistent with the BSS binary predictions, and may support their interpretation as single stars and/or DES systems (Section~\ref{subsec:RV}); however, more rigorous examination is needed to evaluate this, especially since the quantities plotted for the data and models are somewhat different,
and given the significant uncertainties with RV measurements for OBe stars discussed in Sections~\ref{subsec:RVmeas} and \ref{subsec:RV}.

\subsection{BSS Velocity Distributions} \label{sec:OBeVeldist}

We show the cumulative
velocity distributions of OB and OBe stars from BPASS
in Figure \ref{fig:BPASS_VelHist}. 
The models are compared with velocities of all the SMC field OB and OBe stars from \citet{Grant2024} (top panels).
We see that for OBe stars, there is an excess of high-velocity objects traveling with peculiar motions $> 40$ \kms.  We also see that OB stars generally have velocities higher than predicted.  
As shown by \citet{Grant2024}, there is secondary peak around 50 \kms.
Considering that on the order of half of the OBe stars may have velocities that originate from combined dynamical and supernova acceleration \citep{Grant2024, DorigoJones2020}, it is likely that the these patterns in the velocity distribution for both OBe and OB stars are due to dynamical accelerations.  In the case of OBe stars, these correspond to two-step ejections, and for OB stars, there is likely a major contribution to the population from pure DES ejections. 

Our SMC Wing binary OBe stars
are compared to the BPASS BSS models
in bottom row of Figure \ref{fig:BPASS_VelHist}. 
We see that they do not show evidence of the high-velocity excess.  
This may suggest that most identified OBe binaries are not dominated by two-step ejections, and therefore that such objects are limited primarily to single OBe stars.
However, the observed 
OB binaries (median velocity = 41 km/s) show higher velocities than the model (median velocity = 25 km/s)
which again likely indicates
the expected  presence of DES ejections and non-compact binaries among the OB binaries
\citep[e.g.,][]{Grant2024}.

\subsection{BSS Binary orbits and companions} \label{sec:OBeBSSParam}

In Figures \ref{fig:BPASS_mass_distribution1} and \ref{fig:BPASS_mass_distribution2}, we plot the expected donor mass, remnant mass and binary period for the 
ejected field
stars in our BPASS models. 
The binaries comprise $40\%$ of the BPASS field OBe population. There appear to be two different populations in Figure~\ref{fig:BPASS_mass_distribution1}: one with lower remnant masses that have both short and long periods (1 -- 1000 days), and one with higher remnant masses that have only longer periods ($\sim 1000$ days). These populations are also seen in Figure \ref{fig:BPASS_mass_distribution2} which shows that OBe BSS binaries in our BPASS models have a bimodal distribution in their periods. 
Additionally, we see evidence of these two populations in the eccentricities of OBe binaries. As mentioned previously, Figure \ref{fig:BPASS_ecc_distribution} shows that the predicted eccentricities for OBe binaries 
have strong peaks at $e\sim 0.1$ and $e\sim 0.5$, as well as a weaker peak around $e\sim 0.7$.  These map onto the three peaks in the lower plot in Figure~\ref{fig:BPASS_mass_distribution1}, with the lowest eccentricity corresponding to the highest remnant masses and vice versa.
Figure \ref{fig:BPASS_mass_distribution2} shows that there are 
many more long-period systems expected among the
OBe BSS binaries. 
These would be harder to detect in our sample, which is strongly biased toward shorter period systems with $P$ on the order of 1 -- 10 days.

The bimodality in orbital properties corresponds to whether systems have NS versus BH companions. 
NS systems typically have mass-gainer OBe stars on the order of 15 $M_\odot$, and periods on the order of 5 days, but extending to much larger values; while BH systems typically host OBe stars on the order of 30 $M_\odot$ and periods on the order of 1000 days.
We note that some OBe binaries are known to have very long periods, such as AzV 493 ([M2002] 77616), which has a period of at least 2656 days (7.28 years) \citep{77616paper}, X Per (250 days) and RX
J0146.9+6121 (330 days) \citep{Reig2011}. 
Our BPASS model calculates that $43\%$ of the OBe BPASS binaries should have NS companions while $57\%$ of them should have BH companions. 

Our OBe observations similarly show evidence of two populations.
Figure \ref{fig:BPASS_mass_distribution1} overplots the constraints on our observed objects derived in Section~\ref{sec:binprop}.
Based on the mass estimates in Table \ref{table:OBebinaryresults}, there are 3 OBe binaries with NS companions ([M2002] 77458, 77851, and 82711). All of them are identified as HMXBs in the literature, 
with confirmed neutron stars. 
Appendices \ref{App:77458}, \ref{App:77851} and \ref{App:82711} give details on the individual objects. These masses and periods also correspond to the NS companion regime from BPASS in Figure \ref{fig:BPASS_mass_distribution1}, as shown by the blue squares. 

While we do not have period estimates for 
the rest
of our observed OBe BSS binaries, we can place them in Figure \ref{fig:BPASS_mass_distribution1} by using our estimated $M_{\rm 2,min}$ values from Table \ref{table:OBebinaryresults}.
We caution that the companion mass estimates are lower limits, thus they
could be placed in either the NS regime or BH regime in Figure~\ref{fig:BPASS_mass_distribution1}.
One of them, [M2002] 77616 (AzV 493) (Appendix~\ref{App:77616}), is likely in the BH regime since our current lower mass limit (1 $M_{\odot}$) places this target outside of the high-probability contours in Figure~\ref{fig:BPASS_mass_distribution1}.
For the observed OBe star mass, the companion mass must be in the BH regime in order to reside on the higher probability contours.  

There is one additional OBe binary,
[M2002] SMC-76773 ($M_2 > 8.95 M_{\odot}$), that has
a companion whose estimated mass exceeds the $3 M_{\odot}$ neutron star threshold, placing it more solidly in the BH regime
 (Appendix~\ref{App:76773}). 
In Figure \ref{fig:BPASS_mass_distribution1}, this object is also found in the BH mass regime for OBe BSS companions. 
 However, this may be a spurious candidate, since the large $\Delta$RV values are driven by a single observation that is an RV outlier (Appendix ~\ref{App:76773}).
According to our BPASS model results, the periods of 
BH binaries,
such as these objects, 
typically have values $\sim 1000$ days, but
our analysis in Section~\ref{sec:binprop} 
for most BSS binaries are lower limits
and insensitive to such long periods.
  
For OB stars, Table \ref{table:BPASSpop} shows that OB BSS binaries comprise $55\%$ of the model's OB field population. 
Figure \ref{fig:BPASS_mass_distribution1} shows that OB BH systems have periods that are on the order of a few to 1000 days, 
while those for binaries with NS companions range from 1 - 100 days. 
Figure \ref{fig:BPASS_mass_distribution2} shows that the majority of the OB system periods lie within the overlap in range, 10 - 100 days.
Table \ref{table:BPASSpop} also shows that 90\% of OB BSS binary companions are predicted to be BHs. 
However, we caution that the observed OB and OBe binaries in the table include objects that are pre-SN, which are not included in the BPASS models.
For our 4 observed OB BSS binaries ([M2002] 76371, 80573, 82444 and 83510) we are only able to determine lower limits for companion masses (Table \ref{table:OBbinaryresults}). 
We calculated $M_{\rm 2,max}$ values that assume the companions are  non-compact stars (Section~\ref{sec:OBmasses}), however, if these 
indeed turn out to be BSS objects, the companions could be BHs with higher masses.
Recalling that our survey is biased toward detecting
binaries with shorter periods, it may still be likely that we are detecting NS systems, even though they have an expected frequency of only 10\%, because they tend to have short periods.  
Thus, it is unclear what kind of compact companions to expect for these 4 OB binaries.

\section{DES Binaries} \label{sec:OBbinaries}

DES ejections from the parent cluster usually occur from close encounters with a binary or higher-order multiple system. Observations of clusters 
show that 
due to mass segregation, the most massive stars are found near the centers of clusters \citep{LadaLada2003}, which increases the probability of interactions that eject them into the field. Simulations also show that this generates a massive, ``bully binary" in the cluster's center that dominates the cross section for these dynamical interactions \citep{Fujii2011}. This leads the DES runaway population to be weighted towards higher-mass stars, and the runaway frequency to increase with stellar mass \citep{Perets2012}.

Within this scenario, binaries can be ejected as binary runaways, and thus non-compact binaries are a direct probe of DES ejections. \citet{OhKroupa2016} and \citet{OhKroupa2015} performed N-body simulations with different initial conditions for the binary parameters of the massive star population. These simulations show how the fraction and orbital parameters of ejected single and binary systems 
depend on the initial parameters of the binary population in the parent clusters and cluster parameters. 

Additionally, \cite{Oey2018} found that noncompact binaries like EBs and SB2s reach much higher velocities than BSS tracers such as HMXBs, consistent with expectations for
DES ejections \citep[see also][]{Grant2024}. Non-compact binaries are also important since they can be progenitors of two-step ejections, in which the system experiences a SN, reaccelerating the surviving star \citep{Pflamm2010}.

Here, we examine the non-compact binaries in our sample in the context of the DES mechanism, and evaluate
whether their
frequencies and velocities are 
consistent with predictions for binaries produced by DES ejections.  We also probe the importance of the DES mechanism in populating the OB/e binary field stars.

\subsection{DES OB stars and OB binaries} \label{sec:DESOBBin}

Previous studies suggest that field OB stars, in contrast to OBe stars, are dominated by DES ejections \citep{DorigoJones2020, Grant2024}. 
In Table \ref{table:BPASSpop}, our BPASS models indicate that 
BSS field stars correspond to 16\% of the total OB population.
However, we see that 
twice as many field stars (35\%) are observed.
We might therefore attribute the excess field stars to DES ejections.
In Figure \ref{fig:BPASS_VelHist}, we plot the velocity distribution of our BPASS model OB stars along with all the SMC
OB field stars from \cite{Grant2024}. We see clear differences in the cumulative velocity distributions, where there is 
an excess of high-velocity objects, especially among the OB stars. 
Since DES objects have higher average velocities than BSS objects, this is consistent with the contribution of DES objects, as well as likely contribution of dynamical kicks to many BSS objects \citep[e.g.,][]{Grant2024}. 

In particular, non-compact binaries must all be DES ejections.
We have 16 unique OB binaries and candidates out of 34 targets, representing a total of $\sim 47\%$ of our total OB stars. Of these, 10 are confirmed binaries, indicating a lower limit of $\sim 29\%$. 
This range of values is consistent with the N-body simulations of \citet{OhKroupa2016}, which have model-dependent frequencies of $20 - 60$\%, with large dispersions.
Allowing targets to share multiple classifications (Table \ref{table:OBbinaryresults}), they can be divided as: 14 RV binaries (9 circular orbit binaries, 3 eccentric binaries and the remaining 2 identified solely through the $F$-test), 2 EBs, and 5 SB2s. There are no HMXBs in this sample. 

Our confirmed DES binaries are the
EB and SB2 binaries (marked with triangles in Figure~\ref{fig:binaryID1}).
These include 2 EBs: one confirmed EB that is also an SB2 candidate ([M2002] 81258) and our best TESS EB candidate ([M2002] 83232). 
For the remaining 4 binaries with 
non-compact companions,
3 are candidate SB2s ([M2002] 77368, 77816, and 83073), and one is a confirmed SB2  with a B star companion ([M2002] 76253). All the binaries that have RV information ([M2002] 77368, 77816, 81258, 83073, and 83232) 
appear to have
have low eccentricity
in Figure \ref{fig:binaryID1}.
We therefore consider these all to be likely pre-SN binaries.

The target [M2002] 81941 presents an interesting case for the companion. Based on our mass estimates, the companion should have a mass $> 5.3 M_{\odot}$. However, with our current detection limit, we should have detected a star of at least this mass in our spectra (Appendix~\ref{App:81941}). 
Thus, the companion is a good black hole candidate.

Figure \ref{fig:binaryID1} (left panels) demonstrates that the SMC Wing OB field binaries show a preference for low-eccentricity orbits.  Systems with orbital periods on the order of 10 days 
correspond to orbital speeds of $\sim 50 - 200$ \kms\ for OB binaries 
in Figure~\ref{fig:binaryID1}.  These systems experience strong tides and circularize quickly \citep[e.g.,][]{deMink2009, Eldrige2009}; it is therefore unsurprising that these tight binaries have low $e$.  However, if indeed most of the other OB binaries also favor circular orbits, this would be significant.  \citet{OhKroupa2016} show that dynamical ejections in general do not greatly affect the eccentricities of the ejected systems, and therefore, either our OB binaries are able to circularize on long timescales as predicted by \citep[e.g.,][]{Hurley2002}, or they were born with relatively circular orbits.
Their simulation for a binary population with primordial circular orbits is also consistent with a somewhat higher binary ejection frequency, which may be suggested by our observations above. 
Unfortunately, as noted earlier, it is more difficult to evaluate eccentricty for longer-period binaries.

\subsection{Non-Compact OBe Binaries} \label{sec:OBeDES}

As discussed in Section \ref{sec:OBeVeldist},
DES ejections likely contribute substantially to the kinematics of the field OBe star population \citep[e.g.,][]{Grant2024, DorigoJones2020}, including to field non-compact OBe binaries. 
These objects represent a unique class that is 
observed after mass transfer has started, but before the SN event.
Indeed, EBs and SB2s constitute a small subset of our OBe binaries.

Our sample includes 1 confirmed EB system from OGLE, [M2002] 73355, which represents 2\% of the field SMC Wing population. 
We additionally have
2 candidate OBe EBs ([M2002] 81634, 83171), and 2 
candidate OBe SB2s ([M2002] 72535, 83224), which, when combined with the confirmed EB, represent 9\% of our sample. 
Our BPASS models predict that 2\% of the total OB population should correspond to OBe pre-SN binaries (Table~\ref{table:BPASSpop}), but we note that this prediction is for an entire population including clusters, whereas our sample consists of only field objects.
However, it is interesting that our observed frequency of field-star, pre-SN OBe systems seems to hint at being higher than the prediction for the total population.  Since OB/e stars are preferentially found in clusters, it is noteworthy that {\it any} non-compact OBe binaries are field objects, given how rare they are.  Our results could imply that these systems are strongly biased toward the field, presumably due to DES ejections.
We note that at least one additional 
SMC OBe star that is an OGLE EB is known outside our sample ([M2002] 30744), and it is also a field star \cite{Dallas2022}.

\subsection{Stripped-star Binaries} \label{sec:strippedstars}

It is possible that massive binaries with undetected companions have stripped helium-star secondaries. These are lower-mass siblings of Wolf-Rayet stars that are not luminous enough to drive the optically thick stellar winds that are a key feature of that stellar class.  Such stars are typically far hotter than OB stars with surface temperatures that approach 100,000~K. \citet{2018A&A...615A..78G} showed that they are very difficult to observe when they are orbiting around an OB star that has likely accreted much of the hydrogen envelope that was lost, and thus they may be present 
in our sample, in binaries with only one visible star. 
Examples of the stripped helium stars were recently discovered in the LMC and SMC \citep{2023Sci...382.1287D,2023ApJ...959..125G}, although none of these known stripped stars are in our sample of SMC Wing OB/e binaries.

Stripped stars are expected to be rare. 
Once a stripped star is formed, it has at most 10\% of its lifetime remaining, while,
in contrast, the rejuvenated companion may have an extended main sequence lifetime. 
It is therefore more likely that the observed binaries with unseen companions contain the post-SN compact remnant of a stripped star. 
The predicted BPASS frequency for OB and OBe binaries with stripped star companions is 0.012 and 0.005, respectively, for the total OB/e population (Table~\ref{table:BPASSpop}).
This corresponds to about 2\% and 24\% of the OB and OBe pre-SN binaries, and 1.5\% and 7.5\% of all OB and OBe binaries, respectively.  Thus, it is extremely unlikely that any of our OB RV binaries host a helium star.  However, 
for our total of 17 observed OBe binaries, we could expect that one is a pre-SN, helium-star system.
While our sample excludes binaries in clusters, the stripping of helium stars take place late in the lifetime of the mass donor, leaving time to eject the system by the DES mechanism.  Thus, the bias against cluster objects may not be a strong selection effect against finding stripped-star binaries.

A possible parameter to help identify helium-star systems is the eccentricity.
Systems still hosting a stripped star would be expected to have circular orbits, since
tidal forces during the binary interactions could not have avoided circularizing the system. In our sample, considering that OBe binaries are often eccentric, we can identify candidate systems as those with low eccentricity.

We have at least 2 RV binary OBe stars
that are not known HMXBs, EBs, or SB2s, and 4 more candidate such objects.  Thus, any of these could host a helium star companion;  
Table~\ref{table:BPASSpop} shows that there are $\sim5\times$ more BH companions expected than He stars.  
Of the 8 total OBe systems mentioned, the objects whose $\Delta$RV data are consistent with circular orbits are especially interesting:
3 candidate binaries ([M2002] 71652, 76654, 82328) and 1 confirmed binary [M2002] 76773. 
These candidates generally have $M_{\rm 2,min}$ in the NS mass range.

\subsection{DES vs BSS} \label{sec:DESvsBSS}

\begin{deluxetable*}{lcccc}
\tablecaption{DES and BSS Walkaway and Runaway Ratios for Field SMC Wing Binaries\tablenotemark{a} \label{table:DESvsBSS}}
\tablewidth{700pt}
\tabletypesize{\scriptsize}
\tablehead{
\colhead{Ratio } & \colhead{Conf. + cand.} &\colhead{Confirmed} &\colhead{Conf. + cand.} &\colhead{Model \tablenotemark{b}}
\\
\colhead{ } & \colhead{New alloc.} &\colhead{New alloc.} &\colhead{Simple alloc.} &\colhead{Simple alloc.}
} 
\startdata
Total W/R &        1.5    &  1.8 & 1.5 & 1.7 \\
OB W/R &         1.6   &  1.6 &1.6& 1.5 \\
OBe W/R &        1.4    &  1.9 & 1.4& 3.8\\
DES W/R &        1.1    &  1.8 & 1.6 & 1.5\\
BSS W/R &        2.1    &   1.7 & 1.4 & 3.8\\
Runaway DES/BSS &        1.4    &  0.57 &1.3 & 2.7 \\
Walkaway DES/BSS &        0.71   & 0.60 &0.70& 0.50\\
\enddata
\tablenotetext{a}{
Columns 2 and 3 allocate DES and BSS objects according to our revised scheme
(Section \ref{sec:DESvsBSS}) and Column 4 follows the simple allocation that
all OB and OBe stars are DES and BSS ejections, respectively; this is also assumed in the model in Column~5.}
\tablenotetext{b}{From Tables 3 and 4 of \cite{Grant2024} (see text). We consider the DES contributions from pre-SN binaries; and for BSS contributions, the bound BSS targets and half of 
the bound the two-step population.}
\end{deluxetable*}

The binary kinematics provide fundamental constraints on the relative contributions of DES and BES ejections. 
To compare our observational results against models, we assign our SMC Wing binaries as runaways (R) and walkaways (W), using the
transverse velocity threshold of $24$ km/s for runaways, 
following \citet{Grant2024}.  
This corresponds to
a 30 km/s 3-D space velocity threshold.
As discussed by \cite{DorigoJones2020}, the parent RIOTS4 field star sample 
has a selection bias against walkaways, 
which tend to remain closer to their parent clusters;  
we adopt their correction factor of 2.4 to adjust the total number of walkaways to account for this effect.

The SMC Wing sample has a total of 31 binaries with Gaia velocities, with 19 runaways (9 OB, 10 OBe) and 12 walkaways (6 OB, 6 OBe). If we consider only confirmed binaries, we have a total of 19 objects with Gaia velocities: 11 runaways (6 OB, 5 OBe) and 8 walkaways (4 OB, 4 OBe). We use these targets to compare the ratio of walkaways to runaways (W/R).
Previous work by \citet{Grant2024} and \citet{DorigoJones2020} had simply assigned DES ejections to OB stars and BSS ejections to OBe stars, 
allowing initial estimates of the relative contributions of each mechanism to the field star population. 
As discussed above, this allocation is overly simplistic.
Our new criteria for DES binaries 
now include non-compact OBe binaries (Section~\ref{sec:OBeDES}), and
our BSS objects now include
OB targets that are fast rotators and eccentric binaries (Sections~\ref{sec:BSSrot} and \ref{sec:BSSecc}). 
This leaves us with a total of 15 BSS objects (3 OB fast rotators, 1 OB eccentric binary, 3 HMXBs, 8 remaining OBe) and 16 DES objects (6 non-compact OB binaries, 5 non-compact OBe binaries, 5 remaining OB). For confirmed binaries this separates into 12 BSS objects (3 OB fast rotators, 1 OB eccentric binary, 3 HMXBs, 5 remaining OBes) and 7 DES objects (2 non-compact OB, 1 non-compact OBe, 4 remaining OB). 

We obtain the resulting ratios of W/R and DES/BSS for the two different ways of allocating DES versus BSS objects, for
our SMC Wing binaries.  Columns 2 and 3 of Table \ref{table:DESvsBSS} show results for our new allocations when calculated for the subsamples including both confirmed plus candidate binaries, and confirmed binaries only, respectively.  Column 4 gives these values assuming the more simplistic allocation used by, e.g., \citet{Grant2024}, for confirmed plus candidate binaries.  The last column shows the same quantities for the model in that work. This model is updated from the one by \cite{DorigoJones2020} and combines independent DES and BSS simulations from \cite{OhKroupa2016} and \cite{Renzo2019}, respectively, to estimate the expected kinematics produced by ejections from a single stellar population. 
The model is tuned to optimize comparison with the SMC population and
assumes the simple
allocation of OB stars to DES ejections and OBe stars to BSS ejections.
In this work, we only consider ejected binaries in the model and not single stars, i.e.,
pre-SN binaries, post-SN binaries, and bound two-step ejections.
As noted by \cite{Grant2024}, two-step ejections potentially comprise a significant portion of the field stars; we assume here that bound systems comprise half the population of two-step ejections.

Table~\ref{table:DESvsBSS} shows that our new criteria for refining the identification of DES and BSS stars beyond the simple breakdown between OB and OBe stars makes a significant difference.
Whereas \cite{Grant2024} found that observed DES ejections dominate the binary runaways by a factor of 1.3, Table \ref{table:DESvsBSS} shows that this result is sustained only if all the candidate binaries are confirmed.  Indeed if none of them are confirmed, then the BSS objects strongly dominate, suggesting that the relative contributions 
of BSS and DES products
may be fairly similar.  There is no significant change in the observed DES/BSS ratio of 0.6 -- 0.7 for binary walkaways, however, where BSS objects dominate.

When comparing to the model however, Table~\ref{table:DESvsBSS} shows that it overpredicts the observed binary runaway DES/BSS ratio by roughly a factor of 2.  Moreover, the predicted binary BSS W/R ratio is also roughly a factor of 2 larger than the observations.  These both suggest that the BSS binaries are traveling faster than predicted by \citet{Grant2024}, based on the binary population synthesis models of \citet{Renzo2019}.  This may point to the treatment of two-step ejections in the BSS population, where \citet{Grant2024} assumed that 20\% of two-step walkaways are reaccelerated to runaway velocities; we also assumed above that half of two-step systems remain bound post-SN. 
In any case, the observations suggest that for binary systems, walkaways are dominated by BSS objects while runaway systems may be dominated by DES ejections.  This is consistent with expectations that the DES mechanism can accelerate systems to higher velocities.

We note that the model was calibrated to the assumed DES vs BSS branching ratios based on the simple OB vs OBe allocations, and based on the entire SMC population including single stars.  Table~\ref{table:DESvsBSS} shows that the agreement with the model is therefore somewhat better for the assumed simple allocations, but the trend of the discrepancies above remains.
We also caution that our results are affected by small number statistics. For example, our confirmed binary walkaways comprise 5 DES objects and 7 BSS objects;
if one of the BSS objects were identified as a DES object, it 
would change 
the current DES/BSS ratio of 0.71 to 1.0, shifting the balance between the two mechanisms.

\section{Conclusions} \label{sec:conc}

We expand on previous massive binary studies by identifying new RV, EB and SB2 binaries and candidate binaries among 55 RIOTS4 targets of the SMC Wing 
field stars. 
Our sample includes not just OB targets, but also OBe stars.
Together with 
previously confirmed EBs, SB2s and HMXBs, we obtain a more complete estimate of the binary frequency in the SMC Wing.

We identified new RV binaries via a spectroscopic monitoring campaign of the SMC Wing by obtaining multi-epoch observations in the period 2016 June to 2018 August using
the Magellan IMACS and M2FS multi-object spectrographs.
We use the cross-correlation code of
\citet{Becker2015}
to extract RV measurements. 
To identify our RV binaries, we use two methods following \cite{Lamb2016}: 
a comparison of $\Delta$RV(10d) versus $\Delta$RV(tot),
and the $F$-test, which 
evaluates the
probability that observed variations are due to statistical noise \citep{DuquennoyMayor1991}. 
Using these methods, we identify 11 RV binaries and 17 
candidate binaries.
Furthermore, we also use the spectroscopic data to identify 6 new SB2 candidates by visually comparing the spectra of each target taken at different epochs.

We identify new EBs using TESS light curves 
from the field OB sample of \cite{Oey2004}. 
We search for photometric variability in the light curves 
by using two methods 
that compare the large-scale
photometric trends to the noise. 
Periodic variables are identified based on the 
periodograms of their light curves.
From visual inspection of the light curves, we then identify 
3 new EB candidates in the SMC Wing
from the identified periodic variables.

Combining all of these new binary identifications with previously known ones, we obtain a total of 20 unique confirmed binaries and 13 unique candidates for a total of 33 binaries in the SMC Wing. This leads to a binary fraction of $60\%$ if we consider both confirmed binaries and binary candidates and a fraction of $36\%$ if we only consider confirmed binaries. This value is consistent with the results of \cite{Lamb2016} and is likely a substantial lower limit,
since our binary identification is biased against systems with longer periods and lower companion masses.

Using our RV results, we use
the radial velocity semi-amplitude equation \citep{Fischer2014} to obtain lower limits on companion mass $M_2$ (Tables \ref{table:OBbinaryresults} and \ref{table:OBebinaryresults}), while upper limits are based on our spectroscopic detection limits for the case that $M_2$ represents a non-compact star.
Known periods, eccentricities and other properties
for some systems additionally set constraints on $M_2$,
and sometimes we are also able to set additional constraints on orbital parameters.

We compare our observational results to model BSS OB and OBe field populations by using BPASS to generate a synthetic population of BSS ejected field stars and binaries that includes information on periods, companion mass, eccentricities, and velocity distributions. We compare these populations to our SMC Wing binary data, as well as the field population of OB and OBe stars from \cite{Dallas2022}, and velocities from \cite{Grant2024}. We find the following: 
\begin{enumerate}

   \item Our RV measurements yield important information on observed eccentricities that clearly differentiate the OB and OBe populations. Our data suggest that OB binaries on average have quite circular orbits (mean $e=0.08 \pm 0.02$) while OBe binaries 
   may
   have an average $e = 0.45 \pm 0.04$.  The circular orbits suggest that OB binaries are dominated by a pre-SN population, since DES ejections should not significantly change the $e$ distribution \citep{OhKroupa2016}, and the pre-SN nuclear timescale allows circularization if the population is not born with low $e$ \citep[e.g.,]{Eldrige2009}.  Conversely, 
    large
    OBe eccentricities are consistent with originating from SN kicks in our BPASS models, supporting their origin as post-interaction mass-gainers that are also post-SN objects. 
    
    \item BPASS OBe BSS binaries comprise 40\% of the OBe field population.  This is lower than the observed frequency of 48 -- 81\%, which may be affected by selection bias.  BPASS models predict that there are two populations of OBe BSS stars based on the distributions of periods, eccentricities and remnant mass: one with NS companions and periods on the order of 1 -- 1000 days, and one with BH companions and long periods ($\sim 1000$ days). 
    Our models predict that $\sim43$\% of our systems should host NSs while 57\% of them would have BHs. 
    
    \item Our observed OBe systems similarly show evidence of two populations. Our HMXBs match the model's parameters for NS systems, and all of these indeed have confirmed NS companions. 
    Additionally, we have 1, and possibly 3, OBe binaries with BH candidates.
    The remaining OBe binaries have lower limits for their companion mass, meaning that they could have either a NS or BH. We note that BH OBe binaries have much longer periods than our observations can detect, and therefore
    we are biased toward identifying NS OBe binaries.
    
    \item 
    The models show that the high primary mass of the Oe star [M2002] 77616 likely implies that its companion is a BH. 
    Additionally, our analysis of the O9.5 III star, [M2002] 81941, indicates that $M_2 > 5.3 M_{\odot}$. A non-compact companion of this mass should have been detected in our observations, indicating that it is also potentially a BH.
    
    \item For the BPASS OB BSS binaries, which comprise 55\% of the
    predicted
    field OB population, the majority (90\%) are binaries with a BH companion, and they have periods typically of 10 -- 100 days, and a broad range of eccentricities.  As noted above, we have at least one OB binary with a BH candidate.

\item In our SMC Wing binary sample, we find 1 OBe star that is a confirmed non-compact binary, [M2002] 73355, and 2 additional candidates.  This class of objects may represent systems where mass transfer has been initiated, but before the first SN explosion.  Our BPASS models indicate that these are only expected to represent only 2\% of the entire OB/e population.
    
    \item We also find
    that 4 of our OB binaries are likely OB BSS ejections:
    3 fast rotators and 1 eccentric binary. This population is an important subset of binaries that are post-interaction and pre-SN. We currently have only lower limits for $M_2$ for these OB binaries, meaning that their companions could be either NS or BH. Our BPASS models predict that 90\% of such companions are BH, but as noted earlier, our survey is biased against BH systems.
    
    \item 
    We find that binaries hosting stripped-star companions are expected to be rare.  Our BPASS model predicts that about 1.5\% and 7.5\% of OB and OBe binaries, respectively may host stripped He stars; these values apply to the entire OB/e population, including stars in clusters.  We might expect one such OBe system in our sample.

    \item We currently underpredict the number of OBe stars in our sample by a factor of 2. There is a variety of possible reasons, but
    the discrepancy underscores that there are still multiple effects, both observational and theoretical, that remain unaccounted for in our comparisons.
    
\end{enumerate}

The dynamical ejection mechanism is also responsible for a large fraction of our field star sample.
Previous work had assigned DES ejections primarily to OB stars and BSS ejections primarily to OBe stars \citep{DorigoJones2020,Grant2024}. 
Instead of using this simple allocation for DES and BSS binaries, we employ new criteria. For our DES objects, we classify most OB binaries as DES objects
but we also include non-compact OBe binaries. 
For our BSS objects, we classify the remaining OBe binaries as BSS objects, but now also including
OB targets that are fast rotators and eccentric binaries. 
We have a total of 15 BSS objects and 16 DES objects in the total sample and 12 BSS objects and 7 DES objects in the confirmed sample. These new criteria cause significant revisions to the contributions of DES and BSS ejections in the OB and OBe populations.

We use the stellar kinematics, binary data, and also differences between the BPASS predictions and our observations, to 
constrain
the properties of DES ejections in our observed sample. 
\begin{enumerate}    
    \item BPASS models show that 16\% of OB field stars are ejected through BSS. However, we see that twice as many field OB stars (35\%) are observed. We therefore attribute the excess 
    to DES OB ejections. 
    
    \item  OB binaries that are not BSS candidates show a preference for circular orbits,
    rather than the predicted eccentric orbits.  This is 
    consistent with them being dominated by a pre-SN population and are therefore ejected by the DES mechanism.
    
    \item 
    We have 6 OB non-compact binaries, which are 
    pre-SN systems and therefore  DES ejections. 
    
    \item Our sample includes
    1 non-compact OBe binary and 4 candidate non-compact OBe binaries. These objects represent 
    systems that have initiated mass transfer that are also observed before the SN event. Considering both confirmed and candidate binaries, these represent $29 \% $ of the OBe Wing binaries. This may be an unexpectedly large number of non-compact binaries in the OBe population when we would normally expect the probability of finding these to be on the order of 2\% for an entire OB/e population.

\item We find that post-SN, bound BSS binaries appear to be traveling faster than expected, as indicated by the ratio of runaways to walkaways.
This may be due to the contribution of two-step ejections and needs further investigation.  
We confirm that walkaway binaries are dominated by BSS objects while runaways may be dominated by DES ejections.

\end{enumerate}.

An important caveat to our results is that they are currently affected by small number statistics. For example, our confirmed binary walkaways comprise 5 DES objects and 7 BSS objects.
If only one of these BSS objects turns out to be a DES object, it would
change the current observed DES/BSS ratio of 0.71 to 1.0, significantly shifting the balance between the two mechanisms.

Thus, massive star binaries offer powerful insight into
different facets of massive star evolution and populations.
This includes
understanding the interactions and evolution of massive stars within binary systems and within the parent cluster population. Revealing 
how these systems evolve in the field through their ejection mechanisms and the presence of their compact companions 
clarifies our understanding of how binary parameters generate 
a host of crucial processes that drive the evolution of stellar populations and their host galaxies, including
feedback effects, chemical evolution, and the production of various transients including gravitational wave events.

\acknowledgments
We thank Rachel Chen and Yiting Li for help with some data analysis, and we thank Matthew Dallas, Johnny Dorigo Jones, Lee Hartmann, Grant Phillips, and Heloise Stevance for useful discussions of this project.  
Additionally, we thank the referee for important comments.
We are also grateful to Myungshin Im, Dohyeong Kim, Yongmin Yoon, and Yoon Chan Taak for help with observing.
Finally, we thank Jeff Crane, Steve Shectman and Ian Thompson for their help in developing, deploying and supporting M2FS at the Magellan/Clay telescope. Some of the data presented in this paper were obtained from the Mikulski Archive for Space Telescopes (MAST) at the Space Telescope Science Institute
\citep{https://doi.org/10.17909/0cp4-2j79}. 
This work was supported by NSF grant AST-1514838 to M.S.O. and by the University of Michigan, including through a Rackham Graduate School Predoctoral Fellowship to I.V.S.
N.C. acknowledges funding from the Deutsche Forschungs-gemeinschaft (DFG) - CA 2551/1-1,2. M.M. and J.B. acknowledge support from NSF via grant MRI/NSF-0923160 to fund the design, development and deployment of M2FS. Y. K. was supported by the National Research Foundation of Korea (NRF) grant funded by the Korean government (MSIT) (No. 2021R1C1C2091550).

\vspace{5mm}
\facilities{Magellan:6.5m (IMACS and M2FS), OGLE, MAST(TESS)}

\software{COSMOS \citep{COSMOS1,COSMOS2}, 
          IRAF \citep{IRAF}, eleanor \citep{FeinsteinMontet2019}, lightkurve \citep{LightkurveCollaborationCardoso2018}, BPASS v2.2 \citep{2017PASA...34...58E,2018MNRAS.479...75S}
          }

\clearpage
\appendix

\section{Individual RV Measurements} \label{App:RVtable}

This appendix contains all the individual RV measurements from the cross-correlation analyses
of all 55 SMC Wing targets. 
As described in Section \ref{subsec:RVmeas}, these are done using model spectral templates, and for some OBe stars, templates corresponding to a high signal-to-noise individual epoch.
Table \ref{table:RVmeas} gives the ID of the target from \cite{Massey2002}, spectral type (SpT) from \cite{Grant2024} and \cite{Dallas2022}, the R.A. and decl. from \cite{GaiaDR3}, the instrument used (IMACS or M2FS), the multi-object spectrograph field to which the target belongs, the 
Julian Date (JD) of the observation,
and the RV measurement and error we obtain.  
The quoted errors are calculated by the cross-correlation code as described in Section~\ref{subsec:RVmeas}.

Table \ref{table:RVmeas} includes data for both binaries and single stars.  We note that additional RV data for some of our stars may be available in the new survey by \citet{Shenar2024}.

\startlongtable
\begin{deluxetable*}{cccccccccc}
\tablecaption{RIOTS4 Wing RV Measurements\label{table:RVmeas}}
\tablewidth{700pt}
\tabletypesize{\scriptsize}
\tablehead{
\colhead{Massey ID} & \colhead{SpT} & \colhead{RA} & 
\colhead{DEC} & \colhead{Instrument} & 
\colhead{Field} & \colhead{Date} & 
\colhead{RV} & \colhead{RV Err}  \\ 
\colhead{[M2002] SMC} & \colhead{} & \colhead{hrs} & \colhead{deg} & 
\colhead{} & \colhead{} & \colhead{JD} &
\colhead{km/s} & \colhead{km/s}
} 
\startdata
71652 & B0.5$e_2$ & 17.74243 & -73.30373 & IMACS f/4 & FLDob101 & 2457554.215 & 221 &  17 \\
& & & & IMACS f/4 & FLDob101 & 2457555.424 & 189 & 14 \\
& & & & IMACS f/4 & FLDob101 & 2457728.149 & 176 & 15 \\
& & & & IMACS f/4 & FLDob101 & 2457910.284 & 163 & 15 \\
72210 & B0$e_3$ & 17.85817 & -73.29106 & IMACS f/4 & FLDob101 & 2457554.215 & 186 &  26 \\
& & & & IMACS f/4 & FLDob101 & 2457728.149 & 184 & 14 \\
& & & & IMACS f/4 & FLDob101 & 2457910.284 & 194 & 21  \\
72535 & O8-9:IIIp$e_1$ & 17.93123 & -73.23154 & IMACS f/4 & FLDob101 & 2457554.215 & 189 &  16 \\
& & & & IMACS f/4 & FLDob101 & 2457555.424 & 165 & 18 \\
& & & & IMACS f/4 & FLDob101 & 2457728.149 & 166 & 19 \\
& & & & IMACS f/4 & FLDob101 & 2457910.284 & 160 & 17 \\
& & & & M2FS & SMCField10 & 2457943.872 & 187 & 1 \\
& & & & M2FS & SMCField10 & 2457945.792 & 186 & 3 \\
& & & & M2FS & SMCField10 & 2458350.856 & 199 & 7 \\
& & & & M2FS & SMCField10 & 2458352.820 & 187 & 1 \\
73355 & B0$e_2$ & 18.12389 & -73.29161 & IMACS f/4 & FLDob101 & 2457554.215 & 221 &  15 \\
& & & & IMACS f/4 & FLDob101 & 2457555.424 & 189 & 10 \\
& & & & IMACS f/4 & FLDob101 & 2457728.149 & 197 & 13 \\
& & & & IMACS f/4 & FLDob101 & 2457910.284 & 168 & 19 \\
& & & & M2FS & SMCField10 & 2457942.695 & 203 & 8 \\
& & & & M2FS & SMCField10 & 2457943.872 & 236 & 1 \\
& & & & M2FS & SMCField10 & 2458346.817 & 211 & 2 \\
& & & & M2FS & SMCField10 & 2458350.856 & 247 & 6 \\
& & & & M2FS & SMCField10 & 2458352.820 & 208 & 2 \\
73913 & O9.5 I & 18.25167 & -73.28448 &	IMACS f/4 & FLDob101 & 2457554.215 & 211 & 17 \\
& & & & IMACS f/4 & FLDob101 & 2457555.424 & 171 & 14 \\
& & & & IMACS f/4 & FLDob101 & 2457728.149 & 184 & 16 \\
& & & & IMACS f/4 & FLDob101 & 2457910.284 & 175 & 15 \\
& & & & M2FS & SMCField10 & 2457942.695 & 152 & 4 \\
& & & & M2FS & SMCField10 & 2457943.872 & 158 & 3 \\
& & & & M2FS & SMCField10 & 2457945.792 & 156 & 4 \\ 
& & & & M2FS & SMCField10 & 2458343.837 & 178 & 25 \\
& & & & M2FS & SMCField10 & 2458346.817 & 169 & 4 \\
& & & & M2FS & SMCField10 & 2458350.856 & 151 & 4 \\ 
74828 & Be & 18.48876 & -73.15568 & M2FS & SMCField10 & 2457942.695 & 233 & 12 \\ 
& & & & M2FS & SMCField10 & 2457943.872 & 232 & 12 \\ 
& & & & M2FS & SMCField10 & 2458346.817 & 232 & 12 \\
& & & & M2FS & SMCField10 & 2458350.856 & 233 & 14 \\
& & & & M2FS & SMCField10 & 2458352.820 & 232 & 15 \\
75061 & B1$e_2$ & 18.55496 & -73.34586 &	IMACS f/4 & FLDob101 & 2457554.215 & 238 & 9 \\
& & & & IMACS f/4 & FLDob101 & 2457555.424 & 176 & 16 \\
& & & & M2FS & SMCField10 & 2457942.695 & 235 & 1 \\
& & & & M2FS & SMCField10 & 2457943.872 & 210 & 2 \\
& & & & M2FS & SMCField10 & 2457945.792 & 228 & 2 \\
& & & & M2FS & SMCField10 & 2458346.817 & 280 & 4 \\
& & & & M2FS & SMCField10 & 2458350.856 & 236 & 1 \\
& & & & M2FS & SMCField10 & 2458352.820 & 250 & 2 \\
75126 & O9 V & 18.57137 & -73.26366 &	IMACS f/4 & FLDob101 & 2457554.215 & 204 & 15 \\
& & & & IMACS f/4 & FLDob101 & 2457555.424 & 213 & 21 \\
& & & & IMACS f/4 & FLDob101 & 2457728.149 & 203 & 10 \\
& & & & IMACS f/4 & FLDob101 & 2457910.284 & 214 & 16 \\
& & & & M2FS & SMCField10 & 2457942.695 & 205 & 1  \\
& & & & M2FS & SMCField10 & 2457943.872 & 207 & 1  \\
& & & & M2FS & SMCField10 & 2457945.792 & 206 & 1 \\
& & & & M2FS & SMCField10 & 2458343.837 & 208 & 3 \\
& & & & M2FS & SMCField10 & 2458346.817 & 205 & 1  \\
& & & & M2FS & SMCField10 & 2458352.820 & 206 & 1 \\
75210 & O8.5 V & 18.59407 & -73.22313 & M2FS &	SMCField10	& 2457942.695 & 149 & 2 \\
& & & & M2FS & SMCField10 & 2457943.872 & 150 & 1 \\
& & & & M2FS & SMCField10 & 2457945.792 & 149 & 1 \\
& & & & M2FS & SMCField10 & 2458346.817 & 190 & 2 \\
& & & & M2FS & SMCField10 & 2458352.820 & 191 & 2 \\
75626 & O9III-V & 18.71228 & -73.11373 & IMACS f/4 & FLDob102 & 2457554.272 & 178 & 10 \\
& & & & IMACS f/4 & FLDob102 & 2457726.118 & 173 & 12 \\
& & & & IMACS f/4 & FLDob102 & 2457909.218 & 163 & 18 \\
& & & & M2FS & SMCField10 & 2457942.695 & 168 & 3 \\
& & & & M2FS & SMCField10 & 2457943.872 & 185 & 2 \\
& & & & IMACS f/4 & FLDob102 & 2457944.316 & 184 & 13 \\ 
& & & & M2FS & SMCField10 &  2457945.792 & 183 & 2 \\
& & & & M2FS & SMCField10 &  2458350.856 & 199 & 2 \\
& & & & M2FS & SMCField10 &  2458352.820 & 183 & 3 \\ 
75980 & B0$e_3$ & 18.81077 & -73.47938 & IMACS f/4 & FLDob103 & 2457726.173 & 151 & 21 \\
& & & & IMACS f/4 & FLDob103 & 2457909.274 & 132 & 19 \\ 
& & & & M2FS & SMCField10 &  2457942.695 & 226 & 2 \\
& & & & IMACS f/4 & FLDob103 & 2457944.378 & 201 & 17 \\ 
& & & & M2FS & SMCField10 &  2457945.792 & 227 & 1 \\
& & & & M2FS & SMCField10 &  2458346.817 & 253 & 2 \\
& & & & M2FS & SMCField10 &  2458350.856 & 231 & 3 \\
& & & & M2FS & SMCField10 &  2458352.820 & 232 & 7 \\
76253 & B0.2V+B & 18.88191 & -73.24989 & 
IMACS f/4 & FLDob102 & 2457554.272 & 187 & 14 \\
& & & & IMACS f/4 & FLDob102 & 2457726.118 & 169 & 11 \\
& & & & IMACS f/4 & FLDob102 & 2457909.218 & 162 & 11 \\
& & & & M2FS & SMCField10 & 2457942.695 & 184 & 2 \\
& & & & M2FS & SMCField10 & 2457943.872 & 173 & 1 \\
& & & & IMACS f/4 & FLDob102 & 2457944.316 & 181 & 9 \\ 
& & & & M2FS & SMCField10 & 2458346.817 & 185 & 1 \\ 
& & & & M2FS & SMCField10 & 2458350.856 & 182 & 1 \\
76371 & O9.5III & 18.91337 & -73.39708 & 
IMACS f/4 & FLDob103 & 2457554.325 & 148 & 15 \\
& & & & IMACS f/4 & FLDob103 & 2457726.173 & 190 & 12 \\
& & & & IMACS f/4 & FLDob103 &  2457909.274 & 149 & 15 \\
& & & & M2FS & SMCField10 & 2457942.695 & 212 & 2 \\
& & & & M2FS & SMCField10 & 2457943.872 & 217 & 3 \\ 
& & & & IMACS f/4 & FLDob103 & 2457944.378 & 243 & 8 \\ 
& & & & M2FS & SMCField10 & 2457945.792 & 247 & 2 \\
& & & & M2FS & SMCField10 & 2458346.817 & 125 & 2 \\
& & & & M2FS & SMCField10 & 2458350.856 & 139 & 3 \\
76654 & B$e_3$ & 18.99859 & -73.46270 & M2FS & SMCField10 & 2457942.695 & 185 & 8 \\
& & & & M2FS & SMCField10 & 2457945.792 & 226 & 4 \\
& & & & M2FS & SMCField10 & 2458346.817 & 216 & 1 \\
& & & & M2FS & SMCField10 & 2458350.856 & 214 & 5 \\
& & & & M2FS & SMCField10 & 2458352.820 & 211 & 3 \\
76657 & O9-9.5V & 19.00012 & -73.43178 &
IMACS f/4 & FLDob103 &  2457554.325 & 210 & 18 \\
& & & & IMACS f/4 & FLDob103 & 2457726.173 & 170 & 15 \\
& & & & IMACS f/4 & FLDob103 & 2457909.274 & 171 & 18 \\
& & & & IMACS f/4 & FLDob103 & 2457944.378 & 196 & 9 \\ 
& & & & M2FS & SMCField10 & 2457945.792 & 193 & 9 \\ 
& & & & M2FS & SMCField10 & 2458346.817 & 207 & 21 \\
& & & & M2FS & SMCField10 & 2458350.856 & 201 & 32 \\
& & & & M2FS & SMCField10 & 2458352.820 & 188 & 17 \\
76773 & Be & 19.03589 & -73.17875 & M2FS & SMCField10 & 2457943.872 & 220 & 1 \\
& & & & M2FS & SMCField10 & 2457945.792 & 222 & 1 \\
& & & & M2FS & SMCField10 & 2458343.837 & -243 & 34 \\
& & & & M2FS & SMCField10 & 2458346.817 & 266 & 1 \\
& & & & M2FS & SMCField10 & 2458350.856 & 264 & 2 \\
& & & & M2FS & SMCField10 & 2458352.820 & 262 & 6 \\
77290 & B0.5$e_2$+ & 19.21248 & -73.21475 &
IMACS f/4 & FLDob102 &  2457726.118 & 141 & 15 \\
& & & & IMACS f/4 & FLDob102 & 2457909.218 & 168 & 21 \\ 
& & & & M2FS & SMCField10 & 2457942.695 & 211 & 1 \\
& & & & M2FS & SMCField10 & 2457943.872 & 212 & 5 \\
& & & & IMACS f/4 & FLDob102 & 2457944.316 & 230 & 15 \\ 
& & & & M2FS & SMCField10 & 2457945.792 & 232 & 3 \\
& & & & M2FS & SMCField10 & 2458346.817 & 215 & 1 \\
& & & & M2FS & SMCField10 & 2458350.856 & 216 & 1 \\
77368 & O6V & 19.24006 &	-73.32408 & IMACS f/4 & FLDob103 & 2457554.325 & 163 & 17 \\
& & & & IMACS f/4 & FLDob103 & 2457726.173 & 210 & 18 \\
& & & & IMACS f/4 & FLDob103 &  2457909.274 & 202 & 18 \\
& & & & M2FS & SMCField10 & 2457943.872 & 156 & 9 \\
& & & & IMACS f/4 & FLDob103 & 2457944.378 & 175 & 14 \\ 
& & & & M2FS & SMCField10 & 2457945.792 & 182 & 4 \\
& & & & M2FS & SMCField10 & 2458346.817 & 206 & 4 \\
& & & & M2FS & SMCField10 & 2458350.856 & 202 &6 \\
& & & & M2FS & SMCField10 & 2458352.820 & 188 & 6 \\
77458 & B0.2$e_1$ & 19.27146 &	-73.44334 & IMACS f/4 & FLDob103 & 2457554.325 & 170 & 16 \\
& & & & IMACS f/4 & FLDob103 & 2457726.173 & 161 & 11 \\ 
& & & & IMACS f/4 & FLDob103 & 2457909.274 & 170 & 15 \\ 
& & & & M2FS & SMCField10 & 2457942.695 & 222 & 3 \\
& & & & M2FS & SMCField10 & 2457943.872 & 193 & 16 \\
& & & & IMACS f/4 & FLDob103 & 2457944.378 & 193 & 12 \\ 
& & & & M2FS & SMCField10 & 2458346.817 & 210 & 6 \\
& & & & M2FS & SMCField10 & 2458350.856 & 208 & 6 \\
& & & & M2FS & SMCField10 & 2458352.820 & 171 & 5 \\
77609 & B0.5I & 19.32521 &	-73.20005 & IMACS f/4 & FLDob102 & 2457554.272 & 182 & 14 \\
& & & & IMACS f/4 & FLDob103 & 2457726.118 & 184 & 12 \\ 
& & & & IMACS f/4 & FLDob103 & 2457909.218 & 181 & 10 \\ 
& & & & M2FS & SMCField10 & 2457942.695 & 177 &1 \\
& & & & IMACS f/4 & FLDob103 & 2457944.316 & 172 & 11 \\ 
& & & & M2FS & SMCField10 & 2458343.837 & 179 &12 \\
& & & & M2FS & SMCField10 & 2458346.817 & 174 &1 \\
& & & & M2FS & SMCField10 & 2458350.856 & 172 &2 \\
& & & & M2FS & SMCField10 & 2458352.820 & 173 &1 \\
& & & & M2FS & SMCField10 & 2457943.872 & 177 &1 \\
77616 & O3-5p$e_3$pec & 19.32704 &	-73.29795 & M2FS & SMCField10 & 2457942.695 & 189 & 51 \\
& & & & M2FS & SMCField10 & 2457945.792 & 225 & 36 \\
& & & & M2FS & SMCField10 & 2458350.856 & 160 & 40 \\
& & & & M2FS & SMCField10 & 2458352.820 & 160 & 50 \\
77734 & B1-3II & 19.37324 &	-73.40325 & IMACS f/4 & FLDob103 & 2457554.325 & 190 & 14 \\
& & & & M2FS & SMCField10 & 2457942.695 & 204 &7 \\
& & & & IMACS f/4 & FLDob103 & 2457944.378 & 167 & 19 \\
& & & & M2FS & SMCField10 & 2457945.792 & 218 &2 \\
& & & & M2FS & SMCField10 & 2458346.817 & 211 &1 \\
& & & & M2FS & SMCField10 & 2458350.856 & 204 &30 \\
77816 & B0.2III & 19.40600 &	-73.11156 & M2FS & SMCField10 & 2457942.695 & 196 &6 \\
& & & & M2FS & SMCField10 & 2457943.872 & 176 & 4 \\
& & & & M2FS & SMCField10 & 2457945.792 & 178 & 4 \\
& & & & M2FS & SMCField10 & 2458346.817 & 177 & 3 \\
& & & & M2FS & SMCField10 & 2458350.856 & 183 & 3 \\
77851 & B0.2-1$e_3$+ & 19.41736 &	-73.51404 & IMACS f/4 & FLDob103 & 2457554.325 & 154 & 15 \\
& & & & IMACS f/4 & FLDob103 & 2457726.173 & 165 & 15 \\
& & & & IMACS f/4 & FLDob103 & 2457909.274 &169 & 17 \\
& & & & IMACS f/4 & FLDob103 & 2457944.378 & 191 & 10 \\
79248 & O8.5V & 19.91584 &	-73.24719 & M2FS & SMCField11 & 2457942.779 & 179 &9 \\
& & & & M2FS & SMCField11 & 2457943.690 & 184 & 13 \\
& & & & M2FS & SMCField11 & 2458255.880 & 185 & 14 \\
79976 & B$e_3$+ & 20.19855 &	-73.55794 & M2FS & SMCField11 & 2457943.690 & 205 & 27 \\
& & & & M2FS & SMCField11 & 2457944.872 & 206 & 12 \\
& & & & M2FS & SMCField11 & 2458352.881 & 240 & 2 \\
80412 & B0.7IIe & 20.45631 &	-73.62261 & IMACS f/4 & FLDob106 & 2457555.265 & 191 & 13 \\
& & & & IMACS f/4 & FLDob106 & 2457727.151 & 170 & 14 \\
& & & & IMACS f/4 & FLDob106 & 2457909.382 & 178 & 9 \\
& & & & M2FS & SMCField11 & 2457943.690 & 185 & 7 \\
& & & & M2FS & SMCField11 & 2458255.880 & 183 & 4 \\
80545 & B0.5II & 20.54676 &	-73.44775 & IMACS f/4 & FLDob106 & 2457555.265 & 183 & 13 \\
& & & & IMACS f/4 & FLDob106 & 2457727.151 & 199 & 10 \\
& & & & IMACS f/4 & FLDob106 & 2457909.382 & 197 & 14 \\
& & & & M2FS & SMCField11 & 2457943.690 & 195 & 1 \\
& & & & M2FS & SMCField11 & 2458352.881 & 201 & 2 \\
80573 & B1V & 20.56084 &	-73.14040 & IMACS f/4 & FLDob107 & 2457555.319 & 171 & 15 \\
& & & & IMACS f/4 & FLDob107 & 2457727.194 & 178 & 16 \\
& & & & IMACS f/4 & FLDob107 & 2457910.229 & 183 & 18 \\
& & & & M2FS & SMCField11 & 2457942.779 & 217 & 12 \\
& & & & M2FS & SMCField11 & 2457944.872 & 210 & 20 \\
& & & & M2FS & SMCField11 & 2458352.881 & -4 & 1 \\
80579 & B0.7V & 20.56384 &	-73.23343 & IMACS f/4 & FLDob107 & 2457555.319 & 171 & 15 \\
& & & & IMACS f/4 & FLDob107 & 2457727.194 & 182 & 12 \\
& & & & IMACS f/4 & FLDob107 & 2457910.229 & 174 & 21 \\
& & & & M2FS & SMCField11 & 2457944.872 & 186 & 6 \\
& & & & M2FS & SMCField11 & 2458352.881 & 196 & 14 \\
& & & & M2FS & SMCField11 & 2457942.779 & 193 & 8 \\
& & & & M2FS & SMCField11 & 2457943.690 & 189 & 7 \\
80582 & B1-3IIe & 20.56657 &	-73.53782 & IMACS f/4 & FLDob106 & 2457555.265 & 189 & 17 \\
& & & & IMACS f/4 & FLDob106 & 2457727.151 & 151 & 18 \\
& & & & IMACS f/4 & FLDob106 & 2457909.382 & 170 & 11 \\
81019 & O9.5V & 20.85484 & -73.36697 & IMACS f/4 & FLDob107 & 2457555.319 & 180 & 14 \\
& & & & IMACS f/4 & FLDob107 & 2457727.194 & 164 & 19 \\
& & & & IMACS f/4 & FLDob107 & 2457910.229 & 170 & 17 \\
& & & & M2FS & SMCField11 & 2457942.779 & 168 & 11 \\
& & & & M2FS & SMCField11 & 2457943.690 & 168 & 12 \\
& & & & M2FS & SMCField11 & 2457944.872 & 167 & 6 \\
& & & & M2FS & SMCField11 & 2458255.880 & 173 & 9 \\  
81169 & B0.2V & 20.94970 & -73.22633 & IMACS f/4 & FLDob107 & 2457555.319 & 173 & 15 \\
& & & & IMACS f/4 & FLDob107 & 2457727.194 & 180 & 15 \\
& & & & IMACS f/4 & FLDob107 & 2457910.229 & 177 & 13 \\
& & & & M2FS & SMCField11 & 2457942.779 & 176 & 6 \\ 
& & & & M2FS & SMCField11 & 2457943.690 & 182 & 15 \\ 
& & & & M2FS & SMCField11 & 2458352.881 & 175 & 18 \\  
81258 & B0-1.5V & 21.00766 & -73.29926 & IMACS f/4 & FLDob107 & 2457555.319 & 365 & 24 \\
& & & & IMACS f/4 & FLDob107 & 2457727.194 & 135 & 11 \\
& & & & IMACS f/4 & FLDob107 & 2457910.229 & 9 & 19 \\
& & & & M2FS & SMCField11 & 2457942.779 & 13 & 8 \\ 
& & & & M2FS & SMCField11 & 2457943.690 & 95 & 18 \\ 
& & & & M2FS & SMCField11 & 2458255.880 & 130 & 9 \\ 
81465 & B$e_3$ & 21.11587 & -73.54926 & M2FS & SMCField11 & 2457943.690 & 218 & 1 \\ 
& & & & M2FS & SMCField11 & 2458255.880 & 226 & 1 \\ 
81634 & B1.5V$e_3$ & 21.20950 & -73.57023 & IMACS f/4 & FLDob105 & 2457728.200 & 182 & 11 \\
& & & & IMACS f/4 & FLDob105 & 2457910.339 & 171 & 14 \\
81646 & O8V & 21.21337 & -73.45051 & IMACS f/4 & FLDob105 & 2457554.430 & 205 & 19 \\
& & & & IMACS f/4 & FLDob105 & 2457555.209 & 196 & 19 \\
& & & & IMACS f/4 & FLDob105 & 2457728.200 & 165 & 16 \\
& & & & IMACS f/4 & FLDob105 & 2457910.339 & 181 & 18 \\
& & & & M2FS & SMCField11 &  2457942.779 & 72 & 11 \\ 
& & & & M2FS & SMCField11 &  2457943.783 & 171 & 6 \\ 
& & & & M2FS & SMCField11 & 2457944.789 & 256 & 15 \\ 
& & & & M2FS & SMCField11 &  2457944.872 & 225 & 25 \\ 
& & & & M2FS & SMCField11 &  2457945.881 & 170 & 3 \\ 
& & & & M2FS & SMCField11 & 2458255.880 & 89 & 13 \\ 
& & & & M2FS & SMCField11 &  2458346.885 & 160 & 3 \\ 
& & & & M2FS & SMCField11 &  2458349.755 & 233 & 35 \\ 
& & & & M2FS & SMCField11 & 2458351.715 & 114 & 10 \\ 
81647 & B0.2V & 21.21347 & -73.10006 & IMACS f/4 & FLDob108 & 2457555.371 & 159 & 12 \\
& & & & IMACS f/4 & FLDob108 & 2457728.099 & 153 & 11 \\
& & & & IMACS f/4 & FLDob108 & 2457910.394 & 139 & 13 \\
81696 & B1 V & 21.24092 & -73.49664 & IMACS f/4 & FLDob105 & 2457554.430 & 210 & 21 \\
& & & & IMACS f/4 & FLDob105 & 2457555.209 & 131 & 16 \\
& & & & IMACS f/4 & FLDob105 & 2457728.200 & 135 & 13 \\
& & & & IMACS f/4 & FLDob105 & 2457910.339 & 295 & 13 \\
& & & & M2FS & SMCField11 & 2457944.789 & 273 & 20 \\ 
& & & & M2FS & SMCField11 & 2457944.872 & 254 & 25 \\ 
& & & & M2FS & SMCField11 & 2457945.881 & 57 & 7 \\ 
& & & & M2FS & SMCField11 & 2458346.885 & 209 & 7 \\ 
& & & & M2FS & SMCField11 & 2458349.755 & 265 & 44 \\ 
& & & & M2FS & SMCField11 & 2458351.715 & 281 & 9 \\  
81941 & O9.5III & 21.39888 & -73.18636 & IMACS f/4 & FLDob108 & 2457555.371 & 158 & 15 \\
& & & & IMACS f/4 & FLDob108 & 2457728.099 & 189 & 11 \\
& & & & IMACS f/4 & FLDob108 & 2457910.394 & 5 & 18 \\
& & & & M2FS & SMCField12 &  2457943.783 & 40 & 5 \\  
& & & & M2FS & SMCField12 &  2457944.789 & 126 & 12 \\  
& & & & M2FS & SMCField12 &  2457945.881 & 277 & 4 \\  
& & & & M2FS & SMCField12 &  2458257.873 & 126 & 8 \\  
& & & & M2FS & SMCField12 &  2458346.885 & 116 & 3 \\  
& & & & M2FS & SMCField12 &  2458349.755 & 298 & 11 \\ 
& & & & M2FS & SMCField12 &  2458351.715 & 28 & 4 \\  
82322 & O9.5III & 21.64704 & -73.25453 & IMACS f/4 & FLDob108 & 2457555.371 & 171 & 12 \\
& & & & IMACS f/4 & FLDob108 & 2457728.099 & 173 & 16 \\
& & & & IMACS f/4 & FLDob108 & 2457910.394 & 169 & 18 \\
& & & & M2FS & SMCField12 & 2457943.783 & 166 & 2 \\ 
& & & & M2FS & SMCField12 &  2457944.789 & 166 & 3 \\ 
& & & & M2FS & SMCField12 & 2457945.881 & 162 & 2 \\ 
& & & & M2FS & SMCField12 & 2458346.885 & 169 & 1 \\ 
& & & & M2FS & SMCField12 & 2458349.755 & 171 & 2 \\  
82328 & B0$e_2$ & 21.65680 & -73.49267 & IMACS f/4 & FLDob105 & 2457554.430 & 187 & 23 \\
& & & & IMACS f/4 & FLDob105 & 2457555.209 & 200 & 15 \\
& & & & IMACS f/4 & FLDob105 & 2457728.200 & 188 & 18 \\
& & & & IMACS f/4 & FLDob105 & 2457910.339 & 206 & 17 \\
& & & & M2FS & SMCField12 & 2457943.783 & 185 & 9 \\ 
& & & & M2FS & SMCField12 & 2457945.881 & 190 & 7 \\ 
& & & & M2FS & SMCField12 & 2458346.885 & 195 & 7 \\ 
& & & & M2FS & SMCField12 & 2458349.755 & 224 & 26 \\ 
& & & & M2FS & SMCField12 & 2458351.715 & 227 & 12 \\ 
82408 & B1III & 21.70886 & -73.39600 & IMACS f/4 & FLDob105 & 2457554.430 & 188 & 11 \\
& & & & IMACS f/4 & FLDob105 & 2457555.209 & 183 & 10 \\
& & & & IMACS f/4 & FLDob105 & 2457728.200 & 185 & 10 \\
& & & & IMACS f/4 & FLDob105 & 2457910.339 & 180 & 12 \\
& & & & M2FS & SMCField12 & 2457943.783 & 173 & 2 \\ 
& & & & M2FS & SMCField12 & 2457944.789 & 173 & 4 \\ 
& & & & M2FS & SMCField12 & 2457945.881 & 171 & 2 \\ 
& & & & M2FS & SMCField12 & 2458346.885 & 172 & 1 \\ 
& & & & M2FS & SMCField12 & 2458349.755 & 173 & 5 \\ 
82444 & B0V & 21.73714 & -73.51503 & IMACS f/4 & FLDob105 & 2457554.430 & 195 & 14 \\
& & & & IMACS f/4 & FLDob105 & 2457555.209 & 189 & 17 \\
& & & & IMACS f/4 & FLDob105 & 2457728.200 & 197 & 15 \\
& & & & IMACS f/4 & FLDob105 & 2457910.339 & 190 & 17 \\
& & & & M2FS & SMCField12 & 2457942.892 & 194 & 7 \\ 
& & & & M2FS & SMCField12 & 2457943.783 & 195 & 5 \\ 
& & & & M2FS & SMCField12 & 2457944.789 & 238 & 18 \\ 
& & & & M2FS & SMCField12 & 2457945.881 & 215 & 6 \\ 
& & & & M2FS & SMCField12 & 2458351.715 & 230 & 20 \\ 
82489 & O9:IIIp$e_4$+ & 21.76797 & -73.07746 & IMACS f/4 & FLDob108 & 2457555.371 & 153 & 13 \\
& & & & IMACS f/4 & FLDob108 & 2457728.099 & 146 & 12 \\
& & & & IMACS f/4 & FLDob108 & 2457910.394 & 134 & 19 \\
82711 & B1Ve & 21.94144 & -73.54900 & IMACS f/4 & FLDob105 & 2457555.209 & 115 & 48 \\
& & & & IMACS f/4 & FLDob105 & 2457728.200 & 171 & 17 \\
& & & & IMACS f/4 & FLDob105 &  2457910.339 & 157 & 14 \\
& & & & M2FS & SMCField12 & 2457943.783 & 285 & 3 \\ 
& & & & M2FS & SMCField12 & 2457944.789 & 246 & 5 \\ 
& & & & M2FS & SMCField12 & 2457945.881 & 281 & 2 \\ 
& & & & M2FS & SMCField12 & 2458346.885 & 272 & 1 \\ 
& & & & M2FS & SMCField12 & 2458349.755 & 241 & 12 \\ 
82783 & B0.5V & 21.99023 & -73.17075 & IMACS f/4 & FLDob108 & 2457555.371 & 163 & 14 \\
& & & & IMACS f/4 & FLDob108 & 2457728.099 & 165 & 13 \\
& & & & IMACS f/4 & FLDob108 & 2457910.394 & 167 & 14 \\
& & & & M2FS & SMCField12 & 2457943.783 & 170 & 6 \\ 
& & & & M2FS & SMCField12 & 2457944.789 & 171 & 13 \\
& & & & M2FS & SMCField12 & 2457945.881 & 168 & 4 \\ 
& & & & M2FS & SMCField12 & 2458349.755 & 167 & 6 \\ 
& & & & M2FS & SMCField12 & 2458351.715 & 170 & 4 \\ 
83017 & O9.5III & 22.19844 & -73.30637 & IMACS f/4 & FLDob104 & 2457554.408 & 204 & 9 \\
& & & & IMACS f/4 & FLDob104 & 2457726.230 & 211 & 12 \\
& & & & IMACS f/4 & FLDob104 & 2457909.330 & 223 & 14 \\
& & & & M2FS & SMCField12 & 2457944.789 & 191 & 6 \\ 
& & & & M2FS & SMCField12 & 2457945.881 & 192 & 3 \\ 
& & & & M2FS & SMCField12 & 2458346.885 & 192 & 8 \\ 
& & & & M2FS & SMCField12 & 2458349.755 & 194 & 4 \\ 
& & & & M2FS & SMCField12 & 2458351.715 & 191 & 5 \\ 
83073 & B0.7V & 22.23558 & -73.38664 & M2FS & SMCField12 & 2457942.892 & 174 & 5 \\ 
& & & & M2FS & SMCField12 & 2457943.783 & 171 & 5 \\ 
& & & & M2FS & SMCField12 & 2457944.789 & 175 & 15 \\ 
& & & & M2FS & SMCField12 & 2457945.881 & 169 & 4 \\ 
& & & & M2FS & SMCField12 & 2458346.885 & 167 & 3 \\ 
& & & & M2FS & SMCField12 & 2458351.715 & 172 & 5 \\ 
83171 & B0$e_2$ & 22.30245 & -73.29074 & IMACS f/4 & FLDob104 & 2457554.408 & 157 & 17 \\
& & & & IMACS f/4 & FLDob104 & 2457726.230 & 198 & 16 \\
& & & & IMACS f/4 & FLDob104 & 2457909.330 & 217 & 12 \\
& & & & M2FS & SMCField12 & 2457943.783 & 210 & 2 \\ 
& & & & M2FS & SMCField12 & 2457944.789 & 202 & 4 \\ 
& & & & M2FS & SMCField12 & 2457945.881 & 212 & 1 \\ 
& & & & M2FS & SMCField12 & 2458346.885 & 217 & 2 \\ 
& & & & M2FS & SMCField12 & 2458349.755 & 198 & 5 \\ 
83224 & B1$e_3$ & 22.34478 & -73.26554 & IMACS f/4 & FLDob104 & 2457554.408 & 194 & 16 \\
& & & & IMACS f/4 & FLDob104 & 2457726.230 & 202 & 14 \\
& & & & IMACS f/4 & FLDob104 & 2457909.330 & 183 & 17 \\
& & & & M2FS & SMCField12 & 2457943.783 & 229 & 1 \\
& & & & M2FS & SMCField12 & 2457944.789 & 225 & 1 \\
& & & & M2FS & SMCField12 & 2457945.881 & 226 & 1 \\
& & & & M2FS & SMCField12 & 2458257.873 & 239 & 15 \\
& & & & M2FS & SMCField12 & 2458346.885 & 254 & 3 \\
& & & & M2FS & SMCField12 & 2458349.755 & 257 & 5 \\
83232 & B1.5III & 22.35032 & -73.49490 & IMACS f/4 & FLDob104 & 2457554.408 & 210 & 15 \\
& & & & IMACS f/4 & FLDob104 & 2457726.230 & 158 & 13 \\
& & & & IMACS f/4 & FLDob104 & 2457909.330 & 227 & 16 \\
& & & & M2FS & SMCField12 & 2457943.783 & 161 & 5 \\ 
& & & & M2FS & SMCField12 & 2457945.881 & 217 & 8 \\ 
& & & & M2FS & SMCField12 & 2458346.885 & 239 & 4 \\ 
83480 & B6I[e] & 22.54537 & -73.31561  & M2FS & SMCField12 & 2457942.892 & 250 & 2 \\ 
& & & & M2FS & SMCField12 & 2457944.789 & 246 & 2 \\ 
& & & & M2FS & SMCField12 & 2457945.881 & 247 & 1 \\ 
& & & & M2FS & SMCField12 & 2458257.873 & 251 & 3 \\ 
& & & & M2FS & SMCField12 & 2458346.885 & 248 & 2 \\ 
& & & & M2FS & SMCField12 & 2458349.755 & 243 & 2 \\ 
& & & & M2FS & SMCField12 & 2458351.715 & 237 & 9 \\ 
83510 & O8V & 22.56930 & -73.34753 & IMACS f/4 & FLDob104 & 2457554.408 & 128 & 17 \\
& & & & IMACS f/4 & FLDob104 & 2457726.230 & 153 & 14 \\
& & & & IMACS f/4 & FLDob104 & 2457909.330 & 185 & 16 \\
& & & & M2FS & SMCField12 & 2457942.892 & 124 & 10 \\ 
& & & & M2FS & SMCField12 & 2457943.783 & 137 & 11 \\ 
& & & & M2FS & SMCField12 & 2457944.789 & 241 & 41 \\ 
& & & & M2FS & SMCField12 & 2458346.885 & 117 & 9 \\ 
& & & & M2FS & SMCField12 & 2458349.755 & 123 & 9 \\ 
& & & & M2FS & SMCField12 & 2458351.715 & 121 & 8 \\ 
83678 & O8.5III & 22.70940 & -73.38303 & M2FS & SMCField12 & 2457943.783 & 164 & 3 \\
& & & & M2FS & SMCField12 & 2457944.789 & 168 & 7 \\
& & & & M2FS & SMCField12 & 2458346.885 & 162 & 2 \\
& & & & M2FS & SMCField12 & 2458351.715 & 167 & 2 \\ 
& & & & M2FS & SMCField12 & 2457945.881 & 167 & 3 \\ 
\enddata
\end{deluxetable*}

 \section{Notes on Individual Binaries} \label{App:starnotes}

This Appendix provides individual notes for all 33 SMC Wing binaries that are identified in this work
(Tables \ref{table:OBbinaryresults} and \ref{table:OBebinaryresults}). 
The general procedure to obtain the RV measurements is outlined in Section \ref{subsec:RVmeas}, and
RV measurements for each epoch observed for each target are found in Appendix \ref{App:RVtable}.

For M2FS data, the fitting process is outlined for each target in this Appendix, since this process has more procedural free parameters than that for the IMACS data.
Unless otherwise specified, we obtain RV measurements for M2FS data using a cross-correlation fit against a model spectrum from the PoWR \citep{PoWR} atmosphere library, as described in Section~\ref{subsec:RVmeas}. 
If the model spectrum requires broadening to match the observations, then the broadening value is specified. No broadening is mentioned for targets that do not require this.
For targets that use an individual observation as the RV template, that epoch is identified; the systemic velocity for these objects are obtained through Gaussian fits of the H$\delta$ and H$\gamma$ emission lines as specified in Section \ref{subsec:RVmeas}.  

An overview of the methods and error bars 
for the companion mass estimates for each binary type are found in Section \ref{sec:binprop}.
Below, we include more detailed information on each target's RV measurements, notes on binary parameters, information from external work, and companion mass estimates. Plots for RV curves and companion mass analyses for all the targets are published by 
\citet{VargasSalazar2024}.

Binaries and binary candidates
that are identified by their RV curve 
have calculated companion mass ($M_2$) estimates. For non-compact binaries that are not RV binaries, we are able to estimate their companion mass based on our detection thresholds.
This Appendix discusses whether the companion for each target
could be a non-compact star, 
neutron star (NS), black hole (BH) or stripped He star. Depending on how well-constrained the companion mass is, we are sometimes able to provide constraints on period and inclination of the binary.
Plots showing the constraints on $M_2$ obtained in Section~\ref{sec:binprop} are given by \citet{VargasSalazar2024}.
Some targets also have information on binary parameters from external sources,
but this is not meant to be exhaustive.

\subsection{Target [M2002] 71652} \label{App:71652}

Classification: OBe, RV binary candidate

This target has 4 RV measurements from IMACS. It is identified as an RV candidate by only the $\Delta$RV method. 

This target's primary mass is $16.1  M_{\odot}$ (Table \ref{table:OBebinaryresults}). Our  RV and mass constraints suggest that the lowest companion mass 
estimate
is $0.71 M_{\odot}$. 
If the companion is a compact object, the upper mass limit is unconstrained. If the companion is a non-compact star, its
mass should
be $< 5.4 M_{\odot}$ based on our detection limit. Additionally, this OBe system 
may have
a low eccentricity and therefore 
could
have a stripped star companion.

\subsection{Target [M2002] 72535} \label{App:72535}

Classification: OBe, SB2 candidate

The target's primary has an estimated
mass range of $44 - 49 M_{\odot}$ \citep{Dallas2022}. This target has 8 spectroscopic observations: 4 from IMACS and 4 from M2FS. For the M2FS fits, 
we use the one taken on
JD 2458352.820 as the template RV standard for the RV cross-correlation.
We measure the systemic RV by gaussian fitting to the H$\gamma$ emission line.
The target 
does not exhibit any RV variation according to our RV binary identification methods. 

The target's status as an SB2 candidate is established in part from noting that the 4144 \AA\ line 
may show 
redshifting and blueshifting with respect to the RV standard in 
epochs JD 2457943.872, JD 2457945.792 and JD 2458350.856. If this SB2 were to be confirmed, then our signal-to-noise threshold indicates that a non-compact stellar companion should 
be $\lesssim 7.3 M_{\odot}$. 

\subsection{Target [M2002] 73355} \label{App:73355}

Classification: OBe, fast rotator, EB, RV binary, eccentric binary

This target has 9 spectroscopic observations: 4 from IMACS and 5 from M2FS. For the M2FS data, we perform the cross-correlation fit to the He absorption lines. This is a fast rotator 
with measured
$v\sin i = 300$ km/s \citep{DorigoJones2020}, 
which requires us to broaden the PoWR spectral template for the 7 M2FS observations.
Notably 
the line widths show variability, perhaps due to stellar precession, and so we use
different broadening 
values ($v\sin i$): 
JD 2457943.872 (255 km/s),
JD 2458352.820 (280 km/s), 
JD 2457942.695 (200 km/s), 
JD 2458350.856 (280 km/s), 
JD 2458346.817 (275 km/s). 
We could not extract an RV measurement from JD 2458343.837 and JD 2457945.792 due to
its signal-to-noise and these observations were dropped from the analysis and not included in the total observations. 
The target is identified as an RV binary by both 
the $\Delta$RV and $F$-test
methods. The binary 
may have
an $e>0$ according to our RV results.

This target's primary mass is $24.9  M_{\odot}$ (Table \ref{table:OBebinaryresults}). The EB classification comes from the OGLE survey, therefore this binary has a
non-compact
stellar companion. Since 
we see no evidence that it is
an SB2, the companion 
must 
be below our signal-to-noise threshold with a mass between $2.1 - 4.3 M_{\odot}$.

\subsection{Target [M2002] 73913} \label{App:73913}

Classification: OB, fast rotator, RV binary

This target has 10 spectroscopic observations: 4 from IMACS and 6 from M2FS.
For M2FS data, the spectral lines appear broadened, but there is no recorded $v\sin i$ measurement. 
We determine that the spectra best fit a broadening of
$v\sin i = 200$ km/s. The target is identified as an RV binary by both 
the $\Delta$RV and $F$-test
methods.

This target's primary mass is $34.2  M_{\odot}$ (Table \ref{table:OBbinaryresults}). The nature of the companion is unclear, but since it is an OB binary, it is possible for it to be a
non-compact
star 
with mass 
$1.2 - 5.8 M_{\odot}$. For these constraints, the period would be $\lesssim 40$ days. On the other hand,  
fast rotators are 
BSS binary candidates, implying
that the companion also could be a compact object. Further research is required to determine the true nature of the companion. 

\subsection{Target [M2002] 75061} \label{App:75061}

Classification: OBe, RV binary

This target has 8 spectroscopic observations: 2 from IMACS and 6 from M2FS. 
For M2FS fits, we use the one taken on
JD 2458350.856 as the template RV standard for the RV cross-correlation
for 5 observations.
The observation on JD 2458346.817 uses the regular cross-correlation fit
against the PoWR model template,
since it has weak Balmer emission.
The observation on 
JD 2458343.837 is too noisy to perform a proper fit and this observation was dropped from the analysis. The target is identified as an RV binary by both 
the $\Delta$RV and $F$-test
methods. The binary may have $e>0$.

This target's primary mass is $20.0  M_{\odot}$ (Table \ref{table:OBebinaryresults}). The nature of the companion is unclear, but since it is an OBe binary, it is possible for it to be a compact object. Our  mass constraints determine that the companion's mass is $\gtrsim 0.88 M_{\odot}$ based on our RV measurements
Further investigation is required to confirm the companion's nature. 

\subsection{Target [M2002] 75210} \label{App:75210}

Classification: OB, RV binary candidate

This target has 5 spectroscopic observations from M2FS used in the cross-correlation fits. The target is identified as an RV binary by only the $F$-test. 

This target's primary mass is $29.3  M_{\odot}$ (Table \ref{table:OBbinaryresults}). The nature of the companion is unclear, but since it is an OB binary, it is possible for it to be a
non-compact
star 
with mass 
$0.76 - 3.8 M_{\odot}$. For these parameters, the period should be $\lesssim 50$ days.

\subsection{Target [M2002] 75980} \label{App:75980}

Classification: OBe, RV binary, eccentric binary

This target has 8 spectroscopic observations: 3 from IMACS and 5 from M2FS.
For M2FS fits, we use the one taken on
JD 2457945.792 as the template RV standard for the RV cross-correlation. 
The target is identified as an RV binary by both 
the $\Delta$RV and $F$-test
methods. The binary may have an $e>0$.

This target's primary mass is $26.9  M_{\odot}$ (Table \ref{table:OBebinaryresults}). Our  RV and mass constraints indicate that the lowest companion mass
estimate
is $1.1 M_{\odot}$. 
If the companion is a compact object, the upper mass limit is unconstrained. If the companion is a non-compact star, it should be $\lesssim 9.20 M_{\odot}$, based on our detection limits. 

\subsection{Target [M2002] 76253} \label{App:76253}

Classification: OB, SB2

This target has 8 spectroscopic observations: 4 from IMACS and 4 from M2FS.
From M2FS, the observation for 
JD 2458343.837 is too noisy to extract an RV value and we therefore do not use it in our binary analysis. 
We do not detect
any RV variation using our RV binary identification methods. 

This target's primary mass is $13.4  M_{\odot}$ (Table \ref{table:OBbinaryresults}). The SB2 classification comes from \cite{Lamb2016},
who classify
the companion as a B star. 
Our S/N suggests that the companion is $\sim 3 M_\odot$.

\subsection{Target [M2002] 76371} \label{App:76371}

Classification: OB, RV binary, eccentric binary

This target has 9 RV measurements: 4 from IMACS and 5 from M2FS. 
For the M2FS cross-correlation fits, we broaden the model template spectrum to the observed
$v\sin i = 82$ km/s \citep{DorigoJones2020} for the fits. The target is identified as an RV binary by both 
the $\Delta$RV and $F$-test
methods.
This system is one of the OB binaries with a possible $e>0$. 

This target's primary mass is $26.0  M_{\odot}$ (Table \ref{table:OBbinaryresults}). 
Since it is an OB binary the companion could be a non-compact star with mass of $1.2 - 4.2 M_{\odot}$. For these parameters, the period should be $\lesssim 16$ days. On the other hand, OB binaries in eccentric orbits are 
included in our BSS binary sample, and thus
the companion could be a compact object. Further research is required to determine the true nature of the companion. 

\subsection{Target [M2002] 76654} \label{App:76654}

Classification: OBe, fast rotator, RV binary candidate

This target has 5 spectroscopic observations from M2FS 
and
we use the one taken on
JD 2458346.817 as the template RV standard for the RV cross-correlation. 
However, the one obtained on JD 2457942.695 uses the regular cross-correlation fit 
against the PoWR model template
since its Balmer
emission was low in that epoch. Based on the broadening of the lines, the target appears to be a fast rotator, but there is no $v\sin i$ measurement 
reported
previously. 
We use $v\sin i= 370$ km/s to apply the broadening to the model template. This object is identified as an RV binary with only the $F$-test. 

This target's primary mass is $17.6  M_{\odot}$ (Table \ref{table:OBebinaryresults}). Our  RV and mass constraints suggest that the lowest companion mass estimate 
is $0.53 M_{\odot}$. 
If the companion is a compact object, the upper mass limit is unconstrained. If the companion is a non-compact star, 
its mass
should be $\lesssim 7.7 M_{\odot}$, based on our detection limits. Additionally, this OBe system appears to have a low eccentricity and it therefore 
could potentially have a stripped star companion.

\subsection{Target [M2002] 76773} \label{App:76773}

Classification: OBe, RV binary

This target has 6 spectroscopic observations from M2FS and is fitted using 
the model template for 
the cross-correlation fit since 
the Balmer emission is weak and it 
has a \citet{Lesh1968} classification of
$e_1$.  There is evidence of broadening in the spectral lines, but there is no $v\sin i$ reported. 
We use a value of $v\sin i = 100$ km/s 
to broaden the template
lines. 
The observation on 
JD 2458343.837 is too noisy to extract an RV value and this observation was dropped from the analysis.
The target is identified as an RV binary by both 
the $\Delta$RV and $F$-test
methods.

The target's primary has a mass of $20 - 27 M_{\odot}$ \citep{Dallas2022}. 
Our detection limit would have detected a non-compact stellar companion of $6.2 M_{\odot}$, which is a smaller mass than our dynamical $M_{\rm 2,min}$ estimate of 9.0 $M_\odot$. 
Therefore, given the nature of OBe stars,
it is likely that the companion is a compact object, and the  lower-mass detection limit 
therefore implies that
this a black hole candidate. 
However, this mass limit is driven by a single observation on JD 2458343.837, which is a strong RV outlier.  There is no clear indication that the RV measurement is incorrect; however, this mass limit may be a substantial overestimate.
This target 
may be
a circular OBe binary, a characteristic shared by binaries with stripped star companions (Section \ref{sec:strippedstars}).
Further investigation is required to determine the true nature of the companion.

\subsection{Target [M2002] 77290} \label{App:77290}

Classification: OBe, RV binary, eccentric binary

This target has 8 spectroscopic observations: 3 from IMACS and 5 from M2FS. 
For M2FS fits, we use the one taken on
JD 2457945.792  as the template RV standard for the RV cross-correlation. 
The target is identified as an RV binary by both 
the $\Delta$RV and $F$-test
methods. The binary may have an $e>0$  according to our RV results.  

This target's primary mass is $18.8  M_{\odot}$ (Table \ref{table:OBebinaryresults}). Our  RV and mass constraints suggest a lowest companion mass 
estimate
of $0.74 M_{\odot}$. 
If the companion is a compact object, the upper mass limit is unconstrained. If the companion is a non-compact star, it should
be $\lesssim 5.10 M_{\odot}$ based on our detection limits. The analysis for the target's companion mass estimates are shown in Figure \ref{fig:OBemassex}.

\subsection{Target [M2002] 77368} \label{App:77368}

Classification: OB, fast rotator, SB2 candidate, RV binary

This target has 9 spectroscopic observations: 4 from IMACS and 5 from M2FS.
This target is a fast rotator 
and, for M2FS fits, the model template lines are broadened to the observed
$v\sin i = 286$ km/s \citep{DorigoJones2020}. The target is identified as an RV binary by both 
the $\Delta$RV and $F$-test
methods.

This target's primary mass is $39.3  M_{\odot}$ (Table \ref{table:OBbinaryresults}). 
The target's status as an SB2 candidate is established in part from noting that the 4200 \AA\ and 4387 \AA\ lines may show Doppler shifting
with respect to the RV standard in 
epochs JD 2458346.817  and JD 2458350.856. If this SB2 were to be confirmed, then our signal-to-noise threshold indicates that a non-compact stellar companion 
would have a mass near the upper end of the $M_2$ estimated range of
$3.9 - 23.7 M_{\odot}$. 
For these constraints, the period should be $\lesssim 16$ days. The target's companion mass estimates are shown in Figure \ref{fig:SB2paramex}.

On the other hand, fast-rotating OB binaries are 
included in
our BSS binary sample, which contradicts the SB2 candidacy and suggestst
that this could be a potential DES ejection of a system that is still pre-SN. Further investigation is necessary to confirm the nature of the binary.

\subsection{Target [M2002] 77458} \label{App:77458}

Classification: OBe, fast rotator, HMXB, RV binary

This target is SMC X-1. According to the literature, the companion is an accreting neutron star of 1.04 $M_{\odot}$  \citep{Rawls2011} orbiting around the target on a period of 3.89 days \citep{Clark2000,Falanga2015}. The binary is suggested to have an inclination of 70$^{\circ}$ \citep{Reynolds1993} and a low eccentricity \citep{Falanga2015}.
The system also has a warped disk that generates a superorbital period of 40 -- 100 days \citep{oglivedubus2001,Brumback2023}, and can occult the neutron star companion.

We can use the data in the literature as a test of our methods for estimating $M_2$.
We have 9 spectroscopic observations for this target: 4 from IMACS and 5 from M2FS.
The M2FS data are fit with the cross-correlation method. JD 2457942.695 is fit with the Balmer lines masked. To perform the fit, the lines are broadened to $v\sin i = 151$ km/s \citep{DorigoJones2020}. The target is identified as an RV binary by both 
the $\Delta$RV and $F$-test
methods. 

This target's primary mass is $15.4  M_{\odot}$ (Table \ref{table:OBebinaryresults}). Our RV mass estimates indicate that $M_2$ should be $1.5 - 2.1 M_{\odot}$. 
This is in reasonably good agreement with the reported value of 1.04 $M_{\odot}$ above.
\cite{Falanga2015} reported a low eccentricity, which is consistent with our $M_2$ estimates.

\subsection{Target [M2002] 77616} \label{App:77616}

Classification: OBe, fast rotator, RV binary candidate, eccentric binary

This target is also known as AzV 
493
\citep{Azzopardi1975}, the earliest known classical Oe star \citep{GoldenMarx2016}. 
\cite{77616paper} 
find that this system is extreme, not just in mass and spectral type, but also in eccentricity $(e > 0.93)$.
The light curve has a dominant 14.6 yr 
period
with  $\sim 40$ day oscillations of an unknown origin. Its mass and rotation 
are likely 
enhanced through binary interaction 
before
a core-collapse SN of the unseen companion, which currently remains unidentified. 
More information about the  
unusual properties
of this target can be found in \cite{77616paper}.

We have 4 spectroscopic observations of this target from M2FS and we perform the cross-correlation fit on the \heii\ 4200 \AA\ line. Epoch JD 2458343.837 is too noisy to be fit reliably and this observation was dropped from the analysis. This target is a fast rotator and
we use the observed $v\sin i = 300$ km/s \citep{DorigoJones2020}
for broadening the features of the model template. 
The target is identified as an RV binary only by 
the $\Delta$RV
method.

This target's primary mass is $50 M_{\odot}$ \citep{77616paper}. Our 
minimum $M_2$ mass 
estimate is
$1 M_{\odot}$; based on the detection threshold, the maximum $M_2$ for a non-compact companion is 7.1 $M_\odot$.
We have insufficient information to determine the nature of the companion, but \cite{77616paper} suggest that it could be a black hole. 
This is consistent with our discussion in Section~\ref{sec:OBeBSSParam} suggesting a BH companion is implied by the BPASS models based on the mass of the primary.

\subsection{Target [M2002] 77734} \label{App:77734}

Classification: OB, RV binary candidate

This target has 6 RV 
measurements: 2 from IMACS and 4 from M2FS.
The M2FS RVs are obtained 
from the PoWR model
cross-correlation fits.
Epoch JD 2457943.872 is fitted using 
the spectrum taken on 
JD 2458346.817 as the template RV standard for the RV cross-correlation due to low S/N, which made the Balmer lines hard to fit using a PoWR model. The target is identified as an RV binary by only
the $\Delta$RV
method.

The target's primary has a mass range of $12 - 17 M_{\odot}$ \citep{Dallas2022}. 
Based on our mass estimates and detection limit,
and given that the primary is an OB star, 
the companion could be a non-compact star with mass of $0.51 - 6.1 M_{\odot}$ or a compact object,
but we have insufficient information to 
determine its nature.

\subsection{Target [M2002] 77816} \label{App:77816}

Classification: OB, SB2 candidate, RV binary candidate

This target has 5 RV 
measurements from M2FS
obtained primarily from the 
model atmosphere
cross-correlation fits. The model spectral features are broadened by the target's $v\sin i = 135$ km/s \citep{DorigoJones2020} to match 
those of the 
observations. The target is identified as an RV binary by  
only the $F$-test
method.

This target's primary mass is $18.7  M_{\odot}$ (Table \ref{table:OBbinaryresults}). 
The target's status as an SB2 candidate is established in part from noting that the  \hei\ 4144 \AA\ and \hei\ 4120 \AA\ lines may be redshifted relative to the H$\delta$ line for epoch JD 2458346.817. If this SB2 were to be confirmed, then our signal-to-noise threshold indicates that a non-compact stellar companion should be near the upper range of our estimated $M_2 = 3.5 - 4.6 M_{\odot}$. For these constraints, the period should be $\lesssim 100$ days.

\subsection{Target [M2002] 77851} \label{App:77851}

Classification: OBe, HMXB

This target only has 4 RV measurements from IMACS. 
We do not detect
any RV variation using our RV binary identification methods. 

This target is 
an HMXB and pulsar \citep{Schmidtke2013}.
The primary's mass is $23.4  M_{\odot}$ (Table \ref{table:OBebinaryresults}) and the companion is a NS.

\subsection{Target [M2002] 80573} \label{App:80573}

Classification: OB, fast rotator candidate, RV binary candidate

This target has 6 RV 
observations: 3 from IMACS and 3 from M2FS.
The M2FS RV measurements are
obtained using the cross-correlation fits. However, the S/N for
epoch JD 2458352.881 
was too low to use, so this observation was dropped from the analysis.
This target is a fast rotator with $v\sin i = 304$ km/s \citep{DorigoJones2020}. 
The target is identified as an RV binary by both 
the $\Delta$RV and $F$-test
methods.

This target's primary mass is $15.0  M_{\odot}$ (Table \ref{table:OBbinaryresults}). 
Our mass estimate suggests that 
the companion could be a non-compact star with mass of $1.2 - 5.6 M_{\odot}$. For these constraints, the period should be $\lesssim 40$ days. On the other hand, OB binaries that are fast rotators are 
in our BSS sample, and thus
the companion also could be a compact object. 
Further study is required to determine the true nature of the companion.

\subsection{Target [M2002] 81258} \label{App:81258}

Classification: OB, EB, SB2 candidate, RV binary

This target has 6 RV 
measurements: 3 from IMACS and 3 from M2FS.
The M2FS RV data are obtained from the cross-correlation fits. We broaden the
model spectrum 
to match its $v\sin i = 148$ km/s \citep{DorigoJones2020}. The target is identified as an RV binary by both 
the $\Delta$RV and $F$-test
methods.

The RV measurement from JD 2457550.00 
obtained with
IMACS appears to have an unusual value of $373\pm23$ \kms\ that would place this target in the $e>0$ regime of Figure \ref{fig:binaryID1}. 
The rest of the stars in this IMACS multi-object field taken on this date appear to have no unusual RV measurements. However, OGLE 
data indicate 
that this target should have $e \sim 0$ \citep{OGLE}, indicating that this RV measurement 
is likely spurious.
Therefore, we have removed this point 
from our analysis. 

This target is identified as an EB from
the OGLE survey. 
The target's status as an SB2 candidate is established in part from noting that the  H$\delta$ 4102 \AA\ and H$\gamma$ 4340 \AA\ lines may both be redshifted relative to the RV standard in epochs JD 2458352.881 and JD 2457942.779.

This target's primary mass is $11.9  M_{\odot}$ (Table \ref{table:OBbinaryresults}). 
Since the system is an EB, we know that the companion is a non-compact star.
Our signal-to-noise threshold indicates that 
the companion star
is $3.4 - 5.6 M_{\odot}$
(Figure \ref{fig:EBCircex}).

\subsection{Target [M2002] 81465} \label{App:81465}

Classification: OBe, RV binary candidate, eccentric binary

This target only has 2 spectroscopic observations from M2FS and
we use the one taken on
JD 2457943.690  as the template RV standard for the RV cross-correlation. 
It is identified as an RV binary from only the $F$-test, since we cannot carry out the $\Delta$RV test with only 2 RV measurements.

This target's primary mass is $15.7  M_{\odot}$ (Table \ref{table:OBebinaryresults}). Based on our mass estimate of $0.05 - 3.0 M_{\odot}$, the companion's mass
is very small. Therefore, there is a possibility that it is a white dwarf or a NS. However, with only 2 observations, it is likely that we are not adequately sampling $\Delta$ RV. 
This is the only case where we obtain an estimate of $M_{\rm 2,max}$ from our RV analysis that is less than that obtained by the detection limit.
Thus, more observations are needed to confirm the status of this system. 

\subsection{Target [M2002] 81634} \label{App:81634}

Classification: OBe, EB candidate

This target only has 2 RV measurements from IMACS. 
We do not detect 
any RV variation using our RV binary identification methods, possibly due to the lack of data. 

This target's primary mass is $13.1  M_{\odot}$ (Table \ref{table:OBebinaryresults}). 
It is one of our TESS EB candidates whose light curve is shown in Figure \ref{fig:bestEB}, and
we show its periodogram 
in \cite{VargasSalazar2024},
which indicates a period of 0.99 d.  If 
it is confirmed as an EB, 
then the companion is a non-compact star with $M_2 \lesssim 9.30 M_{\odot}$.
We require more RV information to extract further mass constraints.

\subsection{Target [M2002] 81646} \label{App:81646}

Classification: OB, RV binary

This target has 13 RV 
measurements: 4 from IMACS and 9 from M2FS.
The M2FS RV estimates are obtained from the cross-correlation fits. We broaden the model
spectrum template
to match its $v\sin i = 120$ km/s \citep{DorigoJones2020}. The target is identified as an RV binary by both 
the $\Delta$RV and $F$-test
methods.

This target's primary mass is $28.0  M_{\odot}$ (Table \ref{table:OBbinaryresults}). Based on our RV measurements
and given that the primary is an OB star, the companion could be a non-compact star
with mass of $3.4 - 3.7 M_{\odot}$.
The companion could also be a BH, or possibly, a NS.

\subsection{Target [M2002] 81696} \label{App:81696}

Classification: OB, RV binary

This target has 10 RV 
measurements: 4 from IMACS and 6 from M2FS.
The M2FS RV measurements are obtained from the cross-correlation fits. The target is identified as an RV binary by both 
the $\Delta$RV and $F$-test
methods.

This target's primary mass is $15.0  M_{\odot}$ (Table \ref{table:OBbinaryresults}). Based on our RV measurements, 
and given that the primary is an OB star, 
the companion could be a non-compact star with mass of $3.1 - 4.0 M_{\odot}$ 
The companion could also be a BH of unconstrained mass, or possibly, a NS.
For these constraints, the period should be $\lesssim 1$ day with an inclination $i \gtrsim 60^{\circ}$. 

\subsection{Target [M2002] 81941} \label{App:81941}

Classification: OB, RV binary

This target has 10 RV measurements: 3 from IMACS and 7 from M2FS.
The M2FS RV measurements are
obtained from the cross-correlation fits. We broaden the model template spectrum to match its $v\sin i = 127$ km/s \citep{DorigoJones2020}. The target is identified as an RV binary by both 
the $\Delta$RV and $F$-test
methods.

This target's primary mass is $25.0  M_{\odot}$ (Table \ref{table:OBbinaryresults}). It is listed as a pulsating variable in the Gaia DR3 database \citep{GaiaDR3}. Based on our RV mass estimates, the companion should have a mass of at least $5.29 M_{\odot}$. However, our  detection limits should have detected a non-compact star of at least $3.9 M_{\odot}$ in our spectra, indicating that we have an  unseen companion. This could be a BH or possibly a stripped star. 

\subsection{Target [M2002] 82328} \label{App:82328}

Classification: OBe, fast rotator, RV binary candidate

This target has 9 RV 
measurements: 4 from IMACS and 5 from M2FS.
The M2FS RV measurements are
obtained
by performing the 
model cross-correlation fits on just the He lines. 
Since this is a fast rotator reported at $v\sin i = 200$ km/s \citep{DorigoJones2020}, 
we apply this default broadening to the model template.
However, 
the line widths appear to vary, perhaps due to stellar precession, and
different broadening values are
used 
($v\sin i$): 
JD 2457943.783 ($185$ km/s), 
JD 2457945.881 ($220$ km/s), 
JD 2458346.885 ($200$ km/s), 
JD 2458349.755 ($210$ km/s), and
JD 2458351.715 ($220$ km/s). 
We could not extract an RV measurement from JD 2457944.789 due to
its signal-to-noise and this observations is dropped from the analysis and not included in the total number of observations.
The target is identified as an RV binary by 
the $\Delta$RV
method only.

This target's primary mass is $19.5  M_{\odot}$ (Table \ref{table:OBebinaryresults}). Our RV and mass constraints 
yield a
lowest companion mass 
estimate of
$0.57 M_{\odot}$. 
If the companion is a compact object, the upper mass limit is unconstrained and it could be a BH or NS. If the companion is a non-compact star, it should be $\lesssim 4.20 M_{\odot}$ based on our detection limit. Additionally, this OBe system has a low eccentricity and it is therefore also 
possible that it has 
a stripped star companion.

\subsection{Target [M2002] 82444} \label{App:82444}

Classification: OB, fast rotator, RV binary candidate

The target has 9 RV measurements: 4 from IMACS and 5 from M2FS.
The M2FS RV measurements are obtained
from the cross-correlation fits. 
The spectral lines 
of the model template are broadened to $v\sin i = 240$ km/s \citep{DorigoJones2020}. The target is identified as an RV binary by 
the $\Delta$RV
method only.

This target's primary mass is $16.9  M_{\odot}$ (Table \ref{table:OBbinaryresults}). 
Since the target is an OB star, 
the companion could be a non-compact star with mass of $0.62 - 3.2 M_{\odot}$.
For these constraints, the period should be $\lesssim 50$ days.  On the other hand, fast rotators are 
included in our BSS sample, and thus
the companion also could be a compact object. Further research is required to determine the true nature of the companion. 

\subsection{Target [M2002] 82711} \label{App:82711}

Classification: OBe, HMXB, RV binary

This target is also known as XSP 1062 and it is 
an HMXB 
with the supernova remnant still remaining around it
\citep{Haberl2012,HenaultBrunet2012}. The target has a known NS companion, but with no mass measurement. Studies 
show it to have the third longest period for HMXBs in the SMC, at 656 days \citep{SchmidtkeCowUd2012,SchmidtkeCowUd2019,Gvaramadze2021}. 
It is also predicted
to have low eccentricity $e < 0.2$ and an inclination of $73^{\circ}$ \citep{Cappallo2020}. 

We have 8 RV 
measurements: 3 from IMACS and 5 from M2FS.
The M2FS RV measurements are obtained from cross-correlation fits using the observation on
JD 2457945.881 as an RV template instead of the PoWR spectra. We performed the cross-correlation fits with only the H$\gamma$ 4340 \AA\ line instead of both Balmer lines. The target is identified as an RV binary by both 
the $\Delta$RV and $F$-test
methods. 

This target's primary mass is $17.3  M_{\odot}$ (Table \ref{table:OBebinaryresults}). We use the RV measurements and 
constraints on eccentricity and inclination above
from the literature to constrain the companion
mass and period. 
These do not yield reasonable companion mass values
($M_{\rm 2,min} = 67.0 M_{\odot}$). However, if we let the period be a free parameter and use $e < 0.2$ and inclination of $73^{\circ}$, we 
can obtain masses of
$1.5 - 3.0 M_{\odot}$. For these new constraints, the period would have to be $\lesssim 6$ days, contrary to 
the long period reported in the literature.  Since we know that there is a NS companion, this suggests that the long period may not correspond to its orbital period and may be due to some other phenomenon.

\subsection{Target [M2002] 83073} \label{App:83073}

Classification: OB, SB2 candidate

This target has 6 spectroscopic observations from M2FS. It has a reported $v\sin i = 124$ km/s \citep{DorigoJones2020},
which we use to broaden the model spectral template for the cross-correlations. 
We do not detect any RV variation using our RV binary identification methods. 

The target's status as an SB2 candidate is established in part from noting that the  4102 \AA\  line
may be redshifted relative to the RV standard for epochs JD 2458346.885 and JD 2458351.715. Additionally, for epoch JD 2458351.715, there may be a redshift in the \hei\ 4120 \AA\ line relative to H$\delta$. 

This target's primary mass is $13.1  M_{\odot}$ (Table \ref{table:OBbinaryresults}). If this SB2 were to be confirmed, then our signal-to-noise threshold indicates that a non-compact stellar companion should have mass $\sim 3.10 M_{\odot}$. 

\subsection{Target [M2002] 83171} \label{App:83171}

Classification: OBe, fast rotator, EB candidate, RV binary, eccentric binary

This target has 8 spectroscopic observations: 3 from IMACS and 5 from M2FS. The target is considered to be a fast rotator with a $v\sin i = 238$ km/s \citep{DorigoJones2020} and this broadening is used in the M2FS cross-correlation fits.
We use the spectrum taken on
JD 2457945.881 as the template RV standard for the RV cross-correlation. 
The target is identified as an RV binary by both 
the $\Delta$RV and $F$-test
methods.

This target's primary mass is $24.4  M_{\odot}$ (Table \ref{table:OBebinaryresults}). Our TESS analysis identifies it as an EB candidate. 
We estimate that the companion mass should be 
$0.92 - 3.2 M_{\odot}$ and the periodogram indicates that the system should have a period of $\sim 2.0$ days. The target's companion mass estimates are shown in Figure \ref{fig:EBCircex} and its periodogram is given by \cite{VargasSalazar2024}.

Alternative explanations for the photometric behavior shown in the light curve are that it may originate from the circumstellar disk of the OBe star rather than from eclipse by a stellar companion \citep[e.g.,][]{Gaudin2024,Coe2015,Maggi2013} or due to geometric distortions caused by a compact object or He star. 
Thus although we include this star as an EB candidate, it may not be a strong one.

\subsection{Target [M2002] 83224} \label{App:83224}

Classification: OBe, fast rotator, SB2 candidate, RV binary, eccentric binary

This target has 9 RV measurements: 3 from IMACS and 6 from M2FS.
The target is a fast rotator with a $v\sin i = 159$ km/s \citep{DorigoJones2020} and this broadening is used in the M2FS cross-correlation fits. 
We use the observation taken on
JD 2457945.881 as the template RV standard for the RV cross-correlation of the 4340 \AA\ line. The target is identified as an RV binary by both 
the $\Delta$RV and $F$-test
methods.

The target's status as an SB2 candidate is established in part from possible Doppler shifting of the H$\gamma$ 4340 \AA\ line from the 4387 \AA\ line. This is suggested for most of our observations but may be more prevalent on JD 2458349.755.

This target's primary mass is $15.8  M_{\odot}$ (Table \ref{table:OBebinaryresults}). If this SB2 were to be confirmed, then our S/N threshold indicates that a non-compact stellar companion should be near the upper limit of our estimated  $M_2 = 0.53 - 4.0 M_{\odot}$. For these constraints, the period should be $\lesssim 100$ days.

\subsection{Target [M2002] 83232} \label{App:83232}

Classification: OB, EB candidate, RV binary

This target has 6 spectroscopic observations: 3 from IMACS and 3 from M2FS.
For the M2FS fits, the model template lines are broadened to 
the observed
$v\sin i = 147$ km/s \citep{DorigoJones2020}
for the cross-correlations. 
The target is identified as an RV binary by both 
the $\Delta$RV and $F$-test
methods.

It is listed as a pulsating variable in the Gaia DR3 \citep{GaiaDR3}, and 
we identify it as an EB candidate using our TESS analysis.  Its light curve is shown in Figure~\ref{fig:TESS_R1}.  As discussed in Section~\ref{sec:TESSEB}, it is not an especially strong EB candidate.
This target's primary mass is $12.8  M_{\odot}$ (Table \ref{table:OBbinaryresults}).
If it were to be confirmed as an EB, then the companion is a non-compact star of $1.3 - 2.8 M_{\odot}$. Our light curve's periodogram (\cite{VargasSalazar2024}) indicates a period of 1.68 days.

\subsection{Target [M2002] 83510} \label{App:83510}

Classification: OB, fast rotator, RV binary

This target has 9 spectroscopic observations: 3 from IMACS and 6 from M2FS. 
It is considered a fast rotator and for the M2FS cross-correlation fits
the model template lines are broadened to 
the observed
$v\sin i = 217$ km/s \citep{DorigoJones2020}. The target is identified as an RV binary by both 
the $\Delta$RV and $F$-test
methods.

This target's primary mass is $20.9  M_{\odot}$ (Table \ref{table:OBbinaryresults}). If the companion is a non-compact star, it should have a mass of $1.9 - 2.9 M_{\odot}$.
For these parameters, the period should be $\lesssim 2$ days with an inclination $i \gtrsim 45^{\circ}$.
On the other hand, fast rotators are included in our BSS binary sample, thus the companion also could be a
NS.  
Further research is required to determine the true nature of the companion.

\bibliography{Vargas_Salazar2024}{}
\bibliographystyle{aasjournal}

\end{document}